\documentclass[10pt]{article}
\usepackage{authblk}
\usepackage{a4wide}
\usepackage{url}
\usepackage{appendix}

\usepackage{graphicx}
\usepackage{amsmath}
\usepackage{amsthm}
\usepackage{amssymb}
\usepackage{makeidx}

\usepackage[usenames]{color}
\usepackage[inference]{semantic}
\usepackage{subfig}
\usepackage{url}
\usepackage{calc}
\usepackage{pstricks}
\usepackage{floatrow}
\usepackage{nicefrac}
\usepackage{msc}
\usepackage[normalem]{ulem}
\usepackage{centernot}
\usepackage{multibib}
\newcites{web}{Website References}
\newcites{own}{Publications}
\newcites{add}{*} 
 
\usepackage{hyperref}

\index{verification|see{formal verification}}
\index{model|see{formal model}}
\index{requirement|see{security requirement}}

\newcommand{\note}[2]{
  \noindent {\color{blue} {\sf {\bf [#1]} #2} }
}
\renewcommand{\note}[2]{\error{There are still notes in the document}}

\newtheorem{definition}{Definition}
\newtheorem{example}{Example}
\newtheorem{theorem}{Theorem}

\def\addsymbol #1: #2#3{$#1$ \> \parbox{5in}{#2 \dotfill \pageref{#3}}\\}
\def\symbol#1{\label{#1}} 

\makeatletter

\newcommand{\Rmnum}[1]{\expandafter\@slowromancap\romannumeral #1@}
\makeatother
\newcommand{\one}{\Rmnum{1}}
\newcommand{\two}{\Rmnum{2}}

\newcommand{\event}[1]{\bar{#1}}
\newcommand{\fu}[1]{{\sf{#1}}}
\newcommand{\mi}[1]{\ensuremath{\mathit{#1}}}
\newcommand{\name}[1]{{\tt{#1}}}
\newcommand{\rul}[1]{\textsc{#1}}
\newcommand{\type}[1]{{\rm{#1}}}
\newcommand{\variable}[1]{\mi{#1}}

\newcommand{\channel}{\name{c}}
\newcommand{\dataset}{\mi{\widetilde{mc}}}

\newcommand{\prs}{\mi{P}}

\newcommand{\dlvprotocol}{\mi{\prs_\mi{DLV08}}}
\newcommand{\patientCred}{\mi{Cred_{\thepatient}}}
\newcommand{\xpatientCred}{{\variable{c\_Cred_{\thepatient}}}}
\newcommand{\xpatientCredph}{{\variable{c_{\thepharm}\_Cred_{\thepatient}}}}
\newcommand{\xpatientCredmpa}{{\variable{c_{\thempa}\_Cred_{\thepatient}}}}
\newcommand{\xpatientCredhii}{{\variable{c_{\thehii}\_Cred_{\thepatient}}}}
\newcommand{\patientIDv}{{\variable Id_{\thepatient}}}
\newcommand{\xpatientPseudompa}{{\variable{c_{\thempa}\_Pnym_{\thepatient}}}}
\newcommand{\xpatientPseudohii}{{\variable{c_{\thehii}\_Pnym_{\thepatient}}}}
\newcommand{\patientHiiv}{{\mi{Hii}}}
\newcommand{\xpatientHiimpa}{{\variable{c_{\thempa}\_\patientHiiv}}}
\newcommand{\patientSSSv}{{\mi{Sss}}}
\newcommand{\xpatientSSSph}{{\variable{c_{\thepharm}\_\patientSSSv}}}
\newcommand{\patientCommit}{\mi{Comt_{\thepatient}}}
\newcommand{\xpatientCommit}{{\variable{c\_Comt_{\thepatient}}}}
\newcommand{\xpatientCommitr}{{\variable{c\_Comt^{r}_{\thepatient}}}}
\newcommand{\xpatientCommitl}{{\variable{c\_Comt^{l}_{\thepatient}}}}
\newcommand{\xpatientCommitph}{{\variable{c_{\thepharm}\_Comt_{\thepatient}}}}
\newcommand{\xpatientCommitmpa}{{\variable{c_{\thempa}\_Comt_{\thepatient}}}}
\newcommand{\patientOpenInfo}{{\name{r}_\mi{\thepatient}}}
\newcommand{\patientspk}{\mi{PtSpk}}
\newcommand{\patientspkr}{\mi{PtSpk^r}}
\newcommand{\patientspkzr}{\mi{PtSpk^{zr}}}
\newcommand{\patientspkf}{\mi{PtSpk^{f}}}
\newcommand{\patientspkzl}{\mi{PtSpk^{zl}}}
\newcommand{\patientspkl}{\mi{PtSpk^l}}
\newcommand{\xpatientspkph}{{\variable rcv_{\thepharm}\_\patientspk}}
\newcommand{\xpatientProof}{{\variable{rcv\_\patientProof}}}

\newcommand{\patientAuth}{\mi{Auth_{\thepatient}}}
\newcommand{\xpatientAuth}{{\variable{rcv\_Auth_{\thepatient}}}}
\newcommand{\patientAuthsss}{\mi{PtAuthSss}}
\newcommand{\xpatientAuthsss}{{\variable{rcv\_\patientAuthsss}}}
\newcommand{\xdoctorPseudo}{{\variable{c\_Pnym_{\thedoctor}}}}
\newcommand{\xdoctorPseudompa}{{\variable{c_{\thempa}\_Pnym_{\thedoctor}}}}
\newcommand{\doctorCred}{\mi{Cred_{\thedoctor}}}

\newcommand{\xdoctorCred}{{\variable{c\_Cred_{\thedoctor}}}}
\newcommand{\xdoctorCredph}{{\variable{c_{\thepharm}\_Cred_{\thedoctor}}}}
\newcommand{\xdoctorCredmpa}{{\variable{c_{\thempa}\_Cred_{\thedoctor}}}}
\newcommand{\doctorCommit}{\mi{Comt_{\thedoctor}}}
\newcommand{\doctorCommitr}{\mi{Comt^{r}_{\thedoctor}}}
\newcommand{\doctorCommitl}{\mi{Comt^{l}_{\thedoctor}}}
\newcommand{\xdoctorCommit}{{\variable{c\_Comt_{\thedoctor}}}}
\newcommand{\xdoctorCommitph}{{\variable{c_{\thepharm}\_Comt_{\thedoctor}}}}
\newcommand{\xdoctorCommitmpa}{{\variable{c_{\thempa}\_Comt_{\thedoctor}}}}
\newcommand{\xdoctorOpenInfo}{{\variable{rcv\_r_{\thedoctor}}}}
\newcommand{\xprescText}{{\variable{c\_\prescTextv}}}
\newcommand{\xprescTextph}{{\variable{c_{\thepharm}\_\prescTextv}}}
\newcommand{\xprescTextmpa}{{\variable{c_{\thempa}\_\prescTextv}}}
\newcommand{\prescID}{\mi{PrescriptID}}
\newcommand{\xprescID}{{\variable{c\_\prescID}}}
\newcommand{\xprescIDph}{{\variable{c_{\thepharm}\_\prescID}}}
\newcommand{\xprescIDmpa}{{\variable{c_{\thempa}\_\prescID}}}
\newcommand{\xprescIDhii}{{\variable{c_{\thehii}\_\prescID}}}
\newcommand{\receptionAck}{\mi{ReceiptAck}}
\newcommand{\mpaID}{{\name{Id}_{\thempa}}}

\newcommand{\pkphpt}{\variable{rcv_{\thepatient}\_pk_\thepharm}}
\newcommand{\pkphmpa}{\variable{rcv_{\thempa}\_pk_\thepharm}}

\newcommand{\xpharmID}{{\variable{c_{\thepatient}\_Id_{\thepharm}}}}
\newcommand{\xpharmIDmpa}{{\variable{c_{\thempa}\_Id_{\thepharm}}}}
\newcommand{\xpharmIDhii}{{\variable{c_{\thehii}\_Id_{\thepharm}}}}
\newcommand{\proc}{\mi{P}}
\newcommand{\ProPt}{\proc_{\thepatient}}
\newcommand{\ProPtpartone}{\proc'_{\thepatient}}
\newcommand{\ProPtparttwo}{\proc''_{\thepatient}}
\newcommand{\ProDr}{\proc_{\thedoctor}}
\newcommand{\ProPh}{\proc_{\thepharm}}
\newcommand{\ProPhpartone}{\proc'_{\thepharm}}
\newcommand{\ProPhparttwo}{\proc''_{\thepharm}}

\newcommand{\ProMPApartone}{\proc_{\thempa}}
\newcommand{\ProMPAparttwo}{\proc'_{\thempa}}
\newcommand{\ProHII}{\proc_{\thehii}}
\newcommand{\init}{\mi{init}}
\newcommand{\main}{\mi{main}}
\newcommand{\Role}{\mi{R}}

\newcommand{\prescProof}{\mi{PrescProof}}
\newcommand{\xprescProof}{{\variable{rcv\_\prescProof}}}
\newcommand{\xprescProofph}{{\variable{rcv_{\variable{ph}}\_\prescProof}}}
\newcommand{\xprescProofmpa}{{\variable{rcv_{\variable{mpa}}\_\prescProof}}}

\newcommand{\vc}{\mi{vc}}
\newcommand{\C}{\mi{c}}

\newcommand{\invoicev}{\mi{invoice}}
\newcommand{\xinvoice}{{\variable{rcv\_Invoice}}}
\newcommand{\xinvoicempa}{{\variable{rcv_{\thempa}\_Invoice}}}
\newcommand{\doctorAuth}{\mi{Auth_{\thedoctor}}}
\newcommand{\xdoctorAuth}{{\variable{rcv\_Auth_{\thedoctor}}}}

\newcommand{\xpharmAuth}{{\variable{rcv\_Auth_{\thepharm}}}}
\newcommand{\xpharmAuthmpa}{{\variable{rcv_{\thempa}\_Auth_{\thepharm}}}}
\newcommand{\xnonce}{{\variable{c\_msg}}}
\newcommand{\xEnc}{{\variable{c\_Enc}}}
\newcommand{\xEncmpa}{{\variable{c_{\thempa}\_Enc}}}
\newcommand{\xEnchii}{{\variable{c_{\thehii}\_Enc}}}
\newcommand{\xvc}{{\variable{rcv\_vc}}}
\newcommand{\xvcmpa}{{\variable{rcv_{\thempa}\_vc}}}
\newcommand{\xvchii}{{\variable{c_{\thehii}\_vc}}}
\newcommand{\xc}{{\variable{rcv\_c}}}
\newcommand{\xcmpa}{{\variable{rcv_{\thempa}\_c}}}
\newcommand{\xchii}{{\variable{c_{\thehii}\_c}}}
\newcommand{\xReceptionAck}{{\variable{rcv\_ReceiptAck}}}
\newcommand{\xReceptionAckmpa}{{\variable{rcv_{\thempa}\_ReceiptAck}}}
\newcommand{\xReceptionAckhii}{{\variable{rcv_{\thehii}\_ReceiptAck}}}
\newcommand{\ReceptionAck}{\mi{ReceiptAck}}
\newcommand{\xmpaAuth}{{\variable{rcv\_Auth_{\thempa}}}}
\newcommand{\xmpaAuthhii}{{\variable{rcv_{\thehii}\_Auth_{\thempa}}}}
\newcommand{\xhiiAuthmpa}{{\variable{rcv_{\thempa}\_Auth_{\thehii}}}}
\newcommand{\xpkph}{{\mi{rcv\_pk_{\thepharm}}}}

\newcommand{\pkhiipt}{\mi{c_{\thepatient}\_pk_\thehii}}
\newcommand{\npkhiipt}{\mi{c_{\thepatient}\_npk_\thehii}}
\newcommand{\wpkhiipt}{\mi{c_{\thepatient}\_wpk_\thehii}}
\newcommand{\pkmpaph}{\mi{rcv_{\thepharm}\_pk_\thempa}}
\newcommand{\mpaIDph}{\mi{c_{\thepharm}\_Id_{\thempa}}}
\newcommand{\pkmpahii}{\mi{c_{\thehii}\_pk_{\thempa}}}
\newcommand{\pkhiimpa}{\mi{c_{\thempa}\_pk_{\thehii}}}
\newcommand{\pkmpapt}{\mi{c_{\thepatient}\_pk_{\thempa}}}
\newcommand{\xmpaID}{\mi{c_{\thepatient}\_Id_{\thempa}}}
\newcommand{\xmpaIDhii}{\mi{rcv_{\thehii}\_Id_{\thempa}}}

\newcommand{\eHealthProtocol}{\mi{\prs_{\!eh}}}
\newcommand{\prescProofl}{\ensuremath{\prescProof^{l}}}
\newcommand{\doctorIDv}{{\mi{Id_{\thedoctor}}}}
\newcommand{\prescTextv}{\mi{presc}}

\newcommand{\pubkey}{\mi{pk_x}}
\newcommand{\vProDr}{\mi{\main_{\thedoctor}}}
\newcommand{\vProDrb}{\mi{\main^b_{\thedoctor}}}
\newcommand{\ProDrc}{\mi{\proc^c_{\thedoctor}}}  
\newcommand{\vProDri}{\mi{\main^i_{\thedoctor}}}
\newcommand{\vProDrl}{\mi{\main^1_{\thedoctor}}}
\newcommand{\vProDrh}{\mi{\main^h_{\thedoctor}}}
\newcommand{\npatientHii}{\mi{nHii}}
\newcommand{\wpatientHii}{\mi{wHii}}

\newcommand{\Prs}{\mi{Q}}

\newcommand{\drA}{\name{d_A}}
\newcommand{\drB}{\name{d_B}}
\newcommand{\ptA}{\name{t_A}}
\newcommand{\ptB}{\name{t_B}}

\newcommand{\doctorID}{{\name{Id}_{\thedoctor}}}
\newcommand{\doctorIDatt}{{\name{Id}^{a}_{\thedoctor}}}
\newcommand{\doctorOpenInfo}{\name{r}_{\thedoctor}}
\newcommand{\doctorOpenInfor}{\name{r}^{r}_{\thedoctor}}
\newcommand{\doctorOpenInfol}{\name{r}^{l}_{\thedoctor}}
\newcommand{\doctorPseudo}{{\name{Pnym}_{\thedoctor}}}
\newcommand{\vdoctorPseudo}{{\variable{Pnym}_{\thedoctor}}}

\newcommand{\hiiID}{\name{Id}_{\thehii}}
\newcommand{\hiiSk}{\name{sk}_{\thehii}}
\newcommand{\patientAcc}{{\name{Acc}}}
\newcommand{\patientHii}{{\name{Hii}}}
\newcommand{\patientID}{{\name{Id}_{\thepatient}}}
\newcommand{\patientProof}{\mi{PtProof}}
\newcommand{\patientPseudo}{{\name{Pnym}_{\thepatient}}}
\newcommand{\patientPseudoB}{{\name{Pnym}^{B}_{\thepatient}}}

\newcommand{\patientSSS}{{\name{Sss}}}

\newcommand{\ndoctorID}{{\name{nId}_{\thedoctor}}}
\newcommand{\ndoctorPseudo}{\name{nPnym}_{\thedoctor}}
\newcommand{\npatientAcc}{\name{nAcc}}
\newcommand{\npatientID}{\name{nId}_{\thepatient}}
\newcommand{\npatientPseudo}{\name{nPnym}_{\thepatient}}
\newcommand{\npatientSSS}{\name{nSss}}
\newcommand{\wdoctorID}{{\name{wId}_{\thedoctor}}}
\newcommand{\wdoctorPseudo}{\name{wPnym}_{\thedoctor}}
\newcommand{\wpatientAcc}{\name{wAcc}}
\newcommand{\wpatientID}{\name{wId}_{\thepatient}}
\newcommand{\wpatientPseudo}{\name{wPnym}_{\thepatient}}
\newcommand{\wpatientSSS}{\name{wSss}}

\newcommand{\prescText}{\name{presc}}
\newcommand{\pharmSk}{\name{sk}_{\thepharm}}
\newcommand{\ssoSk}{\name{sk}_{\thesso}}
\newcommand{\mpaSk}{\name{sk}_{\thempa}}
\newcommand{\noncedr}{\name{n}_{\thedoctor}}
\newcommand{\noncedrr}{\name{n}^{r}_{\thedoctor}}
\newcommand{\noncedrl}{\name{n}^{l}_{\thedoctor}}

\newcommand{\nonce}{\name{nonce}}

\newcommand{\privchhiipt}{\name{ch}_{hp}}

\newcommand{\privchmpaph}{\name{ch}_{mp}}
\newcommand{\privchphpt}{\name{ch}_{phpt}}

\newcommand{\privchdrpt}{\name{ch}_{dp}}
\newcommand{\privchptph}{\name{ch}_{ptph}}
\newcommand{\chc}{\name{chc}}
\newcommand{\ch}{\name{ch}}
\newcommand{\prescProofr}{\name{\prescProof^{\mi{r}}}}
\newcommand{\prescProofzr}{\name{\prescProof^{\mi{zr}}}}
\newcommand{\prescProofzl}{\name{\prescProof^{\mi{zl}}}}
\newcommand{\doctorPseudor}{\name{Pnym}^{r}_{\thedoctor}}

\newcommand{\doctorPseudol}{\name{Pnym}^{l}_{\thedoctor}}

\newcommand{\pkmpa}{\name{pk}_\thempa}
\newcommand{\pksso}{\name{pk}_\thesso}
\newcommand{\pkhii}{\name{pk}_\thehii}
\newcommand{\pkph}{\name{pk}_\thepharm}
\newcommand{\pharmID}{\name{Id}_{\thepharm}}
\newcommand{\pa}{\name{p_A}}
\newcommand{\pal}{\name{p^1_A}}
\newcommand{\pah}{\name{p^h_A}}

\newcommand{\pb}{\name{p_B}}
\newcommand{\pbl}{\name{p^1_B}}
\newcommand{\pbh}{\name{p^h_B}}

\newcommand{\pubchannel}{\name{ch}}

\newcommand{\forwardchannel}{{\name{chc}}}

\newcommand{\nbn}{\fu{bn}}
\newcommand{\nbv}{\fu{bv}}
\newcommand{\ndec}{\fu{dec}}
\newcommand{\ndomain}{\fu{domain}}
\newcommand{\nenc}{\fu{enc}}
\newcommand{\nfchannel}{\fu{Channel}}
\newcommand{\nfn}{\fu{fn}}
\newcommand{\nfrm}{\fu{frame}}
\newcommand{\nfv}{\fu{fv}}
\newcommand{\nout}{\fu{out}}
\newcommand{\nreadin}{\fu{in}}
\newcommand{\f}{\fu{f}}

\newcommand{\fsign}{\sf sign}
\newcommand{\fdec}{\ensuremath{\sf dec}}

\newcommand{\nsign}{{\sf sign}}
\newcommand{\commit}{{\sf commit}}
\newcommand{\pk}{{\sf pk}}

\newcommand{\getmsg}{{\sf getmsg}}
\newcommand{\true}{{\sf true}}

\newcommand{\open}{{\sf open}}

\newcommand{\frtname}{{\sf first}}
\newcommand{\pair}{{\sf pair}}
\newcommand{\fourth}{{\sf fourth}}

\newcommand{\invoice}{{\sf invoice}}
\newcommand{\patientspkVer}{\ensuremath{\mbox{\sf Vfy-{spk}}_{\sf PtSpk}}}
\newcommand{\patientProofVer}{\ensuremath{\mbox{\sf Vfy-{zk}}_{\sf PtProof}}}

\newcommand{\patientAuthVer}{\ensuremath{\mbox{\sf Vfy-{zk}}_{\sf Auth_{\thepatient}}}}
\newcommand{\patientAuthsssVer}{\ensuremath{\mbox{\sf Vfy-{zk}}_{\sf PtAuthSss}}}
\newcommand{\drcred}{{\sf drcred}}
\newcommand{\ptcred}{{\sf ptcred}}
\newcommand{\hash}{{\sf hash}}
\newcommand{\fenc}{\ensuremath{\sf enc}}

\newcommand{\inv}[1]{{\sf inv}(#1)}
\newcommand{\zk}{{\sf zk}}

\newcommand{\spk}{{\sf spk}}

\newcommand{\zkver}{\mbox{{\sf Vfy-{zk}}}}
\newcommand{\comt}{{\sf commit}}
\newcommand{\prescProofVer}{\ensuremath{\mbox{\sf Vfy-{spk}}_{\sf PrescProof}}}
\newcommand{\doctorAuthVer}{\ensuremath{\mbox{\sf Vfy-{zk}}_{\sf Auth_{\thedoctor}}}}

\newcommand{\getpublic}{{\sf getpublic}}
\newcommand{\spkver}{\mbox{{\sf Vfy-{spk}}}}
\newcommand{\getpubmsg}{{\sf getSpkVinfo}}
\newcommand{\checkAuth}{\ensuremath{\mbox{\sf Vfy-{sign}}}}
\newcommand{\checkPharmAuth}{\ensuremath{\mbox{\sf Vfy-{sign}}}}
\newcommand{\getsignedmessage}{{\sf getsignmsg}}
\newcommand{\CheckVEncHii}{\ensuremath{\mbox{\sf Vfy-{venc}}_{\sf Hii}}}
\newcommand{\CheckVEncDrnymMpa}{\ensuremath{\mbox{\sf Vfy-{venc}}_{\sf DrnymMpa}}}
\newcommand{\CheckVEncPtnym}{\ensuremath{\mbox{\sf Vfy-{venc}}_{\sf Ptnym}}}
\newcommand{\CheckReceptionAck}{\ensuremath{\mbox{\sf Vfy-{spk}}_{\sf
	ReceiptAck}}}
\newcommand{\checkMpaAuth}{\ensuremath{\mbox{\sf Vfy-{sign}}}}
\newcommand{\checkHiiAuth}{\ensuremath{\mbox{\sf Vfy-{sign}}}}
\newcommand{\host}{{\sf host}}
\newcommand{\key}{{\sf key}}
\newcommand{\g}{{\sf g}}

\newcommand{\vencver}{\mbox{{\sf Vfy-{venc}}}}

\newcommand{\bn}[1]{\nbn(#1)}
\newcommand{\bv}[1]{\nbv(#1)}
\newcommand{\dec}[2]{\ndec(#1, #2)}
\newcommand{\domain}[1]{\ndomain(#1)}
\newcommand{\dsub}[4]{\mi{\{#1/#2,#3/#4\}}}
\newcommand{\enc}[2]{\nenc(#1, #2)}
\newcommand{\fchannel}[1]{\nfchannel\mi{\langle#1\rangle}}
\newcommand{\fn}[1]{\nfn(#1)}
\newcommand{\frm}[1]{\nfrm(#1)}
\newcommand{\fv}[1]{\nfv(#1)}
\newcommand{\out}[2]{\nout(#1, #2)}
\newcommand{\readin}[2]{\nreadin(#1, #2)}

\newcommand{\fst}[1]{{\sf fst}(#1)}
\newcommand{\snd}[1]{{\sf snd}(#1)}
\newcommand{\sign}[2]{\nsign(#1, #2)}

\newcommand{\frt}[1]{\frtname(#1)}

\newcommand{\third}[1]{{\sf third}(#1)}

\newcommand{\atype}{\mi{\omega}}

\newcommand{\nsub}{\sigma}

\newcommand{\signature}{\Sigma}

\newcommand{\induc}{\rightarrow}
\newcommand{\lduc}[1]{\xrightarrow{#1}}
\newcommand{\correspondenceto}{==>}

\newcommand{\context}[1]{{\mathcal{C}}[#1]}
\newcommand{\contexti}[1]{{\mathcal{C}}_i[#1]}

\newcommand{\contextc}{\mi{\mathcal{C}}}
\newcommand{\contexthealth}[1]{\mathcal{C}_\mi{eh}[#1]}

\newcommand{\defi}{:=}
\newcommand{\eqtheory}{E}
\newcommand{\relation}{\mathcal{R}}

\newcommand{\sub}[2]{\{#1/#2\}}
 
\newcommand{\substitution}[2]{\mathit{\{#1/#2\}}}
\newcommand{\msub}[4]{\{#1/#2,\cdots, #3/#4\}}

\newcommand{\vect}[1]{\tilde{#1}}

\newcommand{\hole}{\_}
\newcommand{\longhole}{\rule{0.3cm}{0.5pt}}

\renewcommand{\eqref}[1]{({\tt{#1}})}

\newcommand{\eq}{\approx_{\ell}}
 
\newcommand{\eqe}{=_\eqtheory}
\newcommand{\seq}{\approx_{s}}
\newcommand{\steq}{\equiv}

\newcommand{\structeq}{\equiv}

\newcommand{\hateq}{\mi{\ \hat{=}\ }}

\newcommand{\ALIAS}{\rul{ALIAS}}
\newcommand{\COMM}{\rul{COMM}}
\newcommand{\ELSE}{\rul{ELSE}}
\newcommand{\IN}{\rul{IN}}
\newcommand{\NEWC}{\rul{NEW}-\rul{C}}
\newcommand{\NEWnull}{\rul{NEW}-0}
\newcommand{\NEWPar}{\rul{NEW}-\rul{PAR}}
\newcommand{\OPENATOM}{\rul{OPEN}-\rul{ATOM}}
\newcommand{\OUTATOM}{\rul{OUT}-\rul{ATOM}}
\newcommand{\PAR}{\rul{PAR}}
\newcommand{\PARA}{\rul{PAR}-\rul{A}}
\newcommand{\PARC}{\rul{PAR}-\rul{C}}
\newcommand{\PARnull}{\rul{PAR}-0}
\newcommand{\REPL}{\rul{REPL}}
\newcommand{\REWRITE}{\rul{REWRITE}}
\newcommand{\SCOPE}{\rul{SCOPE}}
\newcommand{\SUBST}{\rul{SUBST}}
\newcommand{\STRUCT}{\rul{STRUCT}}
\newcommand{\THEN}{\rul{THEN}}

\newcommand{\docunlink}{prescription privacy}
\newcommand{\docunlinksh}{presc.~priv.}
\newcommand{\Docunlink}{Prescription privacy}
\newcommand{\docrf}{receipt-freeness}
\newcommand{\docrfsh}{receipt-freeness}
\newcommand{\Docrf}{Receipt-freeness}
\newcommand{\docrfm}{multi-session receipt-freeness}

\newcommand{\docindep}{independency of \docunlink}

\newcommand{\docindepshshort}{ind. of \docunlinksh}
\newcommand{\Docindep}{Independency of \docunlink}
\newcommand{\docrfindep}{independency of \docrf}

\newcommand{\docrfindepshshort}{ind. of enf. \docunlinksh}
\newcommand{\Docrfindep}{Independency of \docrf}

\newcommand{\thepatient}{\mi{pt}}
\newcommand{\thedoctor}{\mi{dr}}
\newcommand{\thepharm}{\mi{ph}}
\newcommand{\thempa}{\mi{mpa}}
\newcommand{\thesso}{\mi{sso}}
\newcommand{\thehii}{\mi{hii}}

\newcommand{\eHealth}{e-health }

\newcommand{\tlet}{\mbox{let}\ }
\newcommand{\letin}{\ \mbox{in} }
\newcommand{\tif}{\mbox{if}\ }
\newcommand{\then}{\ \mbox{then} }

\newcommand{\telse}{\ \mbox{else} }

\newcommand{\fun}{\mi{fun}\ }
\newcommand{\funnosp}{\mi{fun}}
\newcommand{\reduc}{\mi{reduc\ }}
\newcommand{\reducnosp}{\mi{reduc}}

\newcommand{\free}{\mi{free}}

\newcommand{\EndDr}{{\sf EndDr}}
\newcommand{\StartDr}{{\sf StartDr}}
\newcommand{\StartPt}{{\sf StartPt}}
\newcommand{\EndPt}{{\sf EndPt}}
\newcommand{\EndPh}{{\sf EndPh}}
\newcommand{\StartPh}{{\sf StartPh}}
\newcommand{\EndPtph}{{\sf EndPtph}}
\newcommand{\StartPtph}{{\sf StartPtph}}

\newcommand{\choice}[2]{{\sf choice}[#1, #2]}
\newcommand{\private}{\mi{private}}
\newcommand{\tequation}{\mi{equation}}
\newcommand{\eventf}[1]{\event{f}(#1)}
\newcommand{\eventg}[1]{\event{g}(#1)}
\newcommand{\ok}{\mi{\surd}}
\newcommand{\fail}{\mi{\times}}
\newlength{\fminilength}
\newsavebox{\fminibox}
\newenvironment{fmini}[1][\linewidth]
  {\setlength{\fminilength}{#1-1.5ex}
   \vspace{-1ex}\noindent\begin{lrbox}{\fminibox}
   \begin{minipage}{\fminilength}
   \mbox{}\hfill\vspace{-1.5ex}}
  {\end{minipage}\end{lrbox}\vspace{0ex}\hspace{0ex}
   \framebox{\usebox{\fminibox}}}

\newenvironment{specification}
{\noindent\tt\begin{fmini}\begin{tabbing}X\=X12345\=XXXX\=XXXX\=XXXX\=XXXX\=XXXX
\=\+\kill}
{\end{tabbing}\normalfont\end{fmini}}

\begin{document}
\title{Formal Analysis of an E-Health Protocol}
\author{Naipeng Dong $^{\mathrm{a, }}$\thanks{Supported by a grant from the Fonds National de la Recherche
	(Luxembourg).
}
}
\author{Hugo Jonker $^\mathrm{b}$}
\author{Jun Pang $^\mathrm{c, }$
}
\affil{\small
$^a$ School of Computing, National University of Singapore, Singapore\\
$^b$ Department of Computer Science, Open University, The Netherlands\\
$^c$ Faculty of Science, Technology and Communication \\ Interdisciplinary Centre for Security, Reliability and Trust,\\
         University of Luxembourg, Luxembourg
}

\date{}
\maketitle

\begin{abstract}
\newcommand{\keyword}{\textbf{Keyword}}
\noindent
Given the sensitive nature of health data, security and privacy in e-health systems is
of prime importance. It is crucial that an e-health system must ensure that users remain
private -- even if they are bribed or coerced to reveal themselves, or
others: a pharmaceutical company could, for example, bribe a pharmacist
to reveal information which breaks a doctor's privacy. In this paper,
we first identify and formalise several new but important privacy properties
on enforcing doctor privacy. Then we analyse the security and privacy of a complicated and 
practical e-health protocol (DLV08). Our analysis uncovers ambiguities in
the protocol, and shows to what extent these new privacy properties as
well as other security properties (such as secrecy and authentication)
and privacy properties (such as anonymity and untraceability) are satisfied by
the protocol. Finally, we address the found ambiguities which result in
both security and privacy flaws, and propose suggestions for fixing them. 
 
\bigskip \noindent\keyword: E-health systems, formal verification, applied
pi, secrecy, authentication, anonymity, privacy, enforced privacy, untraceability \end{abstract}

\section{Introduction}
\label{sec:intro}

The inefficiency of traditional paper-based health care and advances in
information and communication technologies, in particular cloud
computing, mobile, and satellite communications, constitute the ideal
environment to facilitate the development of widespread electronic
health care (e-health for short) systems. E-health systems are
distributed health care systems using devices and computers which
communicate with each other, typically via the Internet. E-health
systems aim to support secure sharing of information and resources
across different health care settings and workflows among different
health care providers. The services of such systems 
are intended to be more secure, effective, efficient
and timely than the currently existing health care systems. 

Given the sensitive nature of health data, handling this data must meet
strict security and privacy requirements. In traditional health care
systems, this is normally implemented by controlling access to the
physical documents that contain the health care data. Security and
privacy are then satisfied, assuming only legitimate access is possible
and assuming that those with access do not violate security or privacy. 

However, the introduction of e-health systems upends this approach. The
main benefit of e-health systems is that they facilitate digital
exchange of information amongst the various parties involved. This has
two major consequences: first, the original health care data is shared
digitally with more parties, such as pharmacists and insurance
companies; and second, this data can be easily shared by any of those
parties with an outsider. Clearly, the assumption of a trusted network
can no longer hold in such a setting. Given that it is trivial for a
malicious entity to intercept or even alter digital data in transit,
access control approaches to security and privacy are no longer
sufficient. Therefore, we must consider security and privacy of the involved
parties with respect to an outsider, the Dolev-Yao
adversary~\cite{DY83}, who controls the communication network (i.e.,~the
adversary can observe, block, create and alter information). 
Communication security against such an adversary is mainly achieved by
employing cryptographic communication protocols. Cryptography is also
employed to preserve and enforce privacy, which prevents problems such
as prescription bribery.

It is well known that designing such protocols is error-prone: time and
again, flaws have been found in protocols that claimed to be secure
(e.g., electronic voting systems~\cite{BT94,LK00} have been
broken~\cite{HS00,LK02}). Therefore, we must require that security and privacy
claims of an \eHealth protocol are verified before the protocol is used in
practice. Without verifying that a protocol satisfies its security and privacy
claims, subtle flaws may go undiscovered.

In order to objectively verify whether a protocol satisfies its claimed security and
privacy requirements, each requirement must be formally defined as a property. 
Various security and privacy properties have already been defined in the
literature, such as secrecy, authentication, anonymity and
untraceability. We refer to these properties as \emph{regular} security
and privacy properties. While they are necessary to ensure security and
privacy, by themselves these regular properties are not sufficient.
Benaloh and Tuinstra pointed out the risk of subverting a
voter~\cite{BT94} to sell her vote. The idea of coercing or bribing a
party into nullifying their privacy is hardly considered in the
literature of \eHealth systems (notable exceptions
include~\cite{Matyas98,DLVV08}). However, this concept impacts \eHealth
privacy: for example, a pharmaceutical company could bribe doctors to
prescribe only their medicine. Therefore, we cannot only consider
privacy with respect to the Dolev-Yao adversary. To fully evaluate privacy
of \eHealth systems, we must also consider this new aspect of privacy in
the presence of an active coercer -- someone who is bribing or
threatening parties to reveal private information. We refer to this new
class of privacy properties as enforced privacy properties. In
particular, we identify the following regular and enforced privacy 
properties~\cite{DJP12} to counter doctor bribery: \emph{\docunlink}: a
doctor cannot be linked to his prescriptions; \emph{\docrf}: a doctor cannot 
prove his prescriptions to the adversary for preventing doctor bribes; 
\emph{\docindep}: third parties cannot help the adversary to link a doctor to 
the doctor's prescriptions for preventing others to reduce a doctor's
\docunlink; and \emph{\docrfindep}: a doctor and third parties cannot prove 
the doctor's prescriptions to the adversary for preventing anyone from affecting a
doctor's \docrf.

\medskip\noindent{\bf Contributions.}
We identify three enforced privacy properties in \eHealth systems and are
the first to provide formal definitions for them.
In addition, we develop an in-depth applied pi model of the DLV08
\eHealth protocol~\cite{DLVV08}. As this protocol was designed for
practical use in Belgium, it needed to integrate with the existing health
care system. As such, it has become a complicated system with many involved
parties, that relies on complex cryptographic primitives to achieve a
multitude of goals. 
We formally analyse privacy and enforced privacy properties
of the protocol, as well as regular security properties.
We identify ambiguities in the protocol description that cause both security and privacy
flaws, and propose suggestions for fixing them.
The ProVerif code of modelling and full analysis of the DLV08 protocol can be found
in~\cite{DJPa12}.

\medskip\noindent{\bf Remark.}
This article is a revised and extended version of~\cite{DJP12} that appears in the proceedings 
of the 17th European Symposium on Research in Computer Security (ESORICS'12). 
In this version we have added
(1) the full formal modelling of the DLV08 protocol in the applied pi calculus (see Section~\ref{sec:dlvmodel}); 
(2) the detailed analysis of secrecy and authentication properties of the protocol (see Section~\ref{sec:analyse}); and
(3) details of the analysis of privacy properties of the protocol which are not
described in the conference paper~\cite{DJP12} (see Section~\ref{sec:analyse}).
In addition, it contains an overview on privacy and enforced privacy in e-health systems (see Section~\ref{sec:privacy}) and
a brief but complete description of the applied pi calculus (see Section~\ref{sec:appliedpi}).

\section{Privacy and enforced privacy in e-health}
\label{sec:privacy}

Ensuring privacy in e-health systems has been recognised as a necessary
prerequisite for adoption such systems by the general public~\cite{MRS06,KAB09}.
However, due to the complexity of e-health settings, existing privacy
control techniques, e.g.,~formal privacy methods, from domains such as e-voting (e.g.,~\cite{DKR09,JMP09})
and e-auctions (e.g.,~\cite{DJP11}) do not carry over directly.
In e-voting and e-auctions, there is a natural division into two types
of roles: participants (voters, bidders) and authorities (who run the
election/auction). In contrast, e-health systems have to deal with a
far more complex constellation of roles, including doctors, patients,
pharmacists, insurance agencies, oversight bodies, etc. These roles
interact in various ways with each other, requiring private data of one
another, which makes privacy even more complex.

Depending on the level of digitalisation, health care systems have
different security requirements. If electronic devices are only used to
store patient records, then ensuring privacy mainly requires local
access control. On the other hand, if data is communicated over a
network, then communication privacy becomes paramount.
Below, we sketch a typical situation of using a health care system,
indicating what information is necessary where. This will help to gain
an understanding for the interactions and interdependencies between the
various roles.

Typically, a patient is examined by a doctor, who then prescribes
medicine. The patient goes to a pharmacist to get the medicine. The
medicine is reimbursed by the patient's health insurance, and the
symptoms and prescription of the patient may be logged with a medical
research facility to help future research.

This overview hides many details. The patient may possess medical
devices enabling her to undergo the examination at home, after which the
devices digitally communicate their findings to a remote doctor. The
findings of any examination (by doctor visit or by digital devices) need
to be stored in the patients health record, either electronic or on
paper, which may be stored at the doctor's office, on a server in the
network, on a device carried by the patient, or any combination of
these. Next, the doctor returns a prescription, which also needs to be
stored. The pharmacist needs to know what medicine is required, which is
privacy-sensitive information. Moreover, to prevent abuse of medicine,
the pharmacist must verify that the prescription came from an authorised
doctor, is intended for this patient, and was not fulfilled before. On
top of that, the pharmacist may be allowed (or even required) to
substitute medicine of one type for another (e.g., brand medicine for
generic equivalents), which again must be recorded in the patient's
health record. For reimbursement, the pharmacist or the patient
registers the transaction with the patient's health insurer. In
addition, regulations may require that such information is stored (in
aggregated form or directly) for future research or logged
with government agencies.
Some health care systems allow emergency access to health data, which
complicates privacy matters even further.
Finally, although a role may need to have access to privacy-sensitive
data of other roles, this does not mean that he is trusted to ensure the
privacy of those other roles. For instance, a pharmacist may sell his
knowledge about prescription behaviour to a pharmaceutical company.

From the above overview on e-health systems,
we can conclude that existing approaches to ensuring privacy from other domains deal with far
simpler division of roles, and they are
not properly equipped to handle the role diversity present in e-health
systems. Moreover, they do not address the influence of other roles on
an individual's privacy. Therefore, current privacy approaches cannot be lifted
directly, but must be redesigned specifically for the e-health domain.

In the following discussions, we focus on the privacy of the main
actors in health care: patient privacy and doctor privacy. Privacy of
roles such as pharmacists does not impact on the core process in health
care, and is therefore relegated to future work. We do not consider
privacy of roles performed by public entities such as insurance
companies, medical administrations, etc.

\subsection{Related work}
\label{ssec:patient}

The importance of patient privacy in e-health is traditionally seen as
vital to establishing a good doctor-patient relationship. This is even
more pertinent with the emergence of the Electronic Patient
Record~\cite{Anderson96}. A necessary early
stage of e-health is to transform the paper-based health care process
into a digital process. The most important changes in this stage are
made to patient information processing, mainly health care records.
Privacy policies are the de facto standard to properly express privacy
requirements for such patient records. There are three main
approaches to implement these requirements: access
control, architectural design, and the use of cryptography.

\paragraph{Patient privacy by access control.}
The most obvious way to preserve privacy of electronic health care
records is to limit access to these records.
The need for access control is supported by several privacy threats to
personal health information listed by Anderson~\cite{Anderson96}.
Controlling access is not as straightforward as it sounds though: the
need for access changes dynamically (e.g., a doctor only needs access to
records of patients that he is currently treating). Consequently, there
exists a wide variety of access control approaches designed for patient
privacy in the literature, from simple access rules
(e.g.,~\cite{Anderson96}), to consent-based access rules
(e.g.,~\cite{Louwerse98}), role-based access control (RBAC)
(e.g.,~\cite{RCHS03}), organisation based access control
(e.g.,~\cite{KBMBCSBDT03}), etc.

\paragraph{Patient privacy by architectural design.}
E-health systems cater to a number of different roles, including
doctors, patients, pharmacists, insurers, etc. Each such role has its
own sub-systems or components. As such, \eHealth systems can be
considered as a large network of systems, including administrative
system components, laboratory information systems, radiology information
systems, pharmacy information systems, and financial management systems.
Diligent architectural design is an essential step to make such a
complex system function correctly. Since privacy is important in
\eHealth systems, keeping privacy in mind when designing the
architecture of such systems is a promising path towards ensuring
privacy~\cite{SV09}. Examples of how to embed privacy constraints in the
architecture are given by the architecture of wireless sensor networks
in e-health~\cite{KLSSTW10}, proxies that may learn location but not
patient ID~\cite{MKDH09}, an architecture for cross-institution image
sharing in e-health~\cite{CHCK07}, etc.

\paragraph{Cryptographic approaches to patient privacy.}
Cryptography is necessary to ensure private communication
between system components over public channels (e.g.,~\cite{BB96}).
For example, Van der Haak et al.~\cite{HWBDWW03} use digital signatures
and public-key authentication (for access control) to
satisfy legal requirements for cross-institutional exchange of
electronic patient records.
Ateniese et al.~\cite{ACMD03} use pseudonyms to preserve 
patient anonymity, and enable a user to transform statements
concerning one of his pseudonyms into statements concerning one of
his other pseudonyms (e.g., transforming a prescription for the pseudonym
used with his doctor to a prescription for the pseudonym used with the
pharmacist).
Layouni et al.~\cite{LVSDV09} consider communication between health
monitoring equipment at a patient's home and the health care centre.
They propose a protocol using wallet-based credentials (a cryptographic
primitive) to let patients control when and how much identifying
information is revealed by the monitoring equipment.
More recently, De Decker et al.~\cite{DLVV08} propose a health care system
for communication between insurance companies and administrative bodies
as well as patients, doctors and pharmacists. Their system relies on
various cryptographic primitives to ensure privacy,
including zero-knowledge proofs,
signed proofs of knowledge,
and bit-commitments. We will explain this system
in more detail in Section~\ref{sec:dlv08}.

\paragraph{Doctor privacy.} \label{ssec:doctor}
A relatively understudied aspect is that of doctor privacy.
Maty\'{a}\v{s}~\cite{Matyas98} investigates the problem of enabling
analysis of prescription information while ensuring doctor privacy. His
approach is to group doctors, and release the data per group, hiding who
is in the group. He does not motivate a need for doctor privacy,
however. Two primary reasons for doctor privacy have been identified in
the literature: (1) (Ateniese et al.~\cite{ACMD03}) to safeguard doctors
against administrators setting specific efficiency metrics on their
performance (e.g.,~requiring the cheapest medicine be used, irrespective
of the patient's needs). To address this, Ateniese et
al.~\cite{AM02,ACMD03}~propose an anonymous prescription system that
uses group signatures to achieve privacy for doctors; (2) (De Decker et al.~\cite{DLVV08})
to prevent a pharmaceutical company from bribing a doctor to prescribe
their medicine. A typical scenario can be described as follows. A
pharmaceutical company seeks to persuade a doctor to favour a certain
kind of medicine by bribing or coercing. To prevent this, a doctor
should not be able to prove which medicine he is prescribing to this
company (in general, to the adversary). This implies that doctor
privacy must be enforced by e-health systems. De Decker et al. also
note that preserving doctor privacy is not sufficient to prevent
bribery: pharmacists could act as intermediaries, revealing the doctor's
identity to the briber, as pharmacists often have access to
prescriptions, and thus know something about the prescription behaviour
of a doctor. This observation leads us to formulate a new but important requirement of
\emph{\docindep} in this paper: no third party should be able to help the adversary
link a doctor to his prescription.

\subsection{Observations}
\label{ssec:observations}
Current approaches to privacy in e-health,
as witnessed from the literature study in Section~\ref{ssec:patient},
mostly focus on patient
privacy as an access control or authentication problem. Even though
doctor privacy is also a necessity, research into ensuring
doctor privacy is still in its infancy. We
believe that doctor privacy is as important as patient privacy
and needs to be studied in more depth. 
It is also clear from the analysis that privacy in e-health systems needs to
be addressed at different layers: access control ensures privacy at the service
layer; privacy by architecture design addresses privacy concerns at the
system/architecture layer; use of cryptography guarantees privacy
at the communication layer. 
Since e-health systems are complex~\cite{TG09}
and rely on correct communications between many sub-systems,
we study privacy in e-health as a communication problem. In
fact, message exchanges in communication protocols may leak information
which leads to a privacy breach~\cite{Low96,CKS04,DKR09}.

Classical privacy properties, which are well-studied in the literature, 
attempt to ensure that privacy can be \emph{enabled}. However, merely
enabling privacy is insufficient in many cases: for such cases, a system must
\emph{enforce} user privacy instead of allowing the user to pursue it.
One example is doctor bribery. To avoid doctor bribery, we take into
account enforced privacy for doctors.
In addition, we consider that one party's privacy may depend on another
party (e.g., in the case of a pharmacist revealing prescription
behaviour of a doctor). In these cases, others can cause (some) loss of
privacy. Obviously, ensuring privacy in such a case requires more from
the system than merely enabling privacy.
Consequently, we propose and study the following privacy properties for doctors in 
communication protocols in the e-health domain,
in addition to regular security and privacy properties as we mentioned before
in Section~\ref{sec:intro}.
\begin{description}
\item[\docunlink:] A protocol preserves \docunlink\ if the adversary cannot link
		a doctor to his prescriptions.
\item[\docrf:] A protocol satisfies \docrf\ if a doctor cannot prove his 
		prescriptions to the adversary.
\item[\docindep:] A protocol ensures \docindep\ if third parties cannot help 
		the adversary to link a doctor to the doctor's prescriptions.
\item[\docrfindep:] A protocol ensures \docrfindep\ if a doctor cannot prove 
		his prescriptions to the adversary given that third parties 
		sharing information with the adversary.
\end{description}

\section{Formalisation of privacy properties}
\label{sec:property}

In order to formally verify properties of a protocol, the
protocol itself as well as the properties need to be formalised.
In this section, we focus on the formalisation of key privacy properties,
while the formalisation of secrecy and authentication properties
can be considered standard as studied in the literature~\cite{Low96,Blanchet01}.
Thus secrecy and authentication properties are introduced later in the case
study (Section~\ref{ssec:proverif}) and are omitted in this section.

We choose the formalism of the applied pi calculus,
due to its capability in expressing equivalence based properties
which is essential for privacy,
and automatic verification supported by the tool ProVerif~\cite{Blanchet01}.
The applied pi calculus is introduced in Section~\ref{sec:appliedpi}.
Next, in Section~\ref{sec:ehealthprotocol}, we show how to model
e-health protocols in the applied pi calculus. Then, from
Section~\ref{sec:docunlink} to Section~\ref{sec:docrfindep}, we
formalise each of the privacy properties described in the end of Section~\ref{ssec:observations}.
Finally, in Sections~\ref{sec:anonymity} and~\ref{sec:untraceability},
we consider (strong) anonymity and (strong) untraceability in e-health,
respectively. These concepts have been formally studied in
other domains (e.g.,~\cite{SS96,DMR08,BHM08,KT09,ACRR10,KTV10}), and thus are only briefly introduced in this section. 

\subsection{The applied pi calculus}
\label{sec:appliedpi}

The applied pi calculus is a language for modelling and analysing concurrent 
systems, in particular cryptographic protocols.
The following (mainly based on~\cite{AF01,RS10}) briefly introduces its syntax, semantics and equivalence relations.

\subsubsection{Syntax}
The calculus assumes an infinite set of \emph{names}, 
which are used to model communication channels and other atomic data, 
an infinite set of \emph{variables}, which are used to 
model received messages, and a signature $\signature$ consisting of a finite 
set of \emph{function symbols}, 
which are used to model cryptographic primitives. Each function symbol has an arity.
A function symbol with arity zero is a constant.
\emph{Terms} (which are used to model messages) are defined as 
names, variables, or function symbols 
applied to terms (see Figure~\ref{fig:term}). 
\begin{figure}[!ht]
\caption{Terms in the applied pi calculus.}
\begin{specification}
{
\begin{math}
\begin{array}{@{}l@{}l@{}}
M, N, T 
      ::= &\mbox{terms}\\
\quad \name{a}, \name{b}, \name{c}, 
      \name{m}, \name{n}, \ldots 
      &\hspace{24pt} \mbox{names}\\
\quad x, y, z
      &\hspace{24pt}\mbox{variables}\\
\quad \f(M_1, \ldots, M_\ell) & \hspace{24pt}\mbox{function application}
\end{array}
\end{math}
}
\end{specification}
\label{fig:term}
\end{figure}
\begin{example}[function symbols and terms]
Typical function symbols are \nenc\ with arity 2 for encryption, 
\ndec\ with arity 2 for decryption. The term for encrypting $x$ with a key $k$ is $\enc{x}{k}$.
\end{example}
The applied pi calculus assumes a sort system for terms. 
Terms can be of a base type
(e.g., \type{Key} or a universal base type \type{Data}) or type 
\fchannel{\atype} where \atype\ is a type. 
A variable and 
a name can have any type. 
A function symbol can only be 
applied to, and return, terms of base type.
Terms are assumed to be well-sorted and substitutions preserve types. 

Terms are often equipped with an equational theory
$\eqtheory$ -- a set of equations on terms. 
The equational theory is normally used to capture features of 
cryptographic primitives.
The equivalence relation induced by $\eqtheory$ is 
denoted as $\eqe$. 
\begin{example}[equational theory]
The behaviour of symmetric encryption and decryption can be captured by the 
following equation:
\[
\dec{\enc{x}{k}}{k}\eqe x,
\] 
where $x$ and $k$ are variables.
\end{example}

Systems are described as processes: plain processes
and extended processes
(see Figure~\ref{fig:grammar}).
\begin{figure}[!ht]
\caption{Processes in the applied pi calculus.}
\begin{specification}
{
\begin{math}
\begin{array}{@{}l@{}l@{}}
P, Q, R ::=
      & \hspace{-24pt} \mbox{plain processes} \\
\quad 0       & \mbox{null process} \\
\quad P\mid Q & \mbox{parallel composition} \\
\quad !P      & \mbox{replication} \\
\quad \nu \name{n}.P & \mbox{name restriction}\\
\quad \tif M\eqe N \then\ P \telse\ Q\quad & \mbox{conditional} \\
\quad \readin{v}{x}.P  & \mbox{message input} \\
\quad \out{v}{M}.P     & \mbox{message output} \\
\\
A, B, C ::=  
      & \hspace{-24pt}\mbox{extended processes} \\
\quad P         &  \mbox{plain process} \\
\quad A\mid B   &  \mbox{parallel composition} \\
\quad \nu \name{n}.A & \mbox{name restriction} \\
\quad \nu x.A        & \mbox{variable restriction} \\
\quad \sub{M}{x} & \mbox{active substitution}
\end{array}
\end{math}
}
\end{specification}
\label{fig:grammar}
\end{figure}
\noindent
In Figure~\ref{fig:grammar}, $M, N$ are terms, $\name{n}$ is 
a name, $x$ is a 
variable and $v$ is a 
metavariable, standing either for a name 
or a variable. The null process $0$ does nothing. The parallel composition 
$P \mid Q$ represents the sub-process $P$ and the sub-process $Q$ running in 
parallel. The replication $!P$ represents an infinite number of process $P$ 
running in parallel. The name restriction $\nu \name{n}.P$ binds the name 
$\name{n}$ in the process $P$, which means the name $\name{n}$ is secret to the 
adversary. The conditional evaluation $M \eqe N$ represents equality over the 
equational theory 
rather than strict syntactic identity. The message input 
$\readin{v}{x}.P$ reads a message from channel $v$, and binds the message to 
the variable $x$ in the following process $P$. The message output 
$\out{v}{M}.P$ sends the message $M$ on the channel $v$, and then runs the 
process $P$. In both of these cases we may omit $P$ when it is $0$.
Extended processes add variable restrictions and active 
substitutions. 
The variable restriction $\nu x.A $ binds the variable $x$ in the process $A$. 
The active substitution $\sub{M}{x}$ replaces variable $x$ with term
$M$ in any process that it contacts with.
We say a process is sequential if it does not involve using the parallel composition $P\mid Q$,
replication $!P$, conditional, or active substitution. 
That is, a sequential process is either null or constructed
using name/variable restriction, message input/output.
In addition, applying syntactical substitution (i.e., ``$\tlet x=N \letin$'' in ProVerif input language) to 
a sequential process still results in a sequential process.
For simplicity of presentation,
we use $\nu \tilde{a}$ as an abbreviation for $\nu a_1.\nu a_2.\cdots.\nu a_n$,
where $\tilde{a}$ is the sequence of names $a_1, a_2,\cdots,a_n$.
We also use the abbreviation
$P.Q$ to represent the process $\mi{action}_1.\cdots.\mi{action}_n.Q$, where $P:=\mi{action}_1.\cdots.\mi{action}_n$; 
and an $\mi{action}_i$
is of the form $\nu \name{n}$, $\nu x$, $\readin{v}{x}$, $\out{v}{M}$ or $\tlet x=N \letin\ \mi{action}_j$.
The intuition of $P.Q$ is that when a process consists of a sequential sub-process $P$ followed by a sub-process $Q$, we 
write the process in an abbreviated manner as $P.Q$.
In addition, we use the abbreviation $\msub{M_1}{x_1}{M_n}{x_n}$ to 
represent $\sub{M_1}{x_1}\cdots\sub{M_n}{x_n}$.

Names and variables have scopes.
A name is \emph{bound}
if it is under restriction.
A variable is \emph{bound} by restrictions or inputs.
Names and variables are \emph{free} if they are not
delimited by restrictions or by inputs. 
The sets of free names,
free variables, 
bound names and bound variables
of a process $A$ are denoted as $\fn{A}$, $\fv{A}$, $\bn{A}$ and
$\bv{A}$, respectively. 
A term is \emph{ground}
when it does not contain variables. 
A process is \emph{closed}
if it does not contain free variables. 

\begin{example}[processes]
Consider a protocol in which $A$ generates a nonce $\name{m}$, encrypts
the nonce with a secret key $\name{k}$, then sends the encrypted message
to $B$. Denote with $\Prs_A$ the process modelling the behaviour of $A$,
with $\Prs_B$ the process modelling the behaviour of $B$, and the whole
protocol by $\Prs$:
\[\begin{array}{ll}
\Prs_A	& \defi \nu \name{m}.\out{\pubchannel}{\enc{\name{m}}{\name{k}}} \\
\Prs_B	& \defi \readin{\pubchannel}{x} \\ 
\Prs	& \defi  \nu \name{k}.(\Prs_A \mid \Prs_B)
\end{array}
\]
Here, $\pubchannel$ is a free name representing a
public channel. Name $\name{k}$ is bound in process $\Prs$; name
$\name{m}$ is bound in 
process $\Prs_A$. Variable $x$ is bound in process $\Prs_B$.
\end{example}

A \emph{frame} is defined as an extended process built up 
from $0$ and active substitutions by parallel composition and restrictions.
The active substitutions in extended processes allow us to map an 
extended process $A$ to its 
frame $\frm{A}$ by replacing every plain process in $A$ with $0$.
The \emph{domain} of a frame $B$, denoted as
$\domain{B}$, is the set of variables for 
which the frame defines a substitution and which are not under a restriction.

\begin{example}[frames]\label{ex:frame}
The frame of the process $\nu \name{m}. (\out{\pubchannel}{x}\mid 
\sub{\name{m}}{x})$, denoted as $\frm{\nu \name{m}.\linebreak 
(\out{\pubchannel}{x}\mid \sub{\name{m}}{x})}$ is $\nu \name{m}.(0\mid \sub{\name{m}}{x})$. 
The domain of this frame, denoted as 
$\domain{\nu \name{m}. (0\mid \sub{\name{m}}{x})}$ is $\{x\}$.
\end{example}
A \emph{context}
$\context{\hole}$ is defined as a
process with a hole, which may be filled with any process. 
An evaluation context
is a context whose hole is not under a replication, a 
condition, an input or an output. 
\begin{example}[context]
Process $\nu \name{k}.(\Prs_A \mid\hole)$ is an evaluation context.
When we fill the hole with process $\Prs_B$, we obtain the process $\nu
\name{k}.(\Prs_A \mid \Prs_B)$, which is the process $Q$.
\end{example}
%

\subsubsection{Operational semantics}

The operational semantics of the applied pi calculus is defined by: 
1)~structural equivalence ($\steq$),
2)~internal reduction ($\induc$), and
3)~labelled reduction ($\lduc{\alpha}$) of processes.

1) Intuitively, two processes are structurally equivalent if they model the same 
thing but differ in structure. Formally, structural equivalence
of processes is the smallest equivalence relation on extended process that is closed by 
$\alpha$-conversion on names and variables, 
by application of evaluation contexts as shown in Figure~\ref{fig:structualequivalence}.
\begin{figure}[!ht]
\caption{Structural equivalence in the applied pi calculus.}
\begin{specification}
{
\begin{math}
\begin{array}{@{}l@{}r@{\ }c@{\ }l@{}l}
\PARnull & A \mid 0 & \steq & A \\
\PARA    & A \mid (B \mid C) & \steq & (A \mid B) \mid C \\
\PARC    & A \mid B & \steq & B \mid A \\
\REPL    & !P & \steq & P \mid\, !P \\
\SUBST   & \sub{M}{x} \mid A & \steq & \sub{M}{x} \mid A\sub{M}{x} \\

\NEWnull & \nu u. 0 & \steq & 0 \\
\NEWC    & \nu u. \nu v. A & \steq & \nu v. \nu u. A \\
\NEWPar \qquad  & A \mid \nu v. B & \steq & \nu v. (A \mid B) 
         & \tif v \not\in \fn{A} \cup \fv{A} \\
\ALIAS   & \nu x. \sub{M}{x} & \steq & 0 \\
\REWRITE & \sub{M}{x} & \steq & \sub{N}{x} 
         & \tif M \eqe N
\end{array}
\end{math}
}
\end{specification}
\label{fig:structualequivalence}
\end{figure}

2) Internal reduction is the 
smallest relation on extended processes 
closed under structural equivalence, 
application of evaluation of contexts as shown in Figure~\ref{fig:internalrealtion}.
\begin{figure}[!ht]
\caption{Internal reduction in the applied pi calculus.}
\begin{specification}
{
\begin{math}
\begin{array}{@{}l@{}l@{}}
\COMM \quad & \out{\channel}{x}. P \mid \readin{\channel}{x}. Q\ \induc\ P \mid Q \\
\THEN       & \tif N \eqe N \then\ P \telse\ Q\ \induc\ P \\
\ELSE \quad & \tif M \eqe N \then\ P \telse\ Q\ \induc\ Q\\
& \quad \text{for ground terms} \ M, N \ \text{where} \ M \not\eqe N 
\end{array}
\end{math}
}
\end{specification}
\label{fig:internalrealtion}
\end{figure}

3) The labelled reduction models the environment interacting with the processes.
It defines a relation $A\lduc{\alpha} A'$ as in Figure~\ref{fig:labelledreduction}.
The label $\alpha$ is either reading a term 
from the process's environment, or sending a name
or a variable of base type to the environment. 
\begin{figure}[!ht]
\caption{Labelled reduction in the applied pi calculus.}
\begin{specification}
{
\begin{math}
\begin{array}{@{}l@{}c@{}}
\IN        & \readin{\channel}{x}. P \lduc{\readin{\channel}{M}} P\sub{M}{x} \\
\OUTATOM   & \out{\channel}{v}. P \lduc{\out{\channel}{v}} P \\
\OPENATOM \qquad & 
\frac{\displaystyle A \lduc{\out{\channel}{v}} A' \quad v \neq \channel}
     {\displaystyle \nu v. A \lduc{\nu v. \out{\channel}{v}} A'}\\
&\\
\SCOPE & 
\frac{\displaystyle A \lduc{\alpha} A' \quad v\ \text{does not occur in}\ \alpha}
     {\displaystyle \nu v. A \lduc{\alpha} \nu v. A'} \\
&\\     
\PAR   & 
\frac{\displaystyle A \lduc{\alpha} A' \quad \bv{\alpha} \cap 
      \fv{B}=\bn{\alpha} \cap \fn{B} = \emptyset}
     {\displaystyle A \mid B \lduc{\alpha} A' \mid B} \\
&\\     
\STRUCT \quad & 
\frac{\displaystyle A \steq B \quad B \lduc{\alpha} B' \quad A' \steq B'}
     {\displaystyle A \lduc{\alpha} A'}
\end{array}
\end{math}
}
\end{specification}
\label{fig:labelledreduction}
\end{figure}

\subsubsection{Equivalences}

The applied pi calculus defines \emph{observational equivalence} and 
\emph{labelled bisimilarity} to model the
indistinguishability of two processes by the adversary. It is proved
that the two relations coincide, when active substitutions are of base type~\cite{AF01,Liu11}. 
We mainly use the
labelled bisimilarity for the convenience of proofs. 
Labelled bisimilarity
is based on \emph{static equivalence}: 
labelled bisimilarity compares the dynamic behaviour of processes, while 
static equivalence compares their static states (as represented by their 
frames).
\begin{definition}[static equivalence]
Two terms $M$ and $N$ are equal in the frame $B$, written as $(M \eqe N)B$, iff
there exists a set of restricted names $\vect{\name{n}}$ and a substitution 
$\nsub$ such that 
$B \steq \nu \vect{\name{n}}. \nsub$, $M\nsub \eqe N\nsub$ and 
$\vect{\name{n}} \cap (\fn{M} \cup \fn{N}) = \emptyset$.  

Closed frames $B$ and $B'$ 
are statically equivalent, 
denoted as $B \seq B'$, if \\
(1) $\domain{B}=\domain{B'}$; \\
(2) $\forall$ terms $M,N$: $(M \eqe N)B$ iff
$(M \eqe N)B'$.

Extended processes $A$, $A'$ are statically 
equivalent, denoted as $A \seq A'$,
if their frames are statically equivalent:
$\frm{A} \seq \frm{A'}$.
\label{def:static_equivalence}
 \end{definition}
\begin{example}[equivalence of frames~\cite{AF01}]
The frame $B$ and the frame $B'$, are equivalent.
However, the two frames are not equivalent to frame $B''$,
because the adversary can discriminate $B''$ by testing 
$y\eqe \f(x)$.  
\[
\begin{array}{lcl}
B&\defi&\nu M. \sub{M}{x}\mid \nu N. \sub{N}{y}\\
B'&\defi&\nu M. (\sub{\f(M)}{x}\mid \sub{\g(M)}{y})\\
B''&\defi&\nu M. (\sub{M}{x}\mid \sub{\f(M)}{y})
\end{array}
\]
where $\f$ and $\g$ are two function symbols without equations.
\end{example}
\begin{example}[static equivalence]
Process $\sub{M}{x}\mid\Prs_1$ is statically equivalent
to process $\sub{M}{x}\mid\Prs_2$ where $\Prs_1$ and $\Prs_2$ are two 
closed plain process,
because the frame of the two processes are
statically equivalent, i.e., $\sub{M}{x}\seq\sub{M}{x}$.
\end{example}	
\begin{definition}[labelled bisimilarity]
\emph{Labelled bisimilarity} $(\eq)$ is the largest
symmetric relation $\relation$ on closed extended processes,
such that
$A\, \relation\, B$ implies:\\
(1) $A \seq B$;\\
(2) if $A \induc A'$ then $B \induc^* B'$ and 
    $A'\, \relation\, B'$ for some $B'$;\\
(3) if $A \lduc{\alpha} A'$ and
	$\fv{\alpha} \subseteq \domain{A}$ and
	$\bn{\alpha} \cap \fn{B}=\emptyset$; then
	$B\induc^*\lduc{\alpha}\induc^* B'$
	and $A'\, \relation\, B'$ for some $B'$,
	where * denotes zero or more.
\label{def:labelled_bisimilarity}	
\end{definition}

\subsection{E-health protocols}
\label{sec:ehealthprotocol}

In the existing e-voting and (sealed bid) e-auction protocols, where bribery and coercion have been
formally analysed using the applied pi calculus (see e.g.,~\cite{DKR09,DJP11}), the number of
participants is determined \emph{a priori}.
In contrast with these protocols, e-health systems should be
able to handle newly introduced participants (e.g., patients).
To this end, we model user-types and each user-type can be instantiated infinite times.

\paragraph{Roles.} An e-health protocol can be specified by a set of roles, each of which
is modelled as a process, $\proc_1, \ldots, \proc_n$.
Each role specifies the behaviour of the user taking this
role in an execution of the protocol. By instantiating the free variables in
a role process, we obtain the process of a specific user taking
the role.

\paragraph{Users.} Users taking a role can be modelled
by adding settings (identity, pseudonym, encryption key, etc.) to the process representing the role, that is $\init_i.\proc_i$,
where $\init_i$ is a sequential process
which generates names/terms modelling the data of the user (e.g., `$\nu \pharmSk.
\tlet \pkph=\pk(\pharmSk)$' in Figure~\ref{fig:composeph}),
reads in setting data from channels (e.g., `$\readin{\privchhiipt}{\patientHii}$'
in Figure~\ref{fig:composept}), or reveals data to the adversary
(e.g., `$\out{\ch}{\pkph}$' in Figure~\ref{fig:composeph}). A user taking a role
multiple times is captured by
add replication to the role process, i.e., $\init_i.!\proc_i$. A user may also take
multiple roles. When the user uses two different settings in different roles, the user
is treated as two separate users. If the user uses shared setting in multiple
roles, the user process is modelled as the user setting sub-process followed by
the multiple role processes in
parallel, e.g., $\init_{k}.!(\proc_i\mid\proc_j)$ when the user takes two roles $\proc_i$
and $\proc_j$,.

\paragraph{User-types.} Users taking a specific
role, potentially multiple times, belong to a user-type.
Hence, a user-type is modelled as $\Role_i:=\init_i.!\proc_i$. The set of users of a type is captured
by adding replication to the user-type process, i.e., $!\Role_i$.
A protocol with $n$ roles naturally forms $n$ user-types.
In protocols where users are allowed to take multiple
roles with one setting, we consider these users form
a new type. For example, a challenge-response
protocol, which specifies two roles -- a role \emph{Initiator} and a role \emph{Responder},
has three user-types - the \emph{Initiator}, the \emph{Responder} and
users taking both \emph{Initiator} and \emph{Responder}, assuming that a
user taking both roles with the same setting is allowed. A user-type with
multiple roles is modelled as $\init_{k}.!(\proc_i\mid\ldots\mid\proc_j)$, where
$\proc_i,\ldots,\proc_j$ are the roles that a user of this type takes at the same time.
Since each user is an instance of a user-type, the formalisation of user-types
allows us to model an unbounded number of users, by simply adding replication
to the user-types. In fact, in most cases, roles and user-types are identical,
and the user-types that allow a user to take multiple roles
can be considered as a new role as well. Hence, we use roles and user-types interchangeably.

\paragraph{Protocol instances.} Instances of an e-health protocol
$\eHealthProtocol$ with $n$ roles/user-types are modelled in the
following form:
\[
	\eHealthProtocol\defi \nu \dataset. \init. (!\Role_{1} \mid \ldots
	\mid !\Role_{n}),
\]
where process $\nu \dataset$, which is the abbreviation for process
$\nu \name{a}_1. \ldots. \nu \name{a}_n. \nu
\name{c}_1.\ldots.\nu \name{c}_n$ ($ \name{a}_i$ stands for
private names, and $\name{c}_i$ stands for private channels), models the
private names and channels in the protocol; $\init$ is a sequential
process, representing settings of the protocol, such as
generating/computing data and revealing information to the adversary
(see Figure~\ref{fig:composedlv} for example). Essentially, $\nu
\dataset. \init$ models the global settings of an instance and auxiliary
channels in the modelling of the protocol.

\paragraph{Doctor role/user-type.}
More specifically, we have a doctor role/user-type $\Role_{\thedoctor}$
of the form:
\[
\begin{array}{rcl}
\Role_{\thedoctor}&\defi&\left\{
\begin{array}{ll}\nu \doctorID. \init_{\thedoctor}. !\ProDr,& \text{if doctor identity is not revealed by setting}\\
                 \nu \doctorID. \out{\ch}{\doctorID}. \init_{\thedoctor}. !\ProDr,& \text{if doctor identity is revealed by setting}
\end{array}\right.\\
\ProDr&\defi&\nu \prescText. \vProDr.
\end{array}
\]
In the following, we focus on the behaviour of a doctor, since our goal
is to formalise privacy properties for doctors. Each doctor is
associated with an identity ($\nu \doctorID$) and can execute an
infinite number of sessions (modelled by the exclamation mark `!' in
front of $\ProDr$).
In case the doctor identity is revealed in the
initialisation phase, we require that this unveiling does not appear in
process $\init_{\thedoctor}$, for the sake of uniformed formalisations
of the later defined privacy properties. Instead, we model this case as
identity generation ($\nu \doctorID$) immediately followed by unveiling
the identity ($\out{\ch}{\doctorID}$) on the public channel $\ch$.
Note that we reserve the name $\ch$ for the adversary's public channel.
We require $\ch$ to be free to model the public channel
that is controlled by the adversary. The adversary uses this channel by
sending and receiving messages over $\ch$.
In fact, since the doctor identity $\doctorID$ is defined
outside of the process $\init_{\thedoctor}. !\ProDr$, the doctor
identity appearing in the process is a free variable of the process.
Hence, in the case that the doctor identity is revealed, the doctor
process can be simply modelled as
$\Role_{\thedoctor}\defi\init_{\thedoctor}. !\ProDr$, where doctor
identity is a free variable.
To distinguish the free variable in process $\init_{\thedoctor}.
!\ProDr$ from the name $\doctorID$, we use the italic font to represent
the free variable, i.e., $\doctorIDv$.

Within each session, the doctor creates a prescription. Since a prescription
normally contains not only prescribed medicines but also the time/date that
the prescription is generated as well as other identification information,
we consider the prescriptions differ in sessions. In the case that a prescription
can be prescribed multiple times, one can add the replication mark $!$ in front
of $\vProDr$ to model that the prescription $\prescText$ can be prescribed in infinite
sessions, i.e., $\ProDr\defi\nu \prescText. !\vProDr$.
Similarly, we use the italic font of the prescription, $\prescTextv$, to represent
the free variable referring to the prescription in the process $\vProDr$.

\paragraph{Well-formed.} We require that $\eHealthProtocol$ is well-formed, i.e., the process $\eHealthProtocol$ satisfies the
following properties:
\begin{enumerate}
\item $\eHealthProtocol$ is canonical: names and
variables in the process never appear both bound and
free, and each name and variable is bound at most once;
\item data is typed, channels are ground, private channels are never
sent on any channel;
\item $\nu \dataset$ may be null;
\item $\init$ and $\init_{\thedoctor}$ are sequential processes;
\item $\init$, $\init_{\thedoctor}$ and $\vProDr$ can be any process (possibly $0$) such that $\eHealthProtocol$ is a closed plain process.
\end{enumerate}
Furthermore, we use $\contexthealth{\hole}$ to denote a context (a process with a hole) consisting of honest users,
\[
\contexthealth{\hole}\defi \nu \dataset. \init. (!\Role_{1} \mid \ldots \mid !\Role_{n}\mid \longhole).
\]
Dishonest agents are captured by the adversary (Section~\ref{ssub:adversary}) with certain initial knowledge.

\subsection{The adversary}\label{ssub:adversary}

We consider security and privacy properties of e-health protocols with respect
to the presence of active attackers -- the Dolev-Yao adversary. The adversary
\begin{itemize}
\item controls the network -- the adversary can block, read and insert
	messages over the network;
\item has computational power -- the adversary can record messages
	and apply cryptographic functions to messages to obtain new
	messages;
\item has a set of initial knowledge -- the adversary knows the
	participants and public information of all participants, as well
	as a set of his own data;
\item has the ability to initiate conversations -- the adversary can
	take part in executions of protocols.
\item The adversary's behaviour models that of every
	dishonest agent (cf.~Section~\ref{ssec:dishonest}),
	which is achieved by including the initial knowledge of each
	dishonest agent in the adversary's initial knowledge.
\end{itemize}
The behaviour of the adversary is modelled as a process running in
parallel with the honest agents. The adversary does whatever he can to
break the security and privacy requirements. We do not need to model
the adversary explicitly, since he is embedded in the applied pi
calculus as well as in the verification tool. Modelling the honest
users' behaviour is sufficient to verify whether the requirements hold.

\paragraph{Limitations.}
Note that the Dolev-Yao adversary model we use includes the
``perfect cryptography'' assumption. This means that the adversary
cannot infer any information from cryptographic messages for which he
does not possess a key. For instance, the attacker cannot decrypt a
ciphertext without the correct key. Moreover, the adversary does not
have the ability to perform side-channel attacks. For instance,
fingerprinting a doctor based on his prescriptions is
beyond the scope of this attacker model.

\subsection{\Docunlink}\label{sec:docunlink}
\Docunlink{} ensures unlinkability of a doctor and his prescriptions,
i.e., the adversary cannot tell whether a prescription is prescribed by
a doctor. This requirement helps to prevent doctors from being
influenced in the prescriptions they issue.

Normally, prescriptions are eventually revealed to the general public,
for example, for research purposes. In the DLV08 e-health protocol,
prescriptions are revealed to the adversary observing the network.
Therefore, in the extreme situation where there is only one
doctor, the doctor's prescriptions are obviously revealed to the
adversary - all the observed prescriptions belong to the doctor. To
avoid such a case, \docunlink{} requires at least one other doctor
(referred to as the counter-balancing doctor). This ensures that the
adversary cannot tell whether the observed prescriptions belong to the
targetted doctor or the counter-balancing doctor. With this in
mind, unlinkability of a doctor to a prescription is modelled as
indistinguishability between two honest users that swap their
prescriptions, analogously to the formalisation of
vote-privacy~\cite{DKR09}. By adopting the vote-privacy formalisation, \docunlink{} is thus modelled as the
equivalence of two doctor processes: in the first process, an honest
doctor $\drA$ prescribes $\pa$ in one of his sessions and another honest
doctor $\drB$ prescribes $\pb$ in one of his sessions; in the second
one, $\drA$ prescribes $\pb$ and $\drB$ prescribes $\pa$.

\begin{definition}[\docunlink]
A well-formed e-health protocol $\eHealthProtocol$ with a doctor role
$\Role_{\thedoctor}$, satisfies \docunlink\ if
for all possible doctors $\drA$ and $\drB$ ($\drA\neq\drB$) we have
\[
\begin{array}{rl}
&\contexthealth{
\big(\init_{\thedoctor}\substitution{\drA}{\doctorIDv}.
 (!\ProDr\substitution{\drA}{\doctorIDv}
 \mid \vProDr\dsub{\drA}{\doctorIDv}{\pa}{\prescTextv})\big)
\mid\\
&\hspace{4ex}
 \big(\init_{\thedoctor}\substitution{\drB}{\doctorIDv}.
 (!\ProDr\substitution{\drB}{\doctorIDv}
 \mid \vProDr\dsub{\drB}{\doctorIDv}{\pb}{\prescTextv})\big)
}\\
\eq
&\contexthealth{
\big(\init_{\thedoctor}\substitution{\drA}{\doctorIDv}.
 (!\ProDr\substitution{\drA}{\doctorIDv}
  \mid
 \vProDr\dsub{\drA}{\doctorIDv}{\pb}{\prescTextv})\big)
\mid\\
&\hspace{4ex}
\big(\init_{\thedoctor}\substitution{\drB}{\doctorIDv}.
 (!\ProDr\substitution{\drB}{\doctorIDv}
  \mid
\vProDr\dsub{\drB}{\doctorIDv}{\pa}{\prescTextv})\big)
},
\end{array}
\]
where $\pa$ and $\pb$ ($\pa\neq\pb$) are any two possible prescriptions, process $!\ProDr\substitution{\drA}{\doctorIDv}$
and process $!\ProDr\substitution{\drB}{\doctorIDv}$ can be $0$.
\label{def:drpriv}
\end{definition}

Process $\init_{\thedoctor}\substitution{\drA}{\doctorIDv}.
 (!\ProDr\substitution{\drA}{\doctorIDv}
 \mid \vProDr\dsub{\drA}{\doctorIDv}{\pa}{\prescTextv})$ models an instance of a doctor,
with identity $\drA$. The sub-process $\vProDr\dsub{\drA}{\doctorIDv}{\pa}{\prescTextv})$
models a prescribing session in which $\drA$ prescribes $\pa$ for a patient.
The sub-process $!\ProDr\substitution{\drA}{\doctorIDv}$ models other
prescribing sessions of $\drA$.
Similarly, process $\init_{\thedoctor}\substitution{\drB}{\doctorIDv}.
 (!\ProDr\substitution{\drB}{\doctorIDv}
 \mid \vProDr\dsub{\drB}{\doctorIDv}{\pb}{\prescTextv})$ models another doctor $\drB$.
On the right-hand side of the equivalence, the two doctors, $\drA$ and $\drB$,
swap their prescriptions, $\pa$ and $\pb$. The labelled bisimilarity ($\eq$)
captures that any dishonest third party (the adversary)
cannot distinguish the two sides. Doctor $\drB$'s process is called the counter-balancing process.
We require the existence of the counter-balancing doctor $\drB$ and
$\pa\neq\pb$ to avoid the situation in which all patients prescribe the same
prescription, and thus the prescription of all patients are simply revealed.

Note that $\doctorIDv$ and $\prescTextv$ are free names in the processes in the definition.
$\drA$ and $\drB$ are free names in the processes, when the doctor identities are initially public, and
are private names in the processes, when the doctor identities are initially private. Similarly,
$\pa$ and $\pb$ are free names in the processes, when the prescriptions are revealed, and are
private names in the processes, when the prescriptions are kept secret. This holds for the following
definitions as well.

\subsection{\Docrf}
\label{sec:receiptfree}
Enforced privacy properties have been formally defined in e-voting and
e-auctions. Examples include receipt-freeness and coercion-resistance in
e-voting~\cite{DKR09,JMP09}, and receipt-freeness for non-winning
bidders in e-auctions~\cite{DJP11}. De Decker et al.~\cite{DLVV08}
identify the need to prevent a pharmaceutical company from bribing a
doctor to favour their medicine. Hence, a doctor's \docunlink\ must be
enforced by the \eHealth\ system to prevent doctor bribery. This means that
intuitively, even if a doctor collaborates, the adversary cannot be
certain that the doctor has followed his instructions. Bribed users
are not modelled as part of the adversary, as they may lie and are thus
not trusted by the adversary. Due to the domain differences -- in e-voting and
sealed-bid e-auctions, participants are fixed before the execution, whereas in e-health, participants
may be infinitely involving; in e-voting and sealed-bid e-auctions, each participant
executes the protocol exactly once, whereas in e-health, a participant may involve
multiple/infinite times. Thus, the formalisation in e-voting and sealed-bid e-auctions
cannot be adopted. Inspired by formalisations of
receipt-freeness in e-voting~\cite{DKR09} and e-auction~\cite{DJP11}, we
define \docrf\ to be satisfied if there exists a process where the
bribed doctor does not follow the adversary's instruction (e.g.,
prescribing a particular medicine), which is indistinguishable from a
process where she does.

Modelling this property necessitates modelling a doctor who genuinely
reveals all her private information to the adversary. This is
achieved by the process transformation $P^{\chc}$ by Delaune et
al.~\cite{DKR09}. This operation transforms a plain
process $P$ into one which shares all private information over the
channel~$\chc$ with the adversary.
The transformation $P^{\forwardchannel}$ is defined as follows:
Let $\prs$ be a plain process
and $\forwardchannel$ a fresh channel name. $P^{\forwardchannel}$,
the process that shares all of $P$'s secrets, is defined as:
\begin{itemize}
\item \makebox[3cm][l]{$0^{\forwardchannel}$}\ $\hateq \symbol{symbol:hateq}0$,
\item \makebox[3cm][l]{$(P\mid Q)^{\forwardchannel}$}
      $\hateq P^{\forwardchannel}\mid Q^{\forwardchannel}$,
\item \makebox[3cm][l]{$(\nu \name{n}.P)^{\forwardchannel}$}
	$\hateq \left\{
      \begin{array}{lr}
	      \nu \name{n}.\out{\forwardchannel}{\name{n}}.P^{\forwardchannel} &
		\qquad\qquad \text{when $\name{n}$ is a name of base type,} \\
	      \nu \name{n}.P^{\forwardchannel} &
	      	\quad \text{otherwise,}
	\end{array}
	\right.$

\item \makebox[3cm][l]{$(\readin{v}{x}.P)^{\forwardchannel}$}
	$\hateq \left\{
      \begin{array}{lr}
      	\readin{v}{x}.\out{\forwardchannel}{x}.P^{\forwardchannel} &
		\qquad \text{when $x$ is a variable of base type,} \\
        \readin{v}{x}.P^{\forwardchannel} & \text{otherwise,}
     \end{array}
     \right.$

\item \makebox[3cm][l]{$(\out{v}{M}.P)^{\forwardchannel}$}
	$\hateq \out{v}{M}.P^{\forwardchannel}$,
\item \makebox[3cm][l]{$(!P)^{\forwardchannel}$}
	$\hateq !P^{\forwardchannel}$,
\item \makebox[5cm][l]{$(\tif\ M\eqe N \then\ P \telse\ Q)^{\forwardchannel}$}
	$\hateq
       \tif\ M\eqe N \then\ P^{\forwardchannel} \telse\ Q^{\forwardchannel}$.
\end{itemize}
In addition, we also use the transformation $P^{\backslash
\out{\chc}{\cdot}}$~\cite{DKR09}. This models
a process $P$ which hides all outputs on channel $\chc$.
Formally, $P^{\backslash \out{\chc}{\cdot}}\defi\nu \chc.(P \mid\, !\readin{\chc}{x})$.

\begin{definition}[\docrf]
A well-formed e-health protocol $\eHealthProtocol$ with a doctor role
$\Role_{\thedoctor}$, satisfies \docrf\ if for any two doctors $\drA$ and $\drB$ ($\drA\neq\drB$) and any two possible prescriptions $\pa$ and $\pb$ ($\pa\neq\pb$),
there exist processes $\init_{\thedoctor}'$ and $\ProDr'$, such that:
\[
\begin{array}{lrl}
1.\ &&\contexthealth{
 \big(\init_{\thedoctor}'.
 (!\ProDr\substitution{\drA}{\doctorIDv}
\mid
   \ProDr')\big)
\mid \\
&&\hspace{3.9ex}
\big(\init_{\thedoctor}\substitution{\drB}{\doctorIDv}.
 (! \ProDr\substitution{\drB}{\doctorIDv}
\mid
  \vProDr\dsub{\drB}{\doctorIDv}{\pa}{\prescTextv})\big)}\\
&\eq
&\contexthealth{
\big((\init_{\thedoctor}\substitution{\drA}{\doctorIDv})^{\chc}.
(!\ProDr\substitution{\drA}{\doctorIDv}
\mid
(\vProDr\dsub{\drA}{\doctorIDv}{\pa}{\prescTextv})^{\chc})\big)
\mid\\
&&\hspace{3.9ex}
\big(\init_{\thedoctor}\substitution{\drB}{\doctorIDv}.
(!\ProDr\substitution{\drB}{\doctorIDv}
\mid
 \vProDr\dsub{\drB}{\doctorIDv}{\pb}{\prescTextv})\big)};  \vspace{2mm}\\
2.\ &&
\contexthealth{\big(
(\init_{\thedoctor}'.(!\ProDr\substitution{\drA}{\doctorIDv}\mid \ProDr'))^{\backslash \out{\chc}{\cdot}}\big)
\mid \\
&&\hspace{3.9ex}
\big(\init_{\thedoctor}\substitution{\drB}{\doctorIDv}.
 (! \ProDr\substitution{\drB}{\doctorIDv}
\mid
  \vProDr\dsub{\drB}{\doctorIDv}{\pa}{\prescTextv})\big)}\\
&\eq
&
\contexthealth{\big(
\init_{\thedoctor}\substitution{\drA}{\doctorIDv}.
(!\ProDr\substitution{\drA}{\doctorIDv}\mid\vProDr\dsub{\drA}{\doctorIDv}{\pb}{\prescTextv})\big)
\mid \\
&&\hspace{3.9ex}
\big(\init_{\thedoctor}\substitution{\drB}{\doctorIDv}.
 (! \ProDr\substitution{\drB}{\doctorIDv}
\mid
  \vProDr\dsub{\drB}{\doctorIDv}{\pa}{\prescTextv})\big)},
\end{array}
\]
where
$\init_{\thedoctor}'.(!\ProDr\substitution{\drA}{\doctorIDv}\mid \ProDr')$ is a closed plain process,
$\chc$ is a free fresh channel name, process $!\ProDr\substitution{\drA}{\doctorIDv}$
and $!\ProDr\substitution{\drB}{\doctorIDv}$ can be $0$.
\label{def:drf}
\end{definition}

In the definition, the sub-process $!\ProDr\substitution{\drA}{\doctorIDv}$ models the sessions of $\drA$ that are not bribed, and the sub-processes $\init_{\thedoctor}'$ and $\ProDr'$
model the process in which the doctor $\drA$ lies to the adversary about one of
his prescriptions. The real prescription behaviour of $\drA$ is modelled by the second equivalence.
The first equivalence shows that the adversary cannot distinguish whether $\drA$
lied, given a counter-balancing doctor $\drB$.

\paragraph{Remark}
\Docrf{} is stronger than \docunlink{}
(cf.~Figure~\ref{fig:relation_def}). Intuitively, this is true since
\docrf{} is like \docunlink\ except that the adversary may gain more
knowledge. Thus, if a protocol satisfies \docrf{} (the adversary cannot
break privacy with more knowledge), \docunlink{} must also be satisfied
(the adversary cannot break privacy
with less knowledge).
We prove this formally, following the proof
that receipt-freeness is stronger than vote-privacy in~\cite{ACRR10}.
We prove that by applying an evaluation context that hides the channel
$\chc$ on both sides of the first equivalence in
Definition~\ref{def:drf}, we can obtain Definition~\ref{def:drpriv}.

\proof
If a protocol satisfies \docrf, there exists a closed plain process
$\init_{\thedoctor}'.(!\ProDr\substitution{\drA}{\doctorIDv}\mid
\ProDr')$ such that the two equations in Definition~\ref{def:drf} are
satisfied. By applying the evaluation context $\nu
\chc.(\hole\mid!\readin{\chc}{x})$ (defined as $P^{\backslash
\out{\chc}{\cdot}}$ in Section~\ref{sec:receiptfree}) on both sides of
the first equation, we obtain
\[
\begin{array}{lrl}
 &&\contexthealth{
 \big(\init_{\thedoctor}'.
 (!\ProDr\substitution{\drA}{\doctorIDv}
\mid
   \ProDr')\big)
\mid \\
&&\hspace{3.9ex}
\big(\init_{\thedoctor}\substitution{\drB}{\doctorIDv}.
 (! \ProDr\substitution{\drB}{\doctorIDv}
\mid
  \vProDr\dsub{\drB}{\doctorIDv}{\pa}{\prescTextv})\big)}^{\backslash \out{\chc}{\cdot}}\\
&\eq
&\contexthealth{
\big((\init_{\thedoctor}\substitution{\drA}{\doctorIDv})^{\chc}.
(!\ProDr\substitution{\drA}{\doctorIDv}
\mid
(\vProDr\dsub{\drA}{\doctorIDv}{\pa}{\prescTextv})^{\chc})\big)
\mid\\
&&\hspace{3.9ex}
\big(\init_{\thedoctor}\substitution{\drB}{\doctorIDv}.
(!\ProDr\substitution{\drB}{\doctorIDv}
\mid
 \vProDr\dsub{\drB}{\doctorIDv}{\pb}{\prescTextv})\big)}^{\backslash \out{\chc}{\cdot}}
\end{array}
\]
Lemma 1~\cite{ACRR10}: Let $\contextc_1=\nu \vect{a_1}.(\hole\mid B_1)$
and $\contextc_2=\nu \vect{a_2}. (\hole\mid B_2)$ be two evaluation
contexts such that $\vect{a_1}\cap (\fv{B_2}\cup \fn{B_2})=\emptyset$
and $\vect{a_2}\cap(\fv{B_1}\cup\fn{B_1})=\emptyset$. We have that
$\contextc_1[\contextc_2[A]]\steq \contextc_2[\contextc_1[A]]$ for any
extended process $A$.

Using Lemma~1, we can rewrite the left-hand side (1) and the right-hand
side (2) of the equivalence as follows.
\[
\begin{array}{lrl}
 (1) &&\contexthealth{
 \big(\init_{\thedoctor}'.
 (!\ProDr\substitution{\drA}{\doctorIDv}
\mid
   \ProDr')\big)
\mid \\
&&\hspace{3.9ex}
\big(\init_{\thedoctor}\substitution{\drB}{\doctorIDv}.
 (! \ProDr\substitution{\drB}{\doctorIDv}
\mid
  \vProDr\dsub{\drB}{\doctorIDv}{\pa}{\prescTextv})\big)}^{\backslash \out{\chc}{\cdot}}\\
&\steq
 &\contexthealth{
 \big(\init_{\thedoctor}'.
 (!\ProDr\substitution{\drA}{\doctorIDv}
\mid
   \ProDr')\big)^{\backslash \out{\chc}{\cdot}}
\mid \\
&&\hspace{3.9ex}
\big(\init_{\thedoctor}\substitution{\drB}{\doctorIDv}.
 (! \ProDr\substitution{\drB}{\doctorIDv}
\mid
  \vProDr\dsub{\drB}{\doctorIDv}{\pa}{\prescTextv})\big)}\\
  ~\\
(2)&&\contexthealth{
\big((\init_{\thedoctor}\substitution{\drA}{\doctorIDv})^{\chc}.
(!\ProDr\substitution{\drA}{\doctorIDv}
\mid
(\vProDr\dsub{\drA}{\doctorIDv}{\pa}{\prescTextv})^{\chc})\big)
\mid\\
&&\hspace{3.9ex}
\big(\init_{\thedoctor}\substitution{\drB}{\doctorIDv}.
(!\ProDr\substitution{\drB}{\doctorIDv}
\mid
 \vProDr\dsub{\drB}{\doctorIDv}{\pb}{\prescTextv})\big)}^{\backslash \out{\chc}{\cdot}}\\
 &\steq
&\contexthealth{
\big((\init_{\thedoctor}\substitution{\drA}{\doctorIDv})^{\chc}.
(!\ProDr\substitution{\drA}{\doctorIDv}
\mid
(\vProDr\dsub{\drA}{\doctorIDv}{\pa}{\prescTextv})^{\chc})\big)^{\backslash \out{\chc}{\cdot}}
\mid\\
&&\hspace{3.9ex}
\big(\init_{\thedoctor}\substitution{\drB}{\doctorIDv}.
(!\ProDr\substitution{\drB}{\doctorIDv}
\mid
 \vProDr\dsub{\drB}{\doctorIDv}{\pb}{\prescTextv})\big)}
\end{array}
\]

For equation $(1)$, by the second equation in Definition~\ref{def:drf}, we have
\[
\begin{array}{lrl}
 &&\contexthealth{
 \big(\init_{\thedoctor}'.
 (!\ProDr\substitution{\drA}{\doctorIDv}
\mid
   \ProDr')\big)
\mid \\
&&\hspace{3.9ex}
\big(\init_{\thedoctor}\substitution{\drB}{\doctorIDv}.
 (! \ProDr\substitution{\drB}{\doctorIDv}
\mid
  \vProDr\dsub{\drB}{\doctorIDv}{\pa}{\prescTextv})\big)}^{\backslash \out{\chc}{\cdot}}\\
  &\eq
  &\contexthealth{\big(
\init_{\thedoctor}\substitution{\drA}{\doctorIDv}.
(!\ProDr\substitution{\drA}{\doctorIDv}\mid\vProDr\dsub{\drA}{\doctorIDv}{\pb}{\prescTextv})\big)
\mid \\
&&\hspace{3.9ex}
\big(\init_{\thedoctor}\substitution{\drB}{\doctorIDv}.
 (! \ProDr\substitution{\drB}{\doctorIDv}
\mid
  \vProDr\dsub{\drB}{\doctorIDv}{\pa}{\prescTextv})\big)}
  \end{array}
\]

Lemma 2~\cite{ACRR10}: let $P$ be a closed plain process and $\chc$ a channel name such that
$\chc \not\in \fn{P}\cup \bn{P}$. We have $(P^{\chc})^{\backslash \out{\chc}{\cdot}}\eq P$.

For equation $(2)$, using Lemma~2, we obtain that
\[
\begin{array}{lrl}
&&\contexthealth{
\big((\init_{\thedoctor}\substitution{\drA}{\doctorIDv})^{\chc}.
(!\ProDr\substitution{\drA}{\doctorIDv}
\mid
(\vProDr\dsub{\drA}{\doctorIDv}{\pa}{\prescTextv})^{\chc})\big)^{\backslash \out{\chc}{\cdot}}
\mid\\
&&\hspace{3.9ex}
\big(\init_{\thedoctor}\substitution{\drB}{\doctorIDv}.
(!\ProDr\substitution{\drB}{\doctorIDv}
\mid
 \vProDr\dsub{\drB}{\doctorIDv}{\pb}{\prescTextv})\big)}\\
 &\eq
 &\contexthealth{
\big((\init_{\thedoctor}\substitution{\drA}{\doctorIDv}).
(!\ProDr\substitution{\drA}{\doctorIDv}
\mid
(\vProDr\dsub{\drA}{\doctorIDv}{\pa}{\prescTextv}))\big)
\mid\\
&&\hspace{3.9ex}
\big(\init_{\thedoctor}\substitution{\drB}{\doctorIDv}.
(!\ProDr\substitution{\drB}{\doctorIDv}
\mid
 \vProDr\dsub{\drB}{\doctorIDv}{\pb}{\prescTextv})\big)}
\end{array}
\]
By transitivity, we have
\[
\begin{array}{lrl}
&&\contexthealth{\big(
\init_{\thedoctor}\substitution{\drA}{\doctorIDv}.
(!\ProDr\substitution{\drA}{\doctorIDv}\mid\vProDr\dsub{\drA}{\doctorIDv}{\pb}{\prescTextv})\big)
\mid \\
&&\hspace{3.9ex}
\big(\init_{\thedoctor}\substitution{\drB}{\doctorIDv}.
 (! \ProDr\substitution{\drB}{\doctorIDv}
\mid
  \vProDr\dsub{\drB}{\doctorIDv}{\pa}{\prescTextv})\big)}\\
&\eq
&\contexthealth{
\big((\init_{\thedoctor}\substitution{\drA}{\doctorIDv}).
(!\ProDr\substitution{\drA}{\doctorIDv}
\mid
(\vProDr\dsub{\drA}{\doctorIDv}{\pa}{\prescTextv}))\big)
\mid\\
&&\hspace{3.9ex}
\big(\init_{\thedoctor}\substitution{\drB}{\doctorIDv}.
(!\ProDr\substitution{\drB}{\doctorIDv}
\mid
 \vProDr\dsub{\drB}{\doctorIDv}{\pb}{\prescTextv})\big)},
\end{array}
\]
which is exactly Definition~\ref{def:drpriv}.\qed

The difference between this formalisation and
receipt-freeness in e-voting~\cite{DKR09} and in e-auctions~\cite{DJP11}
is that in this definition only a part of the
doctor process (the initiation sub-process and a prescribing session) shares
information with the adversary. In e-voting, each voter only votes once. In the
contrast, a doctor prescribes multiple times for various patients. As patients
and situations of patients vary, a doctor cannot prescribe medicine from the
bribing pharmaceutical company all the time. Therefore, only part of the doctor
process shares information with the adversary. Note that we model only one bribed
prescribing session, as it is the simplest scenario. This definition can be
extended to model multiple prescribing sessions being bribed, by replacing
sub-process $(\init_{\thedoctor}\substitution{\drA}{\doctorIDv})$ with the
sub-process modelling multiple doctor sessions. Note that the extended
definition requires multiple sessions of the counter-balancing
doctor or multiple counter-balancing doctors.

Assume $h$ sessions of $\drA$ are bribed, denoted as
$\vProDrb:=\vProDrl\mid\ldots\mid \vProDrh$,
where $\vProDri:=(\vProDr\substitution{\drA}{\doctorIDv})^{\chc}$.
An arbitrary instance of the bribed
sessions is denoted as $\vProDrb\substitution{(\pal, \ldots, \pah)}{\prescTextv}:=
\vProDrl\substitution{\pal}{\prescTextv}\mid\ldots\mid \vProDrh\substitution{\pah}{\prescTextv}$.
Assume there is a counter-balancing process $\ProDrc$
of the bribed sessions. The process $\ProDrc$ has $h$
corresponding sessions from one or more honest doctors. We use
$\ProDrc\substitution{(\pal, \ldots, \pah)}{\prescTextv}$ to denote the
counter-balancing process where the prescriptions of the $h$
sessions are $\pal, \ldots, \pah$, respectively. Following
definition~\ref{def:drf}, the \docrfm\ can be defined as follows.

\begin{definition}[\docrfm]
A well-formed e-health protocol $\eHealthProtocol$ with a doctor role
$\Role_{\thedoctor}$, satisfies \docrfm, if for any doctor $\drA$ with $h$
bribed sessions, denoted as $\vProDrb:=\vProDrl\mid\ldots\mid \vProDrh$,
where $\vProDri:=(\vProDr\substitution{\drA}{\doctorIDv})^{\chc}$,
for any instantiation of the prescriptions in the bribed sessions
$\vProDrb\substitution{(\pal, \ldots, \pah)}{\prescTextv}$, there exist
processes $\init_{\thedoctor}'$ and $\ProDr'$, such that
\[
\begin{array}{lrl}
1.\ &&\contexthealth{
 \big(\init_{\thedoctor}'.
 (!\ProDr\substitution{\drA}{\doctorIDv}
\mid \ProDr')\big) \mid
\ProDrc\substitution{(\pal, \ldots, \pah)}{\prescTextv}}\\
&\eq
&\contexthealth{
\big((\init_{\thedoctor}\substitution{\drA}{\doctorIDv})^{\chc}.
(!\ProDr\substitution{\drA}{\doctorIDv}
\mid
\vProDrb\substitution{(\pal \ldots, \pah)}{\prescTextv}\big)
\mid\\
&&\hspace{3.9ex}
\ProDrc\substitution{(\pbl, \ldots, \pbh)}{\prescTextv}};  \vspace{2mm}\\
2.\ &&
\contexthealth{\big(
(\init_{\thedoctor}'.(!\ProDr\substitution{\drA}{\doctorIDv}\mid \ProDr'))^{\backslash \out{\chc}{\cdot}}\big)\mid
\ProDrc\substitution{(\pal, \ldots, \pah)}{\prescTextv}}\\
&\eq
&
\contexthealth{\big(
\init_{\thedoctor}\substitution{\drA}{\doctorIDv}.
(!\ProDr\substitution{\drA}{\doctorIDv}\mid\vProDr\dsub{\drA}{\doctorIDv}{\pbl}{\prescTextv}\mid \ldots \mid \\
&&\hspace{3.9ex}
\vProDr\dsub{\drA}{\doctorIDv}{\pbh}{\prescTextv})\big)
\mid
\ProDrc\substitution{(\pal, \ldots, \pah)}{\prescTextv}},
\end{array}
\]
where
$\init_{\thedoctor}'.(!\ProDr\substitution{\drA}{\doctorIDv}\mid \ProDr')$ is a closed plain process,
$\chc$ is a free fresh channel name, process $!\ProDr\substitution{\drA}{\doctorIDv}$
and $!\ProDr\substitution{\drB}{\doctorIDv}$ can be $0$, $\ProDrc$ is less than
$h$ doctor processes running in parallel and $\ProDrc\substitution{(\pal, \ldots, \pah)}{\prescTextv}$
denotes that in some sessions of the doctor processes, the prescriptions are instantiated
with $\pal, \ldots, \pah$.
\label{def:drfm}
\end{definition}

Definition~\ref{def:drf} (\docrf) is a specific instance of this definition where
only one session of the targeted doctor is bribed. When multiple sessions are bribed, to ensure
\docrfm, it requires the existence of more than one counter-balancing doctor sessions, and thus
this extended definition is stronger than \docrf\ (Definition~\ref{def:drf}), meaning that if a protocol
satisfied the \docrfm\ where multiple sessions are bribed, then the protocol
satisfies \docrf\ where only one session is bribed. The intuition is that if
there exists a lying process for multiple bribed sessions such that \docrfm\ is satisfied,
by hiding the communications to the adversary in the lying process except one session, we can obtain the lying
process such that \docrf\ is satisfied. More concretely, when \docrf\ is satisfied, \docrfm\
may not be satisfied. For example, when there are exactly two users and two nonces generated by each user,
if the revealed information is only the two nonces, a bribed user can lie to the adversary about his nonce,
since the link between the user and the nonce is private.
However, if the user is bribed on two sessions, i.e., the link between him and his two nonces, then
at least one nonce has to be generated by the user, hence, multi-session receipt-freeness
(two sessions in particular) is not satisfied.

Remark that we restrict the way a bribed user collaborates with the adversary:
we only model forwarding information to the adversary. The
scenario that the adversary provides prescriptions for a bribed doctor,
similar to in coercion in e-voting~\cite{DKR09}, is not modelled.
Although providing ready-made prescriptions is theoretically possible,
we consider this to not be a practical attack:
correctly prescribing requires professional (sometimes empirical)
medical expertise and heavily depends on examination of the patient. As
such, no adversary can prepare an appropriate prescription without
additional information. Moreover, forwarding a non-appropriate
description carries serious legal consequences for the forwarding
doctor. Therefore, we omit the case where the doctor merely forwards an
adversary-prepared prescription. The adversary could still prepare
other information for the bribed doctor, for example, the randomness of a bit-commitment.
Such adversary-prepared information may lead to a stronger adversary than we are
considering. To model such a scenario, the verifier needs to specify exactly
which information is prepared by the adversary. Formalisation of such scenarios
can follow the formal framework proposed in~\cite{DKR09,DJP13}.

\subsection{\Docindep}
Usually, e-health systems have to deal with a complex constellation of roles: doctors,
patients, pharmacists, insurance companies, medical administration, etc.
Each of these roles has access to different private information and has different privacy concerns.
An untrusted role may be bribed to reveal private
information to the adversary
such that the adversary can break the privacy of another role.
De Decker et al.~\cite{DLVV08}
note that pharmacists may have sensitive data which can be revealed to
the adversary to break a doctor's \docunlink.
To prevent a party from revealing sensitive data that affects a
doctor's privacy,
e-health protocols are required to satisfy \emph{\docindep}. The DLV08
protocol, for example, requires \docunlink\ independent of pharmacists~\cite{DLVV08}.
Intuitively, \docindep\ means that even if another party $\Role_i$ reveals their information
(i.e., $\Role_i^{\chc}$), the adversary is not able to break a
doctor's \docunlink.

\begin{definition}[\docindep]
A well-formed e-health protocol $\eHealthProtocol$ with a doctor role
$\Role_{\thedoctor}$, satisfies
\emph{\docunlink\ independent of role $\Role_i$}, if for all possible doctors
$\drA$ and $\drB$ ($\drA\neq\drB$) we have
\[
\begin{array}{lll}
&\contexthealth{
!{\Role_{i}}^{\chc}\mid&\hspace{-2ex}
\big(\init_{\thedoctor}\substitution{\drA}{\doctorIDv}.
(!\ProDr\substitution{\drA}{\doctorIDv}
\mid
\vProDr\dsub{\drA}{\doctorIDv}{\pa}{\prescTextv})\big)
\mid\\
&&\hspace{-2ex}
 \big(\init_{\thedoctor}\substitution{\drB}{\doctorIDv}.
 (!\ProDr\substitution{\drB}{\doctorIDv}
 \mid \vProDr\dsub{\drB}{\doctorIDv}{\pb}{\prescTextv})\big)}
\\
\eq
&\contexthealth{
!{\Role_{i}}^{\chc}\mid&\hspace{-2ex}
\big(\init_{\thedoctor}\substitution{\drA}{\doctorIDv}.
(!\ProDr\substitution{\drA}{\doctorIDv}
\mid
\vProDr\dsub{\drA}{\doctorIDv}{\pb}{\prescTextv})\big)
\mid\\
&&\hspace{-2ex}
\big(\init_{\thedoctor}\substitution{\drB}{\doctorIDv}.
 (!\ProDr\substitution{\drB}{\doctorIDv}
 \mid \vProDr\dsub{\drB}{\doctorIDv}{\pa}{\prescTextv})\big)}.
\end{array}
\]
where $\pa$ and $\pb$ ($\pa\neq\pb$) are any two possible prescriptions, $\Role_i$ is a non-doctor role, process $!\ProDr\substitution{\drA}{\doctorIDv}$
and $!\ProDr\substitution{\drB}{\doctorIDv}$ can be $0$.
\label{def:dpi}
\end{definition}

\noindent
Note that we assume a worst-case situation in which role $\Role_i$
genuinely cooperates with the adversary. For example, the pharmacist
forwards all information obtained from channels hidden from the
adversary. The equivalence requires that no matter how role $\Role_i$
cooperates with the adversary, the adversary cannot link a doctor to the
doctor's prescriptions. The cooperation between pharmacists and the
adversary is modelled in the same way as the cooperation between bribed
doctors and the adversary, i.e., $!{\Role_{i}}^{\chc}$. We do not model
the situation where the adversary prepares information for the
pharmacists, as we focus on doctor privacy -- information sent out by
the pharmacist does not affect doctor privacy, so there is no reason to
control this information. Instead of modelling the pharmacists as
compromised users, our modelling allows the definition to be easily
extended to model new properties which capture situations where
pharmacists lie to the adversary due to, for example, coalition between
pharmacists and bribed doctors. In addition, although we do not model
delivery of medicine, pharmacists do need to adhere to regulations in
providing medicine. Thus, an adversary who only controls the network
cannot impersonate a pharmacist.


Just as \docrf{} is stronger than \docunlink{}, \docindep\ is stronger
than \docunlink\ (cf.~Figure~\ref{fig:relation_def}). Intuitively, this
holds since the adversary obtains at least as much information in
\docindep{} as in \docunlink. Formally,
one can derive Definition~\ref{def:drpriv} from Definition~\ref{def:dpi}
by hiding channel $\chc$ on the left-hand side as well as the right-hand
side of the equivalence in Definition~\ref{def:dpi}.

\proof
Consider a protocol that satisfies \docindep. This protocol thus satisfies
definition~\ref{def:dpi}. By applying the evaluation context $\nu
\chc.(\hole\mid!\readin{\chc}{x})$ to both the left-hand side and the
right-hand side of Definition~\ref{def:dpi}, we obtain
\[
\begin{array}{lll}
&\contexthealth{
!{\Role_{i}}^{\chc}\mid&\hspace{-2ex}
\big(\init_{\thedoctor}\substitution{\drA}{\doctorIDv}.
(!\ProDr\substitution{\drA}{\doctorIDv}
\mid
\vProDr\dsub{\drA}{\doctorIDv}{\pa}{\prescTextv})\big)
\mid\\
&&\hspace{-2ex}
 \big(\init_{\thedoctor}\substitution{\drB}{\doctorIDv}.
 (!\ProDr\substitution{\drB}{\doctorIDv}
 \mid \vProDr\dsub{\drB}{\doctorIDv}{\pb}{\prescTextv})\big)}^{\backslash \out{\chc}{\cdot}}
\\
\eq
&\contexthealth{
!{\Role_{i}}^{\chc}\mid&\hspace{-2ex}
\big(\init_{\thedoctor}\substitution{\drA}{\doctorIDv}.
(!\ProDr\substitution{\drA}{\doctorIDv}
\mid
\vProDr\dsub{\drA}{\doctorIDv}{\pb}{\prescTextv})\big)
\mid\\
&&\hspace{-2ex}
\big(\init_{\thedoctor}\substitution{\drB}{\doctorIDv}.
 (!\ProDr\substitution{\drB}{\doctorIDv}
 \mid \vProDr\dsub{\drB}{\doctorIDv}{\pa}{\prescTextv})\big)}^{\backslash \out{\chc}{\cdot}}.
\end{array}
\]
According to Lemma 1, we have
\[
\begin{array}{lll}
&\contexthealth{
!{\Role_{i}}^{\chc}\mid&\hspace{-2ex}
\big(\init_{\thedoctor}\substitution{\drA}{\doctorIDv}.
(!\ProDr\substitution{\drA}{\doctorIDv}
\mid
\vProDr\dsub{\drA}{\doctorIDv}{\pa}{\prescTextv})\big)
\mid\\
&&\hspace{-2ex}
 \big(\init_{\thedoctor}\substitution{\drB}{\doctorIDv}.
 (!\ProDr\substitution{\drB}{\doctorIDv}
 \mid \vProDr\dsub{\drB}{\doctorIDv}{\pb}{\prescTextv})\big)}^{\backslash \out{\chc}{\cdot}}
\\
\steq &
\contexthealth{
!{{\Role_{i}}^{\chc}}^{\backslash \out{\chc}{\cdot}}\mid&\hspace{-2ex}
\big(\init_{\thedoctor}\substitution{\drA}{\doctorIDv}.
(!\ProDr\substitution{\drA}{\doctorIDv}
\mid
\vProDr\dsub{\drA}{\doctorIDv}{\pa}{\prescTextv})\big)
\mid\\
&&\hspace{-2ex}
 \big(\init_{\thedoctor}\substitution{\drB}{\doctorIDv}.
 (!\ProDr\substitution{\drB}{\doctorIDv}
 \mid \vProDr\dsub{\drB}{\doctorIDv}{\pb}{\prescTextv})\big)}
\end{array}
\]
\[
\begin{array}{lll}
&\contexthealth{
!{\Role_{i}}^{\chc}\mid&\hspace{-2ex}
\big(\init_{\thedoctor}\substitution{\drA}{\doctorIDv}.
(!\ProDr\substitution{\drA}{\doctorIDv}
\mid
\vProDr\dsub{\drA}{\doctorIDv}{\pb}{\prescTextv})\big)
\mid\\
&&\hspace{-2ex}
\big(\init_{\thedoctor}\substitution{\drB}{\doctorIDv}.
 (!\ProDr\substitution{\drB}{\doctorIDv}
 \mid \vProDr\dsub{\drB}{\doctorIDv}{\pa}{\prescTextv})\big)}^{\backslash \out{\chc}{\cdot}}\\
 \steq
 &\contexthealth{
!{{\Role_{i}}^{\chc}}^{\backslash \out{\chc}{\cdot}}\mid&\hspace{-2ex}
\big(\init_{\thedoctor}\substitution{\drA}{\doctorIDv}.
(!\ProDr\substitution{\drA}{\doctorIDv}
\mid
\vProDr\dsub{\drA}{\doctorIDv}{\pb}{\prescTextv})\big)
\mid\\
&&\hspace{-2ex}
\big(\init_{\thedoctor}\substitution{\drB}{\doctorIDv}.
 (!\ProDr\substitution{\drB}{\doctorIDv}
 \mid \vProDr\dsub{\drB}{\doctorIDv}{\pa}{\prescTextv})\big)}
\end{array}
\]
According to Lemma 2, we have
\[
\begin{array}{lll}
& \contexthealth{
!{{\Role_{i}}^{\chc}}^{\backslash \out{\chc}{\cdot}}\mid&\hspace{-2ex}
\big(\init_{\thedoctor}\substitution{\drA}{\doctorIDv}.
(!\ProDr\substitution{\drA}{\doctorIDv}
\mid
\vProDr\dsub{\drA}{\doctorIDv}{\pa}{\prescTextv})\big)
\mid\\
&&\hspace{-2ex}
 \big(\init_{\thedoctor}\substitution{\drB}{\doctorIDv}.
 (!\ProDr\substitution{\drB}{\doctorIDv}
 \mid \vProDr\dsub{\drB}{\doctorIDv}{\pb}{\prescTextv})\big)}\\
\eq &
\contexthealth{
!\Role_{i}\mid&\hspace{-2ex}
\big(\init_{\thedoctor}\substitution{\drA}{\doctorIDv}.
(!\ProDr\substitution{\drA}{\doctorIDv}
\mid
\vProDr\dsub{\drA}{\doctorIDv}{\pa}{\prescTextv})\big)
\mid\\
&&\hspace{-2ex}
 \big(\init_{\thedoctor}\substitution{\drB}{\doctorIDv}.
 (!\ProDr\substitution{\drB}{\doctorIDv}
 \mid \vProDr\dsub{\drB}{\doctorIDv}{\pb}{\prescTextv})\big)}
\end{array}
\]
\[
\begin{array}{lll}
&\contexthealth{
!{{\Role_{i}}^{\chc}}^{\backslash \out{\chc}{\cdot}}\mid&\hspace{-2ex}
\big(\init_{\thedoctor}\substitution{\drA}{\doctorIDv}.
(!\ProDr\substitution{\drA}{\doctorIDv}
\mid
\vProDr\dsub{\drA}{\doctorIDv}{\pb}{\prescTextv})\big)
\mid\\
&&\hspace{-2ex}
\big(\init_{\thedoctor}\substitution{\drB}{\doctorIDv}.
 (!\ProDr\substitution{\drB}{\doctorIDv}
 \mid \vProDr\dsub{\drB}{\doctorIDv}{\pa}{\prescTextv})\big)}\\
\eq &
\contexthealth{
!\Role_{i}\mid&\hspace{-2ex}
\big(\init_{\thedoctor}\substitution{\drA}{\doctorIDv}.
(!\ProDr\substitution{\drA}{\doctorIDv}
\mid
\vProDr\dsub{\drA}{\doctorIDv}{\pb}{\prescTextv})\big)
\mid\\
&&\hspace{-2ex}
\big(\init_{\thedoctor}\substitution{\drB}{\doctorIDv}.
 (!\ProDr\substitution{\drB}{\doctorIDv}
 \mid \vProDr\dsub{\drB}{\doctorIDv}{\pa}{\prescTextv})\big)}
\end{array}
\]
Therefore, by transitivity, we have
\[
\begin{array}{lll}
&\contexthealth{
!\Role_{i}\mid&\hspace{-2ex}
\big(\init_{\thedoctor}\substitution{\drA}{\doctorIDv}.
(!\ProDr\substitution{\drA}{\doctorIDv}
\mid
\vProDr\dsub{\drA}{\doctorIDv}{\pa}{\prescTextv})\big)
\mid\\
&&\hspace{-2ex}
 \big(\init_{\thedoctor}\substitution{\drB}{\doctorIDv}.
 (!\ProDr\substitution{\drB}{\doctorIDv}
 \mid \vProDr\dsub{\drB}{\doctorIDv}{\pb}{\prescTextv})\big)}\\
\eq &
\contexthealth{
!\Role_{i}\mid&\hspace{-2ex}
\big(\init_{\thedoctor}\substitution{\drA}{\doctorIDv}.
(!\ProDr\substitution{\drA}{\doctorIDv}
\mid
\vProDr\dsub{\drA}{\doctorIDv}{\pb}{\prescTextv})\big)
\mid\\
&&\hspace{-2ex}
\big(\init_{\thedoctor}\substitution{\drB}{\doctorIDv}.
 (!\ProDr\substitution{\drB}{\doctorIDv}
 \mid \vProDr\dsub{\drB}{\doctorIDv}{\pa}{\prescTextv})\big)}
\end{array}
\]
which is exactly Definition~\ref{def:drpriv}.
\qed

Note that the first step in the proof (application of an evaluation
context) cannot be reversed. Therefore, \docunlink{} is weaker than
\docindep.

\subsection{\Docrfindep}\label{sec:docrfindep}
We have discussed two situations where a doctor's prescription behaviour can be revealed
when either the doctor or another different party cooperates with the adversary.
It is natural to consider the conjunction of these two, i.e., a
situation in which the adversary coerces both a doctor
and another party (not a doctor). Since the adversary obtains more
information, this constitutes a stronger attack on doctor's \docunlink.
To address this problem, we define \emph{\docrfindep},
which is satisfied when a doctor's
\docunlink\ is preserved even if both the doctor and another party reveal
their private information to the adversary.

\begin{definition}[\docrfindep]
A well-formed e-health protocol $\eHealthProtocol$ with a doctor role
$\Role_{\thedoctor}$, satisfies
\emph{\docrf\ independent of role $\Role_i$} if for any two doctors $\drA$ and $\drB$ ($\drA\neq\drB$) and any two possible prescriptions $\pa$ and $\pb$ ($\pa\neq\pb$),
there exist processes $\init_{\thedoctor}'$ and $\ProDr'$, such that:
\[
\begin{array}{lrl}
1.\ &&\contexthealth{!\Role_{i}^{\chc}\mid
\big(\init_{\thedoctor}'.
(!\ProDr\substitution{\drA}{\doctorIDv}
\mid
\ProDr')\big)\mid\\
&&\hspace{11ex} \big(\init_{\thedoctor}\substitution{\drB}{\doctorIDv}.
(!\ProDr\substitution{\drB}{\doctorIDv}
\mid
 \vProDr\dsub{\drB}{\doctorIDv}{\pa}{\prescTextv})\big)}\\
&\eq
&\contexthealth{!\Role_{i}^{\chc}\mid
\big((\init_{\thedoctor}\substitution{\drA}{\doctorIDv})^{\chc}.(!\ProDr\substitution{\drA}{\doctorIDv}\mid
(\vProDr\dsub{\drA}{\doctorIDv}{\pa}{\prescTextv})^{\chc})\big)\mid\\
&&\hspace{11ex}  \big(\init_{\thedoctor}\substitution{\drB}{\doctorIDv}.
(!\ProDr\substitution{\drB}{\doctorIDv}
\mid
\vProDr\dsub{\drB}{\doctorIDv}{\pb}{\prescTextv})\big)}; \vspace{2mm}\\
2.\ &&
\contexthealth{!\Role_{i}^{\chc}\mid
\big((\init_{\thedoctor}'.
(!\ProDr\substitution{\drA}{\doctorIDv}\mid\ProDr'))^{\backslash \out{\chc}{\cdot}}\big)\mid\\
&&\hspace{11ex} \big(\init_{\thedoctor}\substitution{\drB}{\doctorIDv}.
(!\ProDr\substitution{\drB}{\doctorIDv}
\mid
 \vProDr\dsub{\drB}{\doctorIDv}{\pa}{\prescTextv})\big)}\\
& \eq &
\contexthealth{!\Role_{i}^{\chc}\mid
\big(\init_{\thedoctor}\substitution{\drA}{\doctorIDv}.
(!\ProDr\substitution{\drA}{\doctorIDv}\mid\vProDr\dsub{\drA}{\doctorIDv}{\pb}{\prescTextv})\big)\mid\\
&&\hspace{11ex}  \big(\init_{\thedoctor}\substitution{\drB}{\doctorIDv}.
(!\ProDr\substitution{\drB}{\doctorIDv}
\mid
\vProDr\dsub{\drB}{\doctorIDv}{\pa}{\prescTextv})\big)},
\end{array}
\]
where
$\init_{\thedoctor}'.(!\ProDr\substitution{\drA}{\doctorIDv}\mid \ProDr')$
is a closed plain process,
$\Role_i$ is a non-doctor role, $\chc$ is a free fresh channel name, process $!\ProDr\substitution{\drA}{\doctorIDv}$
and $!\ProDr\substitution{\drB}{\doctorIDv}$ can be $0$.
\label{def:drfi}
\end{definition}
%
\Docrfindep\ implies \docrf\ and \docindep, each of which also implies
\docunlink{} (cf.~Figure~\ref{fig:relation_def}). The proof follows the
same reasoning as the proofs in~\cite{DJP13}. Intuitively, the adversary
obtains more information with \docrfindep\ (namely, from both doctor and
pharmacist) than with either \docindep\ (from pharmacist only) or
\docrf\ (from doctor only). If the adversary is unable to break a
doctor's privacy using this much information, the adversary will not be
able to break doctor privacy using less information. Therefore, if a
protocol satisfies \docrfindep, then it must also satisfies \docindep\
and \docrf. Similarly, since the adversary obtains more information in
both \docindep\ and in \docrf\ than in \docunlink, if a protocol
satisfies either \docindep\ or \docrf, it must also satisfies
\docunlink.


\subsection{Anonymity and strong anonymity}
\label{sec:anonymity}
Anonymity is a privacy property that protects users' identities. We
model anonymity as indistinguishability of processes initiated by two
different users.

\begin{definition}[doctor anonymity]
A well-formed e-health protocol $\eHealthProtocol$ with a doctor role
$\Role_{\thedoctor}$ satisfies doctor anonymity if for any doctor
$\drA$, there exists another doctor $\drB$ ($\drB\neq\drA$), such that
\[
\begin{array}{rl}
\contexthealth{
\init_{\thedoctor}\substitution{\drA}{\doctorIDv}.
!\ProDr\substitution{\drA}{\doctorIDv}
}
\eq
\contexthealth{
\init_{\thedoctor}\substitution{\drB}{\doctorIDv}.
!\ProDr\substitution{\drB}{\doctorIDv}
}.
\end{array}
\]
\label{def:ano}
\vspace{-4ex}
\end{definition}
A stronger property of anonymity is defined in~\cite{ACRR10}, capturing
the situation that the adversary cannot even find out whether a user
(with identity $\drA$) has participated in a session of the protocol or not.

\begin{definition}[strong doctor anonymity~\cite{ACRR10}]
A well-formed e-health protocol $\eHealthProtocol$ with a doctor role
$\Role_{\thedoctor}$ satisfies strong doctor anonymity, if
\[
\eHealthProtocol \eq
\nu \dataset. \init.
\big(!\Role_{1}\mid \ldots \mid !\Role_{n}\mid
(\init_{\thedoctor}\substitution{\drA}{\doctorIDv}.
!\ProDr\substitution{\drA}{\doctorIDv})\big).
\]
\label{def:sano}
\vspace{-4ex}
\end{definition}
Recall that the unveiling of a doctor's identity (when used) is
performed outside the process $\init_{dr}$ (see
Section~\ref{sec:ehealthprotocol}). Therefore, the above two definitions
do not include generation nor unveiling of doctor identities in the initialization phase.

Obviously, the concept of strong doctor anonymity is intended to be
stronger than the concept of doctor anonymity. We show that it is
impossible to satisfy strong doctor anonymity without satisfying doctor
anonymity (arrow $R_5$ in Figure~\ref{fig:relation_def}).

\proof
Assume that a protocol $\eHealthProtocol$ satisfies strong doctor
anonymity but not doctor anonymity. That is, $\eHealthProtocol$
satisfies Definition~\ref{def:sano}, i.e.,
\[
\eHealthProtocol \eq
\nu \dataset. \init.
\big(!\Role_{1}\mid \ldots \mid !\Role_{n}\mid
(\init_{\thedoctor}\substitution{\drA}{\doctorIDv}.
!\ProDr\substitution{\drA}{\doctorIDv})\big), \hspace{3.2cm} (1)
\]
but there exists no $\drB$ such that the equation in
Definition~\ref{def:ano} is satisfied. That is, $\nexists \drB$ s.t.
\[
\begin{array}{rl}
\contexthealth{
\init_{\thedoctor}\substitution{\drA}{\doctorIDv}.
!\ProDr\substitution{\drA}{\doctorIDv}
}
\eq
\contexthealth{
\init_{\thedoctor}\substitution{\drB}{\doctorIDv}.
!\ProDr\substitution{\drB}{\doctorIDv}
}. \hspace{2cm} (2)
\end{array}
\]
Since
\[
\contexthealth{\hole}\defi \nu \dataset. \init. (!\Role_{1} \mid \ldots \mid !\Role_{n}\mid \longhole),
\]
we have
\[
\contexthealth{
\init_{\thedoctor}\substitution{\drA}{\doctorIDv}.
!\ProDr\substitution{\drA}{\doctorIDv}
} \defi \nu \dataset. \init. (!\Role_{1} \mid \ldots \mid !\Role_{n}\mid (\init_{\thedoctor}\substitution{\drA}{\doctorIDv}.
!\ProDr\substitution{\drA}{\doctorIDv})).
\]
That is, the right-hand side of the equation $(1)$ is exactly the left-hand side of the equation $(2)$.
Therefore, there exists no $\drB$ such that the following equation holds,
\[
\begin{array}{rl}
\eHealthProtocol \eq \contexthealth{
\init_{\thedoctor}\substitution{\drB}{\doctorIDv}.
!\ProDr\substitution{\drB}{\doctorIDv}
}. \hspace{6.5cm} (3)
\end{array}
\]
Since $\eHealthProtocol\defi \nu \dataset. \init. (!\Role_{1} \mid \ldots
	\mid !\Role_{n})$, by letting $\drB$ be an identity of a doctor process in $\eHealthProtocol$,
the equation $(3)$ holds.
There obviously exists a $\drB$
such that the equation $(3)$ holds. This
contradicts the assumption.
\qed

\ \\
Anonymity and strong anonymity may be similarly defined for other roles.
We provide definitions for patient anonymity, anonymity for other roles
is defined analogously.

\begin{definition}[patient anonymity]
A well-formed e-health protocol $\eHealthProtocol$ with a patient role $\Role_{\thepatient}$ satisfies patient anonymity
if for any patient $\ptA$, there exists another patient $\ptB$ ($\ptB\neq\ptA$), such that
\[
\begin{array}{rl}
\contexthealth{
\init_{\thepatient}\substitution{\ptA}{\patientIDv}.
!\ProPt\substitution{\ptA}{\patientIDv}
}
\eq
\contexthealth{
\init_{\thepatient}\substitution{\ptB}{\patientIDv}.
!\ProPt\substitution{\ptB}{\patientIDv}
}.
\end{array}
\]
\label{def:anopt}
\vspace{-4ex}
\end{definition}

\begin{definition}[strong patient anonymity~\cite{ACRR10}]
A well-formed e-health protocol $\eHealthProtocol$ with a patient role $\Role_{\thepatient}$ satisfies strong doctor
anonymity, if
\[
\eHealthProtocol \eq
\nu \dataset. \init.
\big(!\Role_{1}\mid \ldots \mid !\Role_{n}\mid
(\init_{\thepatient}\substitution{\ptA}{\patientIDv}.
!\ProPt\substitution{\ptA}{\patientIDv})\big).
\]
\label{def:sanopt}
\vspace{-4ex}
\end{definition}

As with the case for doctor anonymity, strong patient anonymity is
stronger than patient anonymity ($R_7$ in
Figure~\ref{fig:relation_def}). The proof is analoguous to the proof
above.

\subsection{Untraceability and strong untraceability}
\label{sec:untraceability}
Untraceability is a property preventing the adversary from tracing a user,
meaning that he cannot tell whether two executions are initiated by the same user.
\begin{definition}[doctor untraceability]
A well-formed e-health protocol $\eHealthProtocol$ with a doctor role $\Role_{\thedoctor}$ satisfies doctor
untraceability if, for any two doctors $\drA$ and $\drB\not=\drA$,
\[
\begin{array}{rl}
&\contexthealth{
\init_{\thedoctor}\substitution{\drA}{\doctorIDv}.
 (\ProDr\substitution{\drA}{\doctorIDv}\mid
   \ProDr\substitution{\drA}{\doctorIDv})}\\
\eq
&\contexthealth{
(\init_{\thedoctor}\substitution{\drA}{\doctorIDv}.
 \ProDr\substitution{\drA}{\doctorIDv})\mid
(\init_{\thedoctor}\substitution{\drB}{\doctorIDv}.
 \ProDr\substitution{\drB}{\doctorIDv})}.
\end{array}
\]
\label{def:untra}
\vspace{-4ex}
\end{definition}
A stronger version of untraceability, proposed in~\cite{ACRR10},
captures the adversary's inability to distinguish the situation where
one user executes the protocol multiple times
from each user executing the protocol at most once.
\begin{definition}[strong doctor untraceability~\cite{ACRR10}]
A well-formed e-health protocol $\eHealthProtocol$ with a doctor role
$\Role_{\thedoctor}$ being the $j^{\mathit{th}}$ role, satisfies strong doctor
untraceability, if
\[
\eHealthProtocol
\eq
\nu \dataset. \init.
\big(!\Role_{1}\mid \ldots \mid !\Role_{j-1}\mid !\Role_{j+1} \mid !\Role_{n}
\mid !(\nu \doctorID. \init_{\thedoctor}.\ProDr)\big).
\]
\label{def:suntra}
\vspace{-4ex}
\end{definition}
Similarly, we can define untraceability and strong untraceability for
patient and other roles in a protocol, by replacing the doctor role with
a different role.

\begin{definition}[patient untraceability]
A well-formed e-health protocol $\eHealthProtocol$ with a patient role $\Role_{\thepatient}$ satisfies patient
untraceability if, for any two patients $\ptA$ and $\ptB\not=\ptA$,
\[
\begin{array}{rl}
&\contexthealth{
\init_{\thepatient}\substitution{\ptA}{\patientIDv}.
 (\ProPt\substitution{\ptA}{\patientIDv}\mid
   \ProPt\substitution{\ptA}{\patientIDv})}\\
\eq
&\contexthealth{
(\init_{\thepatient}\substitution{\ptA}{\patientIDv}.
 \ProPt\substitution{\ptA}{\patientIDv})\mid
(\init_{\thepatient}\substitution{\ptB}{\patientIDv}.
 \ProPt\substitution{\ptB}{\patientIDv})}.
\end{array}
\]
\label{def:untrapt}
\vspace{-4ex}
\end{definition}
\begin{definition}[strong patient untraceability~\cite{ACRR10}]
A well-formed e-health protocol $\eHealthProtocol$ with a patient role
$\Role_{\thepatient}$ being the $j^{\mathit{th}}$ role, satisfies strong
doctor untraceability, if
\[
\eHealthProtocol
\eq
\nu \dataset. \init.
\big(!\Role_{1}\mid \ldots \mid !\Role_{j-1}\mid !\Role_{j+1} \mid !\Role_{n}
\mid !(\nu \patientID. \init_{\thepatient}.\ProPt)\big).
\]
\label{def:suntrapt}
\vspace{-4ex}
\end{definition}

\begin{figure}
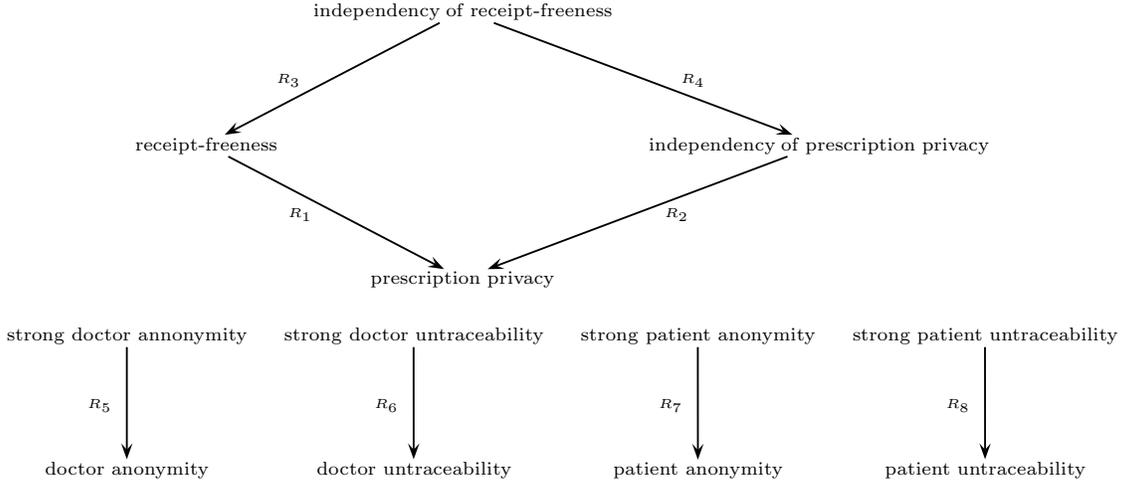

$
\scriptsize
\begin{psmatrix}[colsep=0.5cm]
      &\text{\docrfindep} \\
\text{\docrf}&  & \text{\docindep}\\
      & \text{\docunlink}
\end{psmatrix}
\everypsbox{\scriptstyle}
\psset{nodesep=1pt,arrows=->}
\ncline{1,2}{2,1}\tlput{R_3}
\ncline{1,2}{2,3}\trput{R_4}
\ncline{2,1}{3,2}\tlput{R_1}
\ncline{2,3}{3,2}\trput{R_2}
$\\
\vspace{0.5cm}
$
\scriptsize
\begin{psmatrix}[colsep=0.5cm]
\text{strong doctor annonymity}&\text{strong doctor untraceability}& \text{strong patient anonymity}&\text{strong patient untraceability} \\
\text{doctor anonymity}&\text{doctor untraceability}& \text{patient anonymity}& \text{patient untraceability}\\
\end{psmatrix}
\everypsbox{\scriptstyle}
\psset{nodesep=1pt,arrows=->}
\ncline{1,1}{2,1}\tlput{R_5}
\ncline{1,2}{2,2}\tlput{R_6}
\ncline{1,3}{2,3}\tlput{R_7}
\ncline{1,4}{2,4}\tlput{R_8}
$
\caption{Relations between privacy properties}\label{fig:relation_def}
\end{figure}

Strong (doctor/patient) anonymity is stronger than (doctor/patient)
anonymity, as the two processes (left-hand side and right hand side of
the equivalence) in (doctor/patient) anonymity are instances of the two
processes in strong (doctor/patient) anonymity respectively. Hence, if a
protocol satisfies strong (doctor/patient) anonymity, it also satisfies
(doctor/patient) anonymity. It is evidenced by that the DLV08 protocol
(see Section~\ref{sec:dlv08}) satisfies doctor anonymity (without doctor
ID revealed), but does not satisfy strong doctor anonymity.

Once again, the strong notions of doctor/patient untraceability are
stronger than the standard doctor/patient untraceability (see $R_6$ and
$R_7$, respectively, in Figure~\ref{fig:relation_def}). The proof is
again analoguous to the proof showing strong doctor anonymity is
stronger than doctor anonymity.

Finally, note that the strong versions of anonymity are not comparable
to the strong versions of untraceability (e.g. strong patient anonymity
is not comparable to strong patient untraceability).
Strong anonymity and strong untraceability capture different aspects of
privacy -- anonymity focuses on the link between participants
and their identities, whereas untraceability focuses on the link
between sessions of a participant. This is supported by the case study
in Section~\ref{sec:dlv08} where the strong doctor anonymity and strong
doctor untraceability fail due to different reasons -- the model where
strong doctor anonymity is satisfied, does not satisfy strong doctor
untraceability, and vice versa.

Similarly, \docunlink{} and doctor anonymity are not
comparable either. Doctor anonymity aims to protect the doctor identity
whereas \docunlink{} aims to protect the link between a doctor's
identity and his prescriptions. For instance, a system in which there is
one doctor may satisfy doctor anonymity, if that the doctor's identity
is perfectly protected. However, the system would not satisfy
\docunlink, since there is no counter-balancing doctor.

Conversely, consider a system where two doctors send out their public
keys over public channels and afterwards each doctor sends one
prescription via a private channel to the trusted authority, who finally
outputs both prescriptions. Such a system may satisfy \docunlink, due to
the assumptions of private channel and trusted authority. That is: on
the left hand side of the equation in Definition~\ref{def:drpriv}, the
adversary observes two public keys followed by two prescriptions; on the
right hand side, the adversary observes exactly the same. But this
system does not satisfy anonymity since the adversary can block
communication of participants in $C_{eh}$ and observe the public
channel -- on the left hand side of the equation in
Definition~\ref{def:ano}, the adversary observes the public key of
$\drA$, while on the right hand side, the adversary observes the public
key of $\drB$.

\section{Case study: the DLV08 protocol}
\label{sec:dlv08}

In this section, we apply the above formal definitions for doctor
privacy in a case study as a validation of the definitions. We choose to
analyse the DLV08 e-health protocol proposed by De Decker et
al.~\cite{DLVV08}, as it claims enforced privacy for doctors. However,
our analysis is not restricted to doctor privacy. We provide a rather
complete analysis of the protocol including patient anonymity, patient
untraceability, patient/doctor information secrecy and patient/doctor
authentication as well. The ProVerif code used to perform this analysis is
available from~\cite{DJPa12}.

The DLV08 protocol is a complex health care protocol for the Belgium
situation. It captures most aspects of the current Belgian health
care practice and aims to provide a strong guarantee of privacy for
patients and doctors.
Our analysis of this protocol focuses on the below properties. For those that are
explicitly claimed by DLV08, the corresponding claim identifier in that
paper is given. In addition to those, we analyse secrecy, \docunlink, \docrf,
and \docrfindep, which are implicitly mentioned.
\begin{itemize}
\item Secrecy of patient and doctor information:
      no other party should be able to know a patient or a doctor's
      information, unless the information is intended to be revealed in
      the protocol (for formal definitions, see Section~\ref{ssec:proverif} and Section~\ref{sec:sec}).
\item Authentication (\cite{DLVV08}: S1):
      all parties should properly authenticate each other (for formal definitions, see Section~\ref{ssec:proverif} and Section~\ref{sec:auth}).
\item Patient anonymity (\cite{DLVV08}: P3):
      no party should be able to determine a patient's identity.
\item Patient untraceability (\cite{DLVV08}: P2):
      prescriptions issued to the same patient should not be linkable to
      each other.
\item \Docunlink:
      the protocol protects a doctor's prescription behaviour.
\item \Docrf:
      the protocol prevents bribery between doctors and pharmaceutical
      companies.
\item \Docindep\ (\cite{DLVV08}: P4):
      pharmacists should not be able to provide evidence to
      pharmaceutical companies about doctors' prescription.
\item \Docrfindep:
      pharmacists should not be able to provide evidence to
      pharmaceutical companies about doctors' prescription even if the doctor is bribed.
\end{itemize}
The rest of this section describes the DLV08 protocol in more detail.

\subsection{Roles}
The protocol involves seven roles. We focus on the five roles involved
in the core process: doctor, patient, pharmacist,
medicine prescription administrator (MPA) and health insurance institute
(HII). The other two roles, public safety organisation (PSO) and social
security organisation (SSO), provide properties such as revocability and
reimbursement. As we do not focus on these properties, and as these
roles are only tangentially involved in the core process, we omit these
roles from our model.

The roles interact as follows: a doctor prescribes medicine to a patient;
next the patient obtains medicine from a pharmacist according to the
prescription; following that, the pharmacist forwards the prescription
to his MPA, the MPA checks the prescription and refunds the pharmacist;
finally, the MPA sends invoices to the patient's HII and is refunded.

\subsection{Cryptographic primitives}
To ensure security and privacy properties, the DLV08 protocol employs
several specific cryptographic primitives, besides the classical ones, like encryption.
We briefly introduce these cryptographic primitives.

\paragraph{Bit-commitments.}
The bit-commitments scheme consists of two phases, committing phase and
opening phase. On the committing phase, a message sender commits
to a message. This can be considered as putting the message
into a box, and sending the box to the receiver. Later in the opening
phase, the sender sends the key of the box to the receiver. The receiver
opens the box and obtains the message.

\paragraph{Zero-knowledge proofs.}
A zero-knowledge proof is a cryptographic scheme which is used by
one party (prover) to prove to another party (verifier) that a statement
is true, without leaking secret information of the prover. A zero-knowledge
proof scheme may be either interactive or non-interactive. We consider
non-interactive zero-knowledge proofs in this protocol.

\paragraph{Digital credentials.}
A digital credential is a certificate,
proving that the holder satisfies certain requirements. Unlike paper
certificates (such as passports) which give out the owner's identity,
a digital credential can be used to authenticate the owner anonymously.
For example, a digital credential can be used to prove that a driver is
old enough to drive without revealing the actual age of the driver.

\paragraph{Anonymous authentication.}
Anonymous authentication is a scheme for authenticating a user
anonymously, e.g.,~\cite{BCKL08}. The procedure of anonymous authentication is actually a
zero-knowledge proof, with the digital credential being the public
information of the prover. In the scheme, a user's digital credential is
used as the public key in a public key authentication structure. Using
this, a verifier can check whether a message is signed correctly by the
prover (the person authenticating himself), while the verifier cannot
identify the prover. Thus, this ensures anonymous authentication.

\paragraph{Verifiable encryptions.}
Verifiable encryption is based on zero-knowledge proofs as well. A
prover encrypts a message, and uses zero-knowledge proofs to prove that
the encrypted message satisfies specific properties without revealing
the original message.

\paragraph{Signed proofs of knowledge.}
Signed proofs of knowledge provide a way of using proofs of knowledge as a
digital signature scheme (cf.~\cite{Brands00}). Intuitively, a prover
signs a message using secret information, which can be considered
as a secret signing key. The prover can convince the verifier using proofs of
knowledge only if the prover has the right secret key. Thus it proves the
origination of the message.

\subsection{Setting}\label{ssec:setting}
The initial information available to a participant is as follows.
\begin{itemize}
\item A doctor has an identity ($\doctorID$), a pseudonym ($\doctorPseudo$), and
an anonymous doctor credential ($\doctorCred$) issued by trusted authorities.
\item A patient has an identity ($\patientID$), a pseudonym ($\patientPseudo$),
an HII ($\patientHii$), a social security status ($\patientSSS$), a
health expense account ($\patientAcc$) and an anonymous patient credential
($\patientCred$) issued by trusted authorities.
\item Pharmacists, MPA, and HII are public entities,
each of which has an identity ($\pharmID$, $\mpaID$, $\hiiID$),
a secret key ($\pharmSk$, $\mpaSk$, $\hiiSk$) and
an authorised public key certificate ($\pkph$, $\pkmpa$, $\pkhii$)
issued by trusted authorities.
\end{itemize}
We assume that a user does not take two roles with the same identity. Hence,
one user taking two roles are considered as two individual users.

\subsection{Description of the protocol}
The DLV08 protocol consists of four sub-protocols: doctor-patient sub-protocol,
patient-pharmacist sub-protocol, pharmacist-MPA sub-protocol, and MPA-HII
sub-protocol. We describe the sub-protocols one by one.

\subsubsection{Doctor-patient sub-protocol}
\begin{figure}[ht]
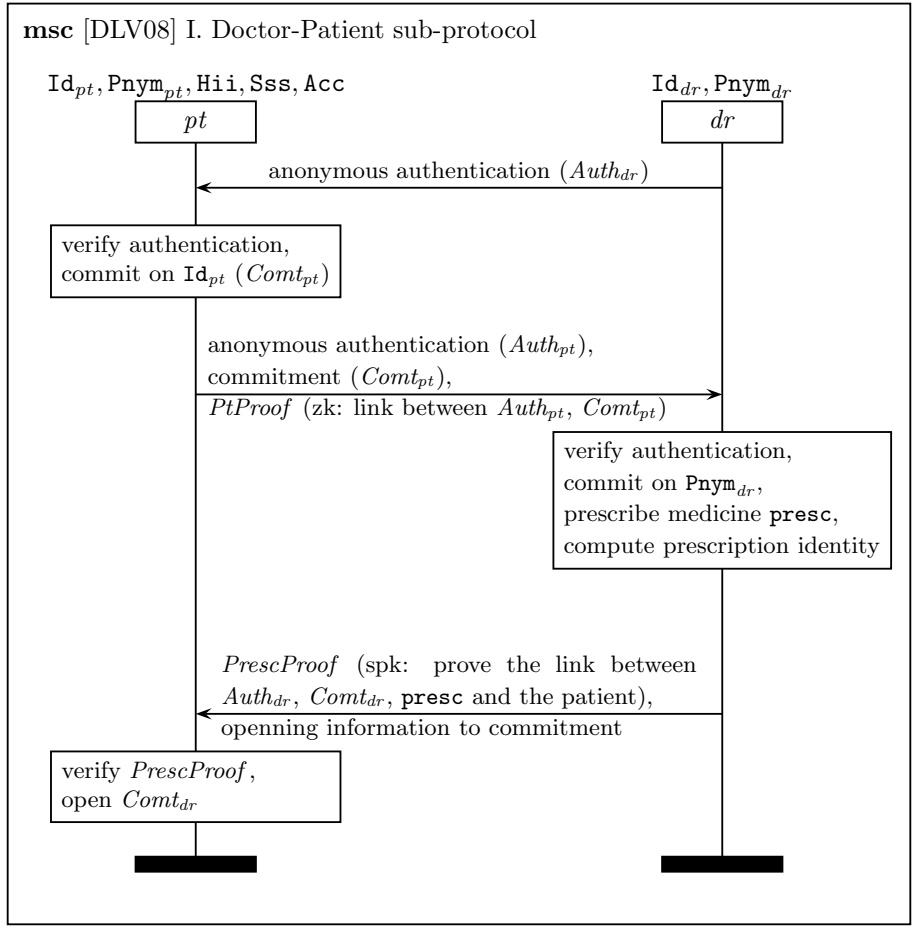

\begin{msc}{[DLV08] I. Doctor-Patient sub-protocol}
\instdist 7.0cm
\envinstdist 2.5cm
\declinst{Pt}{$\patientID, \patientPseudo,
 \patientHii, \patientSSS, \patientAcc$}{$\thepatient$}
\declinst{Dr}{$\doctorID, \doctorPseudo$}{$\thedoctor$}
\mess{{\small anonymous authentication ($\doctorAuth$)}}{Dr}{Pt}
\nextlevel
\action*{\small \parbox{11em}{{\small verify authentication, \\
	commit on $\patientID$ ($\patientCommit$)}}}{Pt}
\nextlevel[4.5]

\mess{\parbox{19em}{{\small anonymous authentication ($\patientAuth$),\\
commitment ($\patientCommit$),\\ $\patientProof$ (zk: link between
$\patientAuth$, $\patientCommit$)}}}{Pt}{Dr}
\nextlevel[1]
\action*{\parbox{12.0em}{{\small verify authentication, \\
	commit on $\doctorPseudo$, \\
	prescribe medicine $\prescText$, \\
        compute prescription identity}}}{Dr}
\nextlevel[7.5]

\mess{\parbox{18em}{{\small $\prescProof$ (spk: prove the link between
$\doctorAuth$, $\doctorCommit$, $\prescText$ and the patient),\\
openning information to commitment}}}{Dr}{Pt}
\nextlevel[1]
\action*{\small \parbox{11em}{ verify $\prescProof$,\\
open $\doctorCommit$}}{Pt}
\nextlevel[2]
\end{msc}
\caption[MSC: Doctor-Patient sub-protocol.]{Doctor-Patient sub-protocol.}
\label{msc:dp}
\end{figure}

The doctor authenticates himself to a patient by anonymous authentication with the authorised doctor
credential as public information. The patient verifies the doctor credential. If the verification
passes, the patient anonymously authenticates himself to the doctor using the patient
credential, sends the bit-commitments
on his identity to the
doctor, and proves to the doctor that the identity used in the
credential is the same as in the bit-commitments. After verifying the
patient credential, the doctor generates a prescription, computes a
prescription identity, computes the doctor bit-commitments. Then the doctor
combines these computed messages with the received patient bit-commitments;
signs these messages using a signed proof of knowledge,
which proves that the
doctor's pseudonym used in the doctor credential is the same as in the doctor
bit-commitments. Together with the proof, the doctor sends the opening information,
which is used to open the doctor bit-commitments.
The communication in the doctor-patient sub-protocol is shown as a message
sequence chart (MSC,~\cite{MB02}) in Figure~\ref{msc:dp}.
\begin{figure}[!h]
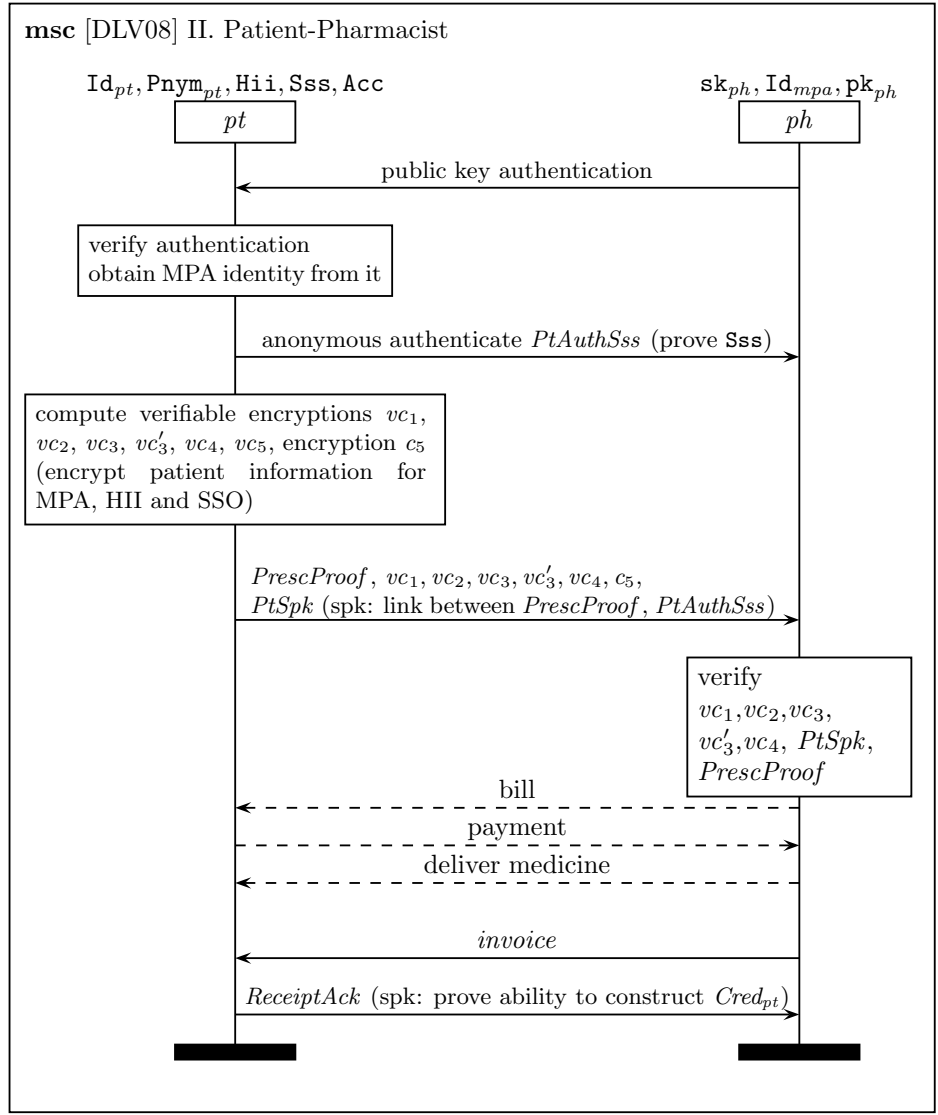

\begin{msc}{[DLV08] II. Patient-Pharmacist}
\instdist 7.5cm
\envinstdist 3.0cm
\declinst{Pt}{$\patientID, \patientPseudo,
 \patientHii, \patientSSS, \patientAcc$}{\thepatient}
\envinstdist 1.8cm
\declinst{Ph}{$\pharmSk, \mpaID, \pkph$}{\thepharm}
\mess{\small public key authentication}{Ph}{Pt}
\nextlevel
\action*{\small \parbox{12em}{verify authentication\\ obtain MPA identity
from it}}{Pt}
\nextlevel[3.5]
\mess{\small anonymous authenticate $\patientAuthsss$
(prove $\patientSSS$)}{Pt}{Ph}
\nextlevel
\action*{\small \parbox{16em}{compute verifiable encryptions
	$\vc_1$, $\vc_2$, $\vc_3$, $\vc'_3$, $\vc_4$, $\vc_5$,
        encryption $\C_5$\\ (encrypt patient information for MPA, HII and
        SSO)
}
}{Pt}
\nextlevel[6]
\mess{\small \parbox{21.5em}{
	$\prescProof$, $\vc_1,\vc_2,\vc_3,\vc'_3,\vc_4,\C_5$,\\
	$\patientspk$ (spk: link between $\prescProof$,
	$\patientAuthsss$)\\
}
}{Pt}{Ph}
\nextlevel
\action*{\parbox{7.7em}{
verify $\vc_1$,$\vc_2$,$\vc_3$,\\$\vc'_3$,$\vc_4$,
$\patientspk$,\\$\prescProof$}}{Ph}
\nextlevel[4]
\mess*{bill}{Ph}{Pt}
\nextlevel
\mess*{payment}{Pt}{Ph}
\nextlevel
\mess*{deliver medicine}{Ph}{Pt}
\nextlevel[2]
\mess{\invoicev}{Ph}{Pt}
\nextlevel[1.5]
\mess{{\small $\receptionAck$ (spk: prove ability to construct $\patientCred$)}}{Pt}{Ph}
\end{msc}
\caption[MSC: Patient-Pharmacist sub-protocol.]{Patient-Pharmacist sub-protocol.}
\label{msc:pp}
\end{figure}
\begin{figure}[!h]
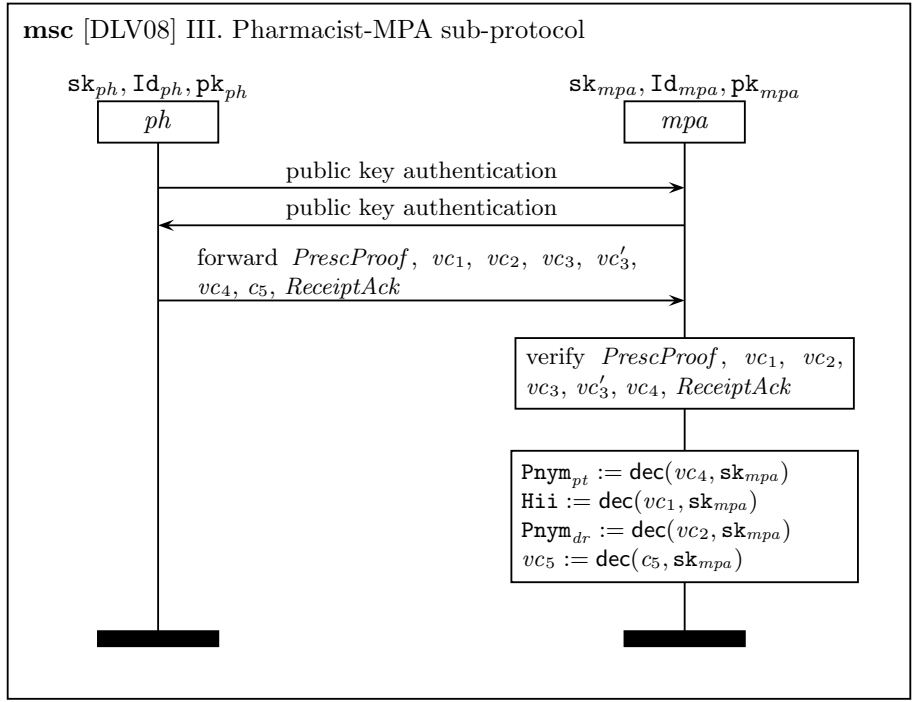

\begin{msc}{[DLV08] III. Pharmacist-MPA sub-protocol}
\instdist 7.0cm
\declinst{Ph}{$\pharmSk, \pharmID, \pkph$}{\thepharm}
\envinstdist 3.0cm
\declinst{MPA}{$\mpaSk, \mpaID, \pkmpa$}{\thempa}
\mess{\small public key authentication}{Ph}{MPA}
\nextlevel[1]
\mess{\small public key authentication}{MPA}{Ph}
\nextlevel[2]
\mess{\small \parbox{18em}{forward
$\prescProof$, $\vc_1$, $\vc_2$, $\vc_3$, $\vc'_3$, $\vc_4$, $\C_5$,
$\receptionAck$\\}
}{Ph}{MPA}
\nextlevel[1]
\action*{\small \parbox{13em}{verify $\prescProof$, $\vc_1$, $\vc_2$, $\vc_3$,
$\vc'_3$, $\vc_4$, $\receptionAck$}}{MPA}
\nextlevel[3]
\action*{\small \parbox{13em}{
	$\patientPseudo\defi\dec{\vc_4}{\mpaSk}$\\
	$\patientHii\defi\dec{\vc_1}{\mpaSk}$\\
	$\doctorPseudo\defi\dec{\vc_2}{\mpaSk}$\\
	$\vc_5\defi\dec{\C_5}{\mpaSk}$
}
}{MPA}
\nextlevel[4]
\end{msc}
\caption[MSC: Pharmacist-MPA sub-protocol.]{Pharmacist-MPA sub-protocol.}
\label{msc:pm}
\end{figure}
\begin{figure}[h]
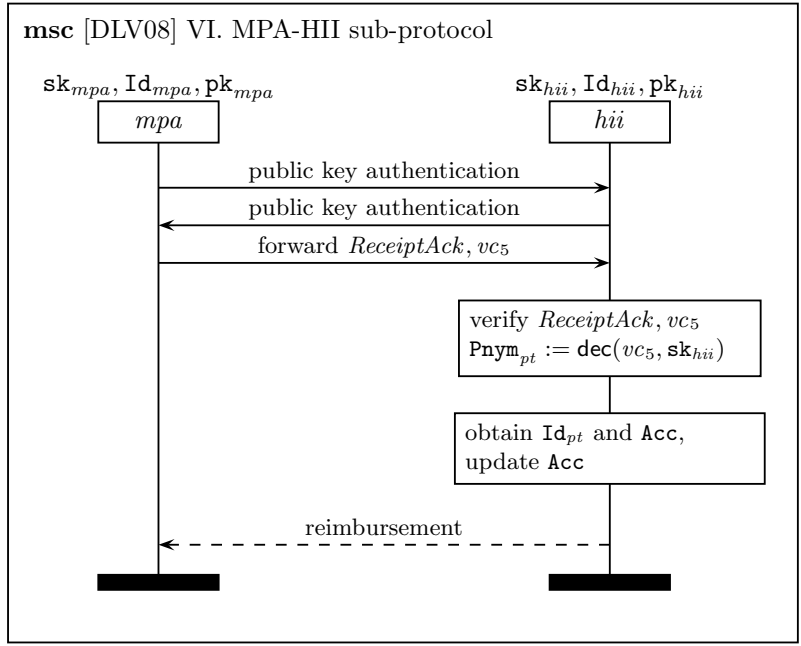

\begin{msc}{[DLV08] VI. MPA-HII sub-protocol}
\instdist 6.0cm
\declinst{MPA}{$\mpaSk, \mpaID, \pkmpa$}{\thempa}
\envinstdist 2.5cm
\declinst{HII}{$\hiiSk, \hiiID, \pkhii$}{\thehii}
\mess{\small public key authentication}{MPA}{HII}
\nextlevel[1]
\mess{\small public key authentication}{HII}{MPA}
\nextlevel[1]
\mess{\small forward $\receptionAck, \vc_5$}{MPA}{HII}
\nextlevel[1]
\action*{\small\parbox{11.5em}{ verify $\receptionAck, \vc_5$\\
 $\patientPseudo\defi\dec{\vc_5}{\hiiSk}$}}{HII}
\nextlevel[3]
\action*{\small \parbox{11.5em}{
obtain $\patientID$ and $\patientAcc$,\\
update $\patientAcc$
}
}{HII}
\nextlevel[3.5]
\mess*{\small reimbursement}{HII}{MPA}
\end{msc}
\caption[MSC: MPA-HII sub-protocol.]{MPA-HII sub-protocol.}
\label{msc:mh}
\end{figure}

\subsubsection{Patient-Pharmacist sub-protocol}

The pharmacist authenticates himself to the patient using public key authentication. The patient
verifies the authentication and obtains, from the authentication, the
pharmacist's identity and the pharmacist's MPA.
Then the patient anonymously authenticates
himself to
the pharmacist, and proves his social security status. Next, the patient
computes verifiable encryptions $\vc_1$, $\vc_2$, $\vc_3$, $\vc'_3$,
$\vc_4$, $\vc_5$, where
\begin{itemize}
\item $\vc_1$ encrypts the patient's HII using the
	MPA's public key and proves that
	the HII encrypted in $\vc_1$ is the same as the one in the
	patient's credential.
\item $\vc_2$ encrypts the doctor's pseudonym using the MPA's public key and
        proves that the
	doctor's pseudonym encrypted in $\vc_2$ is the same as the one in the
	doctor commitment embedded in the prescription.
\item $\vc_3$ encrypts the patient's pseudonym using the public safety
	organisation's public key and proves that the pseudonym
	encrypted in $\vc_3$ is the same as the one in the patient's commitment.
\item $\vc'_3$ encrypts the patient's HII using
	the social security organisation's public key and proves that
	the content encrypted in $\vc'_3$ is the same as the HII in the
	patient's credential.
\item $\vc_4$ encrypts the patient's pseudonym using the MPA's public key and
proves that the
	patient's pseudonym encrypted in $\vc_4$ is the same as the one in the
	patient's credential.
\item $\vc_5$ encrypts the patient's pseudonym using his HII's
public key and proves that the patient's
	pseudonym encrypted in $\vc_5$ is the same as the one in the patient's
	credential.
\item $\C_5$ encrypts $\vc_5$ using the MPA's public key.
\end{itemize}

The patient sends the received prescription to the
pharmacist and proves to the pharmacist that the patient's identity in
the prescription is the same as in the patient credential. The patient
sends $\vc_1, \vc_2, \vc_3, \vc'_3, \vc_4, \C_5$ as well. The pharmacist
verifies the correctness of all the received messages. If every message
is correctly formatted, the pharmacist charges the patient, and delivers
the medicine. Then the pharmacist generates an invoice and sends it to
the patient. The patient computes a receipt $\receptionAck$: signing a
message (consists of the prescription identity, the pharmacist's
identity, $\vc_1$, $\vc_2$, $\vc_3$, $\vc'_3$, $\vc_4$, $\vc_5$) using a
signed proof of knowledge and proving that he knows the patient
credential. This receipt proves that the patient has received his
medicine. The pharmacist verifies the correctness of the receipt.
The communication in the patient-Pharmacist sub-protocol is shown in
Figure~\ref{msc:pp}. Since the payment and medicine delivery procedures are out
of the protocol scope, they are interpreted as dashed arrows in the figure.

\subsubsection{Pharmacist-MPA sub-protocol}

The pharmacist and the MPA first authenticate each other using public
key authentication. Next, the pharmacist sends the received prescription
and the receipt $\receptionAck$, together with $\vc_1$, $\vc_2$,
$\vc_3$, $\vc'_3$, $\vc_4$, $\C_5$, to the MPA. The MPA verifies
correctness of the received information. Then the MPA decrypts $\vc_1$,
$\vc_2$, $\vc_4$ and $\C_5$, which provide the patient's HII, the
doctor's pseudonym, the patient's pseudonym, and $\vc_5$. The
communication in the pharmacist-MPA sub-protocol is shown in
Figure~\ref{msc:pm}. Note that after authentication, two parties often
establish a secure communication channel. However, it is not mentioned
in~\cite{DLVV08} that the pharmacist and the MPA agree on anything.
Nevertheless, this does not affect the properties that we verified,
except authentication between pharmacist and MPA.

\subsubsection{MPA-HII sub-protocol}

The MPA and the patient's HII first authenticate each other using public key
authentication. Then the MPA sends the
receipt $\receptionAck$ to the patient's HII as
well as the verifiable encryption $\vc_5$ which encrypts the patient's
pseudonym with the patient's HII's public key.
The patient's HII checks the correctness of
$\receptionAck$, decrypts $\vc_5$ and obtains the patient's pseudonym.
From the patient pseudonym, the HII obtains the
identity of the patient; then updates the patient's account and pays the
MPA. The MPA pays the pharmacist when he receives the payment.
The communication in the MPA-HII sub-protocol is shown in
Figure~\ref{msc:mh}. Similar to the previous sub-protocol, there is nothing established during the
authentication which can be used in the later message exchanges. Note that in addition
to authentications between MPA and HII, this
affects the secrecy of a patient's pseudonym when the adversary controls dishonest
patients (see Section~\ref{sec:dishonest_secrecy}).

\section{Modelling DLV08}\label{sec:dlvmodel}
We model the DLV08 protocol in the applied pi calculus as introduced in
Section~\ref{sec:appliedpi}. For clarity, we
also borrow some syntactic expressions from ProVerif, such as key words
`\funnosp', `\private\ \funnosp', `\reducnosp' and `\tequation', and
expression $`\tlet x=N \letin\ P'$.
Particularly, `(\private) \funnosp' denotes a constructor which uses terms to
form a more complex term ('\private' means the adversary cannot use it).
`\reducnosp' and `\tequation' are key words used to
construct the equational theory $E$.
`\reducnosp' denotes a destructor which retrieves sub-terms of a constructed term.
For the cryptographic primitives that cannot be captured by destructors, ProVerif
provides `\tequation' to capture the relationship between constructors.
The expression `$\tlet x=N \letin\ P$' is used as syntactical substitutions,
i.e., $P\sub{N}{x}$ in the applied pi calculus. It is an abbreviation of
$`\tlet x=N \letin\ P \telse\ Q'$ when $Q$ is the null process. When $N$ is a
destructor, there are two possible outcomes. If the term $N$ does not fail, then
$x$ is bound to $N$ and process $P$ is taken, otherwise $Q$ (in this case,
the null process) is taken.

Since the description of the protocol in its original paper is not clear in
some details, before modelling the protocol, several ambiguities need to be settled
(Section~\ref{sec:ambiguities}). Next we
explain the modelling of the cryptographic primitives
(Section~\ref{ssec:crypto}), since security and privacy rely heavily on
these cryptographic primitives in the protocol. Then, we illustrate the
modelling of the protocol (Section~\ref{ssec:model}).

\subsection{Underspecification of the DLV08 protocol}
\label{sec:ambiguities}
The DLV08 protocol leaves the following issues unspecified: \\

\begin{tabular}{ll}
\textbf{a1:} & whether a zero-knowledge proof is transferable; \\
\textbf{a2:} & whether an encryption is probabilistic; \\
\textbf{a3:} & whether a patient/doctor uses a fresh identity and/or pseudonym
	for each session; \\
\textbf{a4:} & whether credentials are freshly generated in each session; \\
\textbf{a5:} & what a patient's social security status is and how it can be
	modified; \\
\textbf{a6:} & how many HIIs exist and whether a patient can change his HII; \\
\textbf{a7:} & whether a patient/doctor can obtain a credential by requesting one; \\
\textbf{a8:} & what type of communication channels are used (public or untappable). \\
\end{tabular}
\\ \\
\noindent
To be able to discover potential flaws on privacy, we make the following
(weakest) assumptions in our modelling of the DLV08 protocol: \\

\begin{tabular}{lp{12cm}}
{\bf s1:} &
the zero-knowledge proofs used are non-interactive and transferable; \\
{\bf s2:} &
encryptions are not probabilistic;  \\
{\bf s3:} &
a patient/doctor uses the same identity and pseudonym in every session;
\\
{\bf s4:} &
a patient/doctor has the same credential in every session; \\
{\bf s5:} &
a patient's social security status is the same in every session; \\
{\bf s6:} &
there are many HIIs, different patients may have
different HIIs, and a patient's HII is fixed and cannot be changed; \\
{\bf s7:} &
a patient/doctor's credential can be obtained by requesting one; \\
{\bf s8:} &
the communication channels are public. \\
\end{tabular}

Note that some assumptions may look weak to security experts, for example
the assumption of deterministic encryption.
However, without explicit warning,
deterministic encryption algorithms may be used, which will lead to security flaws.
With this in mind, we assume the weakest assumption when there is ambiguity.
By assuming weak assumptions and showing the security flaws with the assumptions,
we provide security warnings for the implementation of the protocol.

\subsection{Modelling cryptographic primitives}
\label{ssec:crypto}
The cryptographic primitives are modelled in the applied pi calculus using
function symbols and equations.
All functions and equational theory are summarised
in Figures~\ref{fig:dlvfunction},~\ref{fig:eqtheory1} and~\ref{fig:eqtheory2}.
%
\begin{figure}[!h]
\begin{specification}
\begin{math}
\begin{array}{ll@{\hskip 1.5cm}ll@{\hskip 1.5cm}ll@{\hskip 1.5cm}ll}
\rule{0pt}{3.0ex}

\fun& \true/0.&
\fun& \hash/3.&
\fun& \pk/1.&
\fun& \fenc/2.\\
\fun& \comt/2.&
\fun& \fsign/2.&
\fun& \zk/2.&
\fun& \spk/3.\\
\fun& \invoice/1.&
\fun& \key/1.&
\fun& \host/1.\\
 & \multicolumn{3}{c}{\private\ \fun \drcred/2.}&
	\multicolumn{3}{l}{\hspace{-1ex}\private\ \fun \ptcred/5.}
\end{array}
\end{math}
\end{specification}
\caption[Functions in DLV08.]{Functions.}
\label{fig:dlvfunction}
\end{figure}
\begin{figure}[!h]
\begin{specification}
\begin{math}
\begin{array}{l@{\hskip 0.2cm}l@{\hskip 0.3cm}l}
\rule{0pt}{3.0ex}
\reduc & \dec{\enc{m}{\pk(sk)}}{sk}=m. &
	\text{(*asymmetric encryption*)}\\
\reduc & \open(\comt(x,y),y)=x. &
	\text{(*bit commitments*)} \\
\reduc & \checkAuth(\sign{x}{y}, \pk(y))=\true. &
	\text{(*signature verification*)} \\
\reduc & \getsignedmessage(\sign{x}{y},\pk(y))=x. &
	\text{(*message from signature*)} \\
\reduc & \getpublic(\zk(x,y))=y. &
	\text{(*public part of ZK*)} \\
\reduc & \getmsg(\spk(x,y,z))=z. &
	\text{(*message of SPK*)} \\
\reduc & \getpubmsg(\spk(x,y,z))=y. &
	\text{(*public part of SPK*)} \\
\tequation & \key(\host(x))=x. &
       \text{(*an identity has a unique public key*)}\\
\tequation & \host(\key(x))=x.&
       \text{(*a public key belongs to a unique}\\
       && \hfill \text{identity*)}\\
\end{array}
\end{math}
\end{specification}
\caption[Equational theory in DLV08 part \one: non-zero-knowledge part.]
{Equational theory part \one: non-zero-knowledge part.}
\label{fig:eqtheory1}
\end{figure}
\begin{figure}[!h]
\begin{specification}
\begin{math}
\begin{array}{ll}
\rule{0pt}{3.0ex}
\reduc& \doctorAuthVer(\zk((\doctorPseudo, \doctorID), \drcred(\doctorPseudo, \doctorID)),\\
&\hspace{12.2ex}       \drcred(\doctorPseudo, \doctorID))=\true.\\

\reduc& \patientAuthVer(\zk((\patientID, \patientPseudo, \patientHii,
       \patientSSS, \patientAcc), \\
& \hspace{15.5ex}\ptcred(\patientID, \patientPseudo, \patientHii,
       \patientSSS, \patientAcc)),\\
& \hspace{12ex}     \ptcred(\patientID, \patientPseudo, \patientHii,
       \patientSSS, \patientAcc))=\true.\\
\reduc& \patientProofVer(\zk((\patientID, \patientPseudo, \patientHii, \patientSSS, \patientAcc),\\
&\hspace{16ex}      (\commit(\patientID, \patientOpenInfo),\\
& \hspace{17ex}        \ptcred(\patientID, \patientPseudo, \patientHii, \patientSSS, \patientAcc))),\\
& \hspace{13ex}    \commit(\patientID, \patientOpenInfo),\\
& \hspace{13ex}     \ptcred(\patientID, \patientPseudo, \patientHii, \patientSSS, \patientAcc))=\true.\\

\reduc& \prescProofVer(\spk((\doctorPseudo, \doctorOpenInfo, \doctorID),\\
& \hspace{20.5ex}(\commit(\doctorPseudo, \doctorOpenInfo),
                            \drcred(\doctorPseudo, \doctorID)),\\
& \hspace{20.5ex} (\prescText, \prescID, \commit(\doctorPseudo, \doctorOpenInfo),\\
& \hspace{21.5ex}      \commit(\patientID, \patientOpenInfo))),\\
& \hspace{16ex}\drcred(\doctorPseudo, \doctorID),\prescText, \prescID,\\
& \hspace{16ex}    \commit(\doctorPseudo, \doctorOpenInfo),
 \commit(\patientID, \patientOpenInfo))=\true.\\
\reduc& \patientAuthsssVer(\zk(
                (\patientID, \patientPseudo, \patientHii, \patientSSS, \patientAcc),\\
&\hspace{18ex}(\ptcred(\patientID, \patientPseudo, \patientHii, \patientSSS,
	        \patientAcc), \patientSSS),\\
&\hspace{15ex}\ptcred(\patientID, \patientPseudo, \patientHii, \patientSSS,
	\patientAcc), \patientSSS)
                        =\true.\\
\reduc& \patientspkVer(\spk((\patientID, \patientPseudo, \patientHii, \patientSSS, \patientAcc, \patientOpenInfo),\\
& \hspace{17ex}(\ptcred(\patientID, \patientPseudo, \patientHii, \patientSSS,
	                   \patientAcc), \commit(\patientID, \patientOpenInfo)),\\
& \hspace{17.5ex} \nonce),\\
& \hspace{13ex} \ptcred(\patientID, \patientPseudo, \patientHii, \patientSSS,
	                   \patientAcc), \\
& \hspace{13ex}  \commit(\patientID, \patientOpenInfo), \nonce)=\true.\\
\reduc& \CheckVEncHii(\zk((\patientID, \patientPseudo, \patientHii, \patientSSS, \patientAcc),\\
& \hspace{14.5ex} (\ptcred(\patientID, \patientPseudo, \patientHii, \patientSSS, \patientAcc),\\
& \hspace{15.5ex} \enc{\patientHii}{\pubkey})),\\
& \hspace{11.5ex} \ptcred(\patientID, \patientPseudo, \patientHii, \patientSSS, \patientAcc),
 \enc{\patientHii}{\pubkey}, \pubkey)=\true.\\
\reduc& \CheckVEncDrnymMpa(\zk((\doctorPseudo,\doctorOpenInfo),\\
& \hspace{20.8ex} (\spk((\doctorPseudo, \doctorOpenInfo, \doctorID),\\
& \hspace{26.0ex}(\commit(\doctorPseudo, \doctorOpenInfo), \drcred(\doctorPseudo, \doctorID)),\\
& \hspace{26.0ex} (\prescText, \prescID, \\
& \hspace{26.5ex}   \commit(\doctorPseudo, \doctorOpenInfo), \xpatientCommitph)),\\
& \hspace{22ex}\enc{\doctorPseudo}{\pubkey})),\\
& \hspace{18ex} \spk((\doctorPseudo, \doctorOpenInfo, \doctorID),\\
& \hspace{22ex} (\commit(\doctorPseudo, \doctorOpenInfo), \drcred(\doctorPseudo, \doctorID)),\\
& \hspace{22ex} (\prescText, \prescID,\\
& \hspace{23ex} \commit(\doctorPseudo, \doctorOpenInfo), \xpatientCommitph)),\\
& \hspace{18.5ex} \enc{\doctorPseudo}{\pubkey}, \pubkey)=\true.\\
\reduc& \CheckVEncPtnym(\zk((\patientID, \patientPseudo, \patientHii, \patientSSS, \patientAcc),\\
&\hspace{17.5ex}(\ptcred(\patientID, \patientPseudo, \patientHii, \patientSSS, \patientAcc),
		         \enc{\patientPseudo}{\pubkey})),\\
&\hspace{14.5ex}\ptcred(\patientID, \patientPseudo, \patientHii, \patientSSS, \patientAcc),\\
&\hspace{14.5ex}\enc{\patientPseudo}{\pubkey}, \pubkey)=\true.\\
\reduc& \CheckReceptionAck(\spk((\patientID,\patientPseudo,\patientHii,\patientSSS, \patientAcc),\\
& \hspace{20.5ex}  \ptcred(\patientID,\patientPseudo,\patientHii,\patientSSS, \patientAcc),\\
&\hspace{20.5ex}  (\xprescID,\xpharmID,\vc_1,\vc_2,\vc_3,\vc'_3,\vc_4,\C_5)),\\
&\hspace{16.5ex} \ptcred(\patientID,\patientPseudo,\patientHii,\patientSSS, \patientAcc),\\
& \hspace{16.5ex}   \xprescID,\xpharmID,\vc_1,\vc_2,\vc_3,\vc'_3,\vc_4,\C_5)=\true.
\end{array}
\end{math}
\end{specification}
\caption[Equational theory in DLV08 part \two: zero-knowledge part.]
{Equational theory part \two: zero-knowledge part.}
\label{fig:eqtheory2}
\end{figure}

\paragraph{Bit-commitments.}
The bit-commitments scheme is modelled as two functions:
$\comt$, modelling the committing phase, and $\open$, modelling
the opening phase. The function $\comt$ creates a commitment with two
parameters: a message $m$ and a random number $r$. A commitment can only
be opened with the correct opening information $r$, in which case the
message $m$ is revealed.
\[
\begin{array}{rl}
\fun& \comt/2. \\
\reduc& \open(\comt(m,r),r)=m.
\end{array}
\]

\paragraph{Zero-knowledge proofs.}
Non-interactive zero-knowledge proofs can be modelled as function
$\zk(\mi{secrets}, \mi{pub\_info})$
inspired by~\cite{BMU08}. The public verification information $\mi{pub\_info}$ and
the secret information $\mi{secrets}$ satisfy a pre-specified relation. Since the secret
information is only known by the prover, only the prover can construct
the zero-knowledge proof. To verify a zero-knowledge proof is to check whether
the relation between the secret information and the verification
information is satisfied.
Verification of a zero-knowledge proof is modelled as function
$\zkver(\zk(\mi{secrets},\mi{pub\_info}), \mi{verif\_info})$,
with a zero-knowledge proof to be verified
$\zk(\mi{secrets},\linebreak\mi{pub\_info})$ and
the verification information $\mi{verif\_info}$.
Compared to the more generic definitions in~\cite{BMU08}, we
define each zero-knowledge proof specifically, as only a
limited number of zero-knowledge proofs are used in the protocol.
We specify each verification rule in Figure~\ref{fig:eqtheory2}.
Since the $\mi{pub\_info}$ and $\mi{verif\_info}$ happen to be the same in
all the zero-knowledge proofs verifications in this protocol, the
generic structure of verification rule is given as
\[
\zkver(\zk(\mi{secrets},\mi{pub\_info}),\mi{pub\_info})=\true,
\]
where $\true$ is a constant.
The specific function to check a zero-knowledge proof of
type $z$ is denoted as $\zkver_z$, e.g., verification of a patient's anonymous
authentication modelled by function $\patientAuthVer$.

\paragraph{Digital credentials.}
A digital credential is issued by trusted authorities. We assume the
procedure of issuing a credential is perfect, which means that the
adversary cannot forge a credential nor obtain one by impersonation. We
model digital credentials as a private function (declaimed by key word
$\private\ \fun$ in ProVerif)
which is only usable by
honest users. In the DLV08 protocol, a credential can have
several attributes; we model these as parameters of the credential
function.
\[
\begin{array}{l@{\hskip 1cm}l}
\private\ \fun \drcred/2.&
\private\ \fun \ptcred/5.
\end{array}
\]
There are two credentials in the DLV08 protocol: a doctor credential
which is modelled as $\doctorCred\defi \drcred(\doctorPseudo, \doctorID)$,
and a patient credential which is modelled as
$\patientCred\defi\linebreak\ptcred(\patientID, \patientPseudo, \patientHii,
\patientSSS, \patientAcc)$.
Unlike private data, the two private functions cannot be coerced, meaning
that even by coercing, the adversary cannot apply the private functions. Because
a doctor having an anonymous credentials is a basic setting of the protocol, and thus
the procedure of obtaining a credential is not assumed to be bribed or coerced. However,
the adversary can coerce patients or doctors for the credentials and parameters of the private functions.

\paragraph{Anonymous authentication.}
The procedure of anonymous authentication is a zero-knowledge proof
using the digital credential as public information. The anonymous
authentication of a doctor is modelled as
\[\doctorAuth\defi \zk((\doctorPseudo, \doctorID), \drcred(\doctorPseudo,\doctorID)),\]
and the verification of the authentication is modelled as
\[\doctorAuthVer(\doctorAuth,\drcred(\doctorPseudo,\doctorID)).\]
The equational theory for the verification is
\[
\reduc \doctorAuthVer(\zk((\doctorPseudo,\doctorID), \drcred(\doctorPseudo,\doctorID)),\drcred(\doctorPseudo,\doctorID))=\true.
\]
The verification implies that the creator of the authentication is a
doctor who has the credential $\drcred(\doctorPseudo,\doctorID)$. Because only
legitimate doctors can obtain a credential from authorities,
i.e., use the function $\drcred$ to create a credential; and the
correspondence between the parameters of the anonymous
authentication (the first parameter $(\doctorPseudo,\doctorID)$ in $\doctorAuth$) and the parameters of
the credential (parameters $\doctorPseudo$ and$\doctorID$ in $\drcred(\doctorPseudo,\doctorID)$) ensures that the prover can only be
the owner of the credential. Other doctors may be able to use function $\drcred$
but do not know $\doctorPseudo$ and $\doctorID$, and thus cannot create a valid proof.
The adversary can observe a credential
$\drcred(\doctorPseudo,\doctorID)$, but does not know secrets $\doctorPseudo, \doctorID$, and thus cannot forge a
valid zero-knowledge proof. If the adversary forges a zero-knowledge
proof with fake secret information $\doctorPseudo'$ and $\doctorID'$, the fake
zero-knowledge proof will not pass verification. For the same reason, a
validated proof proves that the credential belongs to the creator of the
zero-knowledge proof.
Similarly, an anonymous authentication of a patient is modelled as
\[
\begin{array}{rl}
\patientAuth\defi \zk(&\hspace{-2ex}(\patientID, \patientPseudo, \patientHii,
                     \patientSSS, \patientAcc),
\ptcred(\patientID, \patientPseudo, \patientHii,
                       \patientSSS, \patientAcc)),
\end{array}
\]
and the verification rule is modelled as
\[
\begin{array}{rl}
\reduc \patientAuthVer(&\hspace{-2ex}\zk((\patientID, \patientPseudo, \patientHii,
	\patientSSS, \patientAcc), \\
 & \hspace{1ex}\ptcred(\patientID, \patientPseudo, \patientHii,
	\patientSSS, \patientAcc)),\\
 & \hspace{-2ex}\ptcred(\patientID, \patientPseudo, \patientHii,
	\patientSSS, \patientAcc))=\true.
\end{array}
\]

\paragraph{Verifiable encryptions.}
A verifiable encryption is modelled as a zero-knowledge proof.
The encryption is embedded in the zero-knowledge proof as public information.
The receiver can obtain the cipher text from the proof. For
example, assume a patient wants to prove that he has encrypted a secret $s$ using
a public key $k$ to a pharmacist, while the pharmacist does not know the
corresponding secret key for $k$. The pharmacist cannot open the cipher
text to test whether it uses the public key $k$ for encryption. However, the
zero-knowledge proof can prove that the cipher text is encrypted using
$k$, while not revealing the secret $s$.
The general structure of the verification of a verifiable encryption is
\[
\vencver(\zk(\mi{secrets}, (\mi{pub\_info},
	\mi{cipher})),\mi{verif\_info})=\true,
\]
where $\mi{secrets}$ is private information, $\mi{pub\_info}$ and $\mi{cipher}$
consist public information, $\mi{verif\_info}$ is the verification information.

\paragraph{Signed proofs of knowledge.}
A signed proof of knowledge is a scheme which signs a message, and proves a property of
the signer. For the DLV08 protocol, this proof only concerns
equality of attributes of credentials and commitments (e.g., the identity
of this credential is the same as the identity of that commitment).
To verify a signed proof of knowledge, the verifier must know which
credentials/commitments are considered. Hence, this information must be
obtainable from the proof, and thus is included in the model.
In general, a signed proof of knowledge is modelled as function
\[\spk(\mi{secrets}, \mi{pub\_info}, \mi{msg}),\]
which models a signature using private value(s) \emph{secrets} on the
message \emph{msg}, with public information $\mi{pub\_info}$ as settings.
Similar to zero-knowledge proofs, $\mi{secrets}$ and $\mi{pub\_info}$ satisfy a
pre-specified relation. $\mi{msg}$ can be any message.
What knowledge is proven, depends on the specific instance of the proof
and is captured by the verification functions for the specific proofs.
For example, to prove that a user knows (a) all fields of a (simplified)
credential, (b) all fields of a commitment to an identity, and (c) that
the credential concerns the same identity as the commitment, he
generates the following proof:
\[
\begin{array}{l@{\hskip -0.5ex}l@{\hskip 1cm}l}
\spk(  & (\patientID,\patientPseudo,\patientOpenInfo),& (*\mi{secrets}*) \\
\qquad & (\ptcred(\patientID,\patientPseudo),
 	\comt(\patientID,\patientOpenInfo)), &(*\mi{public\_info}*)\\
       & \mi{msg}
).   &(*\mi{message}*)
\end{array}
\]
These proofs are verified by checking that the signature is correct,
given the signed message and the verification information.
E.g., the above example proof can be verified as follows:
\[
\begin{array}{lll@{\hskip 1cm}l}
\reduc \spkver( & \hspace{-2ex}\spk( &\hspace{-2ex}(\patientID,\patientPseudo,\patientOpenInfo), \\
 & & \hspace{-2ex}(\ptcred(\patientID,\patientPseudo),
	\comt(\patientID,\patientOpenInfo)), \\
 & & \hspace{-2ex}\mi{msg}\ ), &(*\mi{signed\_message}*)\\
 & \multicolumn{2}{l}{\hspace{-2ex}(\ \ptcred(\patientID,\patientPseudo),
	\comt(\patientID,\patientOpenInfo)\ ),} &(*\mi{verify\_info}*)\\

 &\hspace{-2ex}  \mi{msg} &&(*\mi{message}*)\\
 & \multicolumn{2}{l}{\hspace{-2ex}) = \true.}
\end{array}
\]

\paragraph{Other cryptographic primitives.}
Hash functions, encryptions and signing messages are modelled by
functions $\hash$, $\fenc$, and $\fsign$, respectively (see Figure~\ref{fig:dlvfunction}).
Correspondingly, decryption, verifying a signature and retrieving the
message from a signature are modelled as functions $\fdec$, $\checkAuth$ and
$\getsignedmessage$ (see Figure~\ref{fig:eqtheory1}).
Function $\pk$ models the corresponding public key of a secret key,
and function $\invoice$ is used for a pharmacist to generate an invoice for a patient
(see Figure~\ref{fig:dlvfunction}).
Functions
$\getpublic$, $\getpubmsg$ and $\getmsg$
model retrieving public information from a zero-knowledge proof,
from a signed proof of
knowledge, and obtaining the message from a signed proof of knowledge,
respectively (see Figure~\ref{fig:eqtheory1}).
Function $\key$ models the public key of a user's identity and function $\host$
retrieves the owner's identity from a public key (see Figure~\ref{fig:eqtheory1}).

\subsection{Modelling the DLV08 protocol}
\label{ssec:model}
We first show how to model each of the sub-protocols and
then how to compose them to form the full DLV08 protocol.

\paragraph{Modelling the doctor-patient sub-protocol.}
This sub-protocol is used for a doctor, whose steps are labelled
\textbf{d$i$} in Figure~\ref{fig:prodr}, to prescribe medicine for a
patient, whose steps are labelled \textbf{t$i$} in Figure~\ref{fig:propt1}.
\begin{figure}[ht]
\begin{specification}
{
\begin{math}
\begin{array}{lrl}
\rule{0pt}{3.0ex}
& \ProDr& \defi\\

{\bf d1}.&&\out{\ch}{\zk((\doctorPseudo, \doctorID),
                \drcred(\doctorPseudo, \doctorID))}.\\

{\bf d2}.&&\readin{\ch}{(\xpatientAuth,\xpatientProof)}.   \\
{\bf d3}.&&\tlet \xpatientCred=\getpublic(\xpatientAuth) \letin \\
{\bf d4}.&&\tlet (\xpatientCommit,=\xpatientCred)=\getpublic(\xpatientProof) \letin   \\
{\bf d5}.&&\tif \patientAuthVer(\xpatientAuth,\xpatientCred)=\true \then \\
{\bf d6}.&&\tif \patientProofVer(\xpatientProof,(\xpatientCommit,\xpatientCred))=\true \then \\

{\bf d7}.&&\nu \prescText.\\
{\bf d8}.&&\nu \doctorOpenInfo.\\
{\bf d9}.&&\tlet \prescID=\hash(\prescText,\xpatientCommit,
                       \comt(\doctorPseudo,\doctorOpenInfo)) \letin \\

{\bf d10}.&&\out{\ch}{(\spk((\doctorPseudo, \doctorOpenInfo, \doctorID),\\
&&\hspace{12.7ex}             (\comt(\doctorPseudo,\doctorOpenInfo),
                             \drcred(\doctorPseudo, \doctorID)),\\
&&\hspace{12.7ex}             (\prescText,\prescID,
                   \comt(\doctorPseudo,\doctorOpenInfo),\xpatientCommit)),\\
&&\hspace{8.5ex}                   \doctorOpenInfo)}

\end{array}
\end{math}
}
\end{specification}
\caption{The doctor process $\ProDr$.}
\label{fig:prodr}
\end{figure}
\begin{figure}[ht]
\begin{specification}
{
\begin{math}
\begin{array}{lrl}
\rule{0pt}{3.0ex}
& \ProPtpartone&\defi\\

{\bf t1}.&&\readin{\ch}{\xdoctorAuth}.\\
{\bf t2}.&&\tlet \xdoctorCred=\getpublic(\xdoctorAuth) \letin\\
{\bf t3}.&&\tif \doctorAuthVer(\xdoctorAuth,\xdoctorCred)=\true \then \\

{\bf t4}.&&\nu \patientOpenInfo.\\
{\bf t5}.&&\out{\ch}{(\zk((\patientID, \patientPseudo, \patientHii,
                 \patientSSS, \patientAcc), \\
&&\hspace{11.5ex}             \ptcred(\patientID, \patientPseudo, \patientHii,
                      \patientSSS, \patientAcc)), \\
&&\hspace{8.5ex}\zk((\patientID, \patientPseudo, \patientHii,
                   \patientSSS, \patientAcc), \\
&&\hspace{11.5ex}   (\commit(\patientID, \patientOpenInfo),\\
&&\hspace{12.5ex}                      \ptcred(\patientID, \patientPseudo, \patientHii,
                \patientSSS, \patientAcc))))}.\\

{\bf t6}.&&\readin{\ch}{(\xprescProof,\xdoctorOpenInfo)}.\\
{\bf t7}.&&\tlet (\xprescText,\xprescID, \xdoctorCommit,
          =\commit(\patientID, \patientOpenInfo))\\
&&\hspace{2.5ex} =\getmsg(\xprescProof) \letin\\
{\bf t8}.&&\tif \prescProofVer(\xprescProof,(\xdoctorCred,\xprescText,\\
&&\hspace{5ex} \xprescID,
       \xdoctorCommit,\commit(\patientID, \patientOpenInfo)))=\true \then \\

{\bf t9}.&& \tlet \xdoctorPseudo=\open(\xdoctorCommit,\xdoctorOpenInfo) \letin\ \ProPtparttwo

\end{array}
\end{math}
}
\end{specification}
\caption[The patient process part \one: $\ProPtpartone$.]
{The patient process in the doctor-patient sub-protocol $\ProPtpartone$.}
\label{fig:propt1}
\end{figure}

First, the doctor anonymously authenticates to the patient
using credential $\doctorCred$ ({\bf d1}).
The patient reads in the doctor authentication ({\bf t1}),
obtains the doctor credential ({\bf t2}),
and verifies the authentication ({\bf t3}).
If the verification in step ({\bf t3}) succeeds, the patient
anonymously authenticates himself to the doctor
using his credential ({\bf t5}, the first $\zk$ function),
generates a nonce $\patientOpenInfo$ ({\bf t4}),
computes a commitment with the nonce as opening information,
and proves that the patient identity used in the patient credential is the same as
in the commitment, thus linking the patient commitment and the
patient credential ({\bf t5}, the second $\zk$).

The doctor reads in the patient authentication as $\xpatientAuth$ and the
patient proof as\linebreak $\xpatientProof$ ({\bf d2}),
obtains the patient credential from the patient authentication ({\bf d3}),
obtains the patient commitment $\xpatientCommit$ and
the patient credential from the patient proof,
tests whether the credential matches the one embedded in the patient authentication ({\bf d4}),
then verifies the authentication ({\bf d5}) and the patient proof ({\bf d6}).
If the verification in the previous item succeeds, the doctor generates a
prescription $\prescText$\footnote{Note that a medical examination of the patient
is not part of the DLV08 protocol.} ({\bf d7}),
generates a nonce $\doctorOpenInfo$ ({\bf d8}),
computes a prescription identity $\prescID$ ({\bf d9}),
and computes a commitment $\doctorCommit$ using the nonce as opening information ({\bf d10}).
Next, the doctor signs the
message ($\prescText$, $\prescID$, $\doctorCommit$, $\xpatientCommit$) using
a signed proof of knowledge.
This proves the pseudonym
used in the credential $\doctorCred$ is the same as in the commitment
$\doctorCommit$, thus linking the prescription to the credential.
The doctor sends the signed proof of knowledge together with the open
information of the doctor commitment $\doctorOpenInfo$ ({\bf d10}).

The patient reads in the prescription as $\xprescProof$ and the opening information of
the doctor commitment ({\bf t6}),
obtains the prescription $\xprescText$, prescription identity $\xprescID$, doctor
commitment $\xdoctorCommit$, and tests the patient commitment signed in the receiving
message ({\bf t7}). Then the patient verifies the signed proof of prescription ({\bf t8}).
If the verification succeeds, the patient
obtains the doctor's pseudonym $\xdoctorPseudo$ by opening the doctor commitment ({\bf t9})
and continues the next sub-protocol behaving as in process $\ProPtparttwo$.

\smallskip
\noindent\textbf{Rationale for modelling of prescriptions.}
In the description of DLV08 protocol~\cite{DLVV08}, it is unclear
precisely what information is included in a prescription. Depending on
the implementation, a prescription
may contain various information, such as
name of medicines prescribed, amount of medicine prescribed, the
timestamp and organization that wrote the prescription, etc.
Some information in the prescription may reveal privacy of patients and
doctors. For instance, if the identities of patients and doctors are
included in the prescription, then doctor and patient \docunlink\ is
trivially broken. In addition, both (doctor/patient) anonymity and
untraceability would also be trivially broken,
if the prescriptions were revealed to the adversary.
In order to focus only on the logical flaws of the DLV08 protocol and
exclude such dependencies, we assume that the prescriptions in the
protocol are de-identified.
However, this may not be sufficient. Doctors may e.g.~be identifiable by
the way they prescribe, the order in which medicine appear on
prescriptions, etc. Such ``fingerprinting'' attacks would
also trivially break \docunlink.
For our analysis, we assume that a prescription cannot be
linked to its doctor or patient by its content. That is, the
prescription shall not be modelled as a function of doctor or patient
information.
To avoid any of the above concerns, we model prescriptions as
abstract pieces of data: each prescription is represented by a single,
unique name.
First, this modelling captures the assumption that prescriptions from
different doctors are often different even for the same diagnose, due to
different prescription styles.
Second, this allows us to capture an
infinite number of prescriptions in infinite sessions, without
introducing false attacks to the DLV08 protocol that are caused by the
modelling of the prescriptions.


\paragraph{Modelling the patient-pharmacist sub-protocol.}
This sub-protocol is used for a patient, whose steps are labelled
\textbf{t$i$} in Figure~\ref{fig:propt2}, to obtain medicine from a
pharmacist, whose steps are labelled \textbf{h$i$} in
Figure~\ref{fig:proph1}.
\begin{figure}[!ht]
\begin{specification}
{
\begin{math}
\begin{array}{lrl}
\rule{0pt}{3.0ex}
& \ProPtparttwo&\defi\\
{\bf t10}.&& \readin{\ch}{\xpharmAuth}.\\
{\bf t11}.&&\tif \checkPharmAuth(\xpharmAuth,\pkphpt)=\true \then \\
{\bf t12}.&&\tlet (=\xpharmID,\xmpaID)\\
&&\hfill =\getsignedmessage(\xpharmAuth,\pkphpt) \letin\\
{\bf t13}.&&\tlet \pkmpapt=\key(\xmpaID) \letin \\

{\bf t14}.&&\out{\ch}{\zk((\patientID,\patientPseudo,\patientHii,\patientSSS, \patientAcc),\\
&&\hspace{10.7ex} (\ptcred(\patientID,\patientPseudo,\patientHii,\patientSSS, \patientAcc),\patientSSS))}.\\

{\bf t15}.&&\nu \nonce.\\

{\bf t16}.&&\tlet \vc_1=\zk((\patientID,\patientPseudo,\patientHii,\patientSSS, \patientAcc),\\
&&\hspace{14.5ex}(\ptcred(\patientID,\patientPseudo,\patientHii,\patientSSS, \patientAcc),\\
&&\hspace{15.5ex}  \enc{\patientHii}{\pkmpapt})) \letin\\
{\bf t17}.&&\tlet \vc_2=\zk((\xdoctorPseudo,\xdoctorOpenInfo),\\
&&\hspace{14.5ex}          (\xprescProof,\enc{\xdoctorPseudo}{\pkmpapt})) \letin\\
{\bf t18}.&&\tlet \vc_3=\zk((\patientID,\patientPseudo,\patientHii,\patientSSS, \patientAcc),\\
&&\hspace{14.5ex}(\ptcred(\patientID,\patientPseudo,\patientHii,\patientSSS, \patientAcc),\\
&&\hspace{15.5ex}    \enc{\patientPseudo}{\pksso})) \letin\\
{\bf t19}.&&\tlet \vc'_3=\zk((\patientID,\patientPseudo,\patientHii,\patientSSS, \patientAcc),\\
&&\hspace{14.5ex}(\ptcred(\patientID,\patientPseudo,\patientHii,\patientSSS, \patientAcc),\\
&&\hspace{15.5ex}      \enc{\patientHii}{\pksso})) \letin\\
{\bf t20}.&&\tlet \vc_4=\zk((\patientID,\patientPseudo,\patientHii,\patientSSS, \patientAcc),\\
&&\hspace{14.5ex}(\ptcred(\patientID,\patientPseudo,\patientHii,\patientSSS, \patientAcc),\\
&&\hspace{15.5ex}     \enc{\patientPseudo}{\pkmpapt})) \letin\\
{\bf t21}.&&\tlet \vc_5=\zk((\patientID,\patientPseudo,\patientHii,\patientSSS, \patientAcc),\\
&&\hspace{14.5ex}(\ptcred(\patientID,\patientPseudo,\patientHii,\patientSSS, \patientAcc),\\
&&\hspace{15.5ex} \enc{\patientPseudo}{\pkhiipt})) \letin\\
{\bf t22}.&&\tlet \C_5=\enc{\vc_5}{\pkmpapt} \letin\\

{\bf t23}.&&\out{\ch}{(\xprescProof,
            \spk((\patientID, \patientPseudo, \patientHii, \patientSSS, \patientAcc),\\
&&\hfill (\ptcred(\patientID, \patientPseudo, \patientHii, \patientSSS, \patientAcc),
                    \commit(\patientID, \patientOpenInfo)),
                    \nonce),\\
&&\hfill      \vc_1,\vc_2,\vc_3,\vc'_3,\vc_4,\C_5)}.\\

{\bf t24}.&&\readin{\ch}{\xinvoice}.\\
{\bf t25}.&&\tlet \ReceptionAck=\spk((\patientID,\patientPseudo,\patientHii,\patientSSS, \patientAcc),\\
&&\hspace{23.5ex}\ptcred(\patientID,\patientPseudo,\patientHii,\patientSSS, \patientAcc),\\
&&\hfill(\xprescID,\xpharmID,\vc_1,\vc_2,\vc_3,\vc'_3,\vc_4,\C_5)) \letin \\
{\bf t26}.&&\out{\ch}{\ReceptionAck}

\end{array}
\end{math}
}
\end{specification}
\caption[The patient process part \two: $\ProPtparttwo$.]
{The patient process in the patient-pharmacist sub-protocol $\ProPtparttwo$.}
\label{fig:propt2}
\end{figure}
\begin{figure}[!ht]
\begin{specification}
{
\begin{math}
\begin{array}{lrl}
\rule{0pt}{3.0ex}
& \ProPhpartone&\defi\\
{\bf h1}.&&\out{\ch}{\sign{(\pharmID, \mpaIDph)}{\pharmSk}}.\\

{\bf h2}.&&\readin{\ch}{\xpatientAuthsss}.\\
{\bf h3}.&&\tlet (\xpatientCredph,\xpatientSSSph)=\getpublic(\xpatientAuthsss) \letin\\
{\bf h4}.&&\tif \patientAuthsssVer(\xpatientAuthsss, (\xpatientCredph,\xpatientSSSph))\\
&&\hfill =\true \then\\

{\bf h5}.&&\readin{\ch}{(\xprescProofph,\xpatientspkph,\\
&&\hspace{7ex}\xvc_1,\xvc_2,\xvc_3,\xvc'_3,\xvc_4,\xc_5)}.\\
{\bf h6}.&&\tlet (\xdoctorCommitph,\xdoctorCredph)\\
&&\hfill =\getpubmsg(\xprescProofph) \letin \\
{\bf h7}.&&\tlet (\xprescTextph,\xprescIDph,=\xdoctorCommitph,\xpatientCommitph)\\
&&\hfill=\getmsg(\xprescProofph) \letin \\
{\bf h8}.&&\tif \prescProofVer(\xprescProofph,(\xdoctorCredph,\xprescTextph,\\
&&\hfill  \xprescIDph,\xdoctorCommitph,\xpatientCommitph))=\true \then\\

{\bf h9}.&&\tlet \xnonce=\getmsg(\xpatientspkph) \letin \\
{\bf h10}.&&\tif \patientspkVer(\xpatientspkph,\\
&&\hfill(\xpatientCredph,\xpatientCommitph,\xnonce))=\true \then \\

{\bf h11}.&&\tlet (=\xpatientCredph,\xEnc_1)=\getpublic(\xvc_1) \letin \\
{\bf h12}.&&\tif \CheckVEncHii(\xvc_1,(\xpatientCredph,\xEnc_1,\pkmpaph))\\
&&\hfill=\true \then      \\

{\bf h13}.&&\tlet (=\xprescProofph,\xEnc_2)=\getpublic(\xvc_2) \letin \\
{\bf h14}.&&\tif \CheckVEncDrnymMpa(\xvc_2,(\xprescProofph,\\
&&\hfill
\xEnc_2,\pkmpaph))=\true \then \\

{\bf h15}.&&\tlet (=\xpatientCredph,\xEnc_3)=\getpublic(\xvc_3) \letin \\
{\bf h16}.&&\tif \CheckVEncPtnym(\xvc_3,(\xpatientCredph,\xEnc_3,\pksso))\\
&&\hfill=\true \then \\

{\bf h17}.&&\tlet (=\xpatientCredph,\xEnc'_3)=\getpublic(\xvc'_3) \letin \\
{\bf h18}.&&\tif \CheckVEncHii(\xvc'_3,(\xpatientCredph,\xEnc'_3,\pksso))=\true \then \\

{\bf h19}.&&\tlet (=\xpatientCredph,\xEnc_4)=\getpublic(\xvc_4) \letin \\
{\bf h20}.&&\tif \CheckVEncPtnym(\xvc_4,\\
&&\hfill(\xpatientCredph,\xEnc_4,\pkmpaph))=\true \then \\

{\bf h21}.&&\out{\ch}{\inv{\xprescIDph}}.\\

{\bf h22}.&&\readin{\ch}{\xReceptionAck}.\\
{\bf h23}.&&\tif \CheckReceptionAck(\xReceptionAck,(\xpatientCredph,\xprescIDph,\\
&& \hfill         \pharmID,\xvc_1,\xvc_2,\xvc_3,\xvc'_3,\xvc_4,\xc_5))=\true \\
&& \then\ \ProPhparttwo
\end{array}
\end{math}
}
\end{specification}
\caption[The pharmacist process part \one: $\ProPhpartone$.]
{The pharmacist process in the patient-pharmacist sub-protocol $\ProPhpartone$.}
\label{fig:proph1}
\end{figure}

First, the pharmacist authenticates to the patient using a public key authentication ({\bf h1}).
Note that the pharmacist does not authenticate anonymously, and that
the pharmacists's MPA identity is embedded.
The patient reads in the pharmacist authentication $\xpharmAuth$ ({\bf t10})
and verifies the authentication ({\bf t11}).
If the verification succeeds, the pharmacist
obtains the pharmacist's MPA identity from the authentication ({\bf t12}),
thus obtains the public key of MPA ({\bf t13}).
Then the patient anonymously authenticates himself to the pharmacist,
and proves his social security status using the proof $\patientAuthsss$ ({\bf t14}).
The patient generates a nonce which will be used as a message
in a signed proof of knowledge ({\bf t15}),
and computes verifiable encryptions $\vc_1$, $\vc_2$, $\vc_3$, $\vc_3'$, $\vc_4$ and
$\vc_5$ ({\bf t16-t21}).
These divulge
the patient's HII, the doctor's pseudonym, and the patient's
pseudonym to the MPA, the patient's pseudonym to the HII, and the
patient pseudonym and HII to the social safety
organisation, respectively.
The patient encrypts $\vc_5$ with MPA's public key as $\C_5$ ({\bf t22}).
The patient computes a signed proof of knowledge
$$
\begin{array}{rl}
\patientspk=&\spk((\patientID, \patientPseudo, \patientHii, \patientSSS, \patientAcc),\\
&\quad\ \ (\ptcred(\patientID, \patientPseudo, \patientHii, \patientSSS, \patientAcc),
                    \commit(\patientID, \patientOpenInfo)),\\
&\qquad               \nonce)
\end{array}$$ which
proves that the patient identity embedded in the prescription is
the same as in his credential. In the
prescription, this identity is contained in a
commitment. For simplicity, we model the proof using the commitment instead of
the prescription. The link between commitment and
prescription is ensured when the proof is verified ({\bf h10}).

The patient sends the prescription $\xprescProof$, the signed proof $\patientspk$, and
$\vc_1,\vc_2,\vc_3,\linebreak \vc'_3,\vc_4,\C_5$ to the pharmacist ({\bf t23}).
The pharmacist reads in the authentication $\xpatientAuthsss$ ({\bf h2}),
obtains the patient credential and his social security status ({\bf h3}),
verifies the authentication ({\bf h4}).
If the verification succeeds, the pharmacist reads in the patient's prescription $\xprescProofph$,
the signed proof of knowledge $\xpatientspkph$, the verifiable encryptions
$\xvc_1$, $\xvc_2$, $\xvc_3$, $\xvc'_3$, $\xvc_4$, and cipher text $\xc_5$ ({\bf h5});
and verifies $\xprescProofph$ ({\bf h6-h8}),
$\xpatientspkph$ ({\bf h9-h10}), and $\xvc_1$, $\xvc_2$, $\xvc_3$, $\xvc'_3$, $\xvc_4$ ({\bf h11-h20}).
If all the verifications succeed, the pharmacist
charges the patient, and delivers the medicine (neither
are modelled as they are out of DLV08's scope).
Then the pharmacist generates an invoice with the prescription identity
embedded in it and sends the invoice to the patient ({\bf h21}).	

The patient reads in the invoice ({\bf t24}),
computes a receipt: a signed proof of knowledge $\receptionAck$
which proves that he receives the medicine ({\bf t25}); and sends the signed proof of
knowledge to the pharmacist ({\bf t26}).	
The pharmacist reads in the receipt $\xReceptionAck$ ({\bf h22}), verifies its
correctness ({\bf h23}) and continues the next sub-protocol behaving as in $\ProPhparttwo$.

\paragraph{Modelling the pharmacist-MPA sub-protocol.}
The pharmacist-MPA sub-protocol is used for the pharmacist, whose steps are labelled
\textbf{h$i$} in Figure~\ref{fig:proph2} to report the received
prescriptions to the MPA, whose steps are labelled
\textbf{m$i$} in Figure~\ref{fig:prompa1}.
\begin{figure}[!h]
\begin{specification}
\begin{math}
\begin{array}{lrl}
\rule{0pt}{3.0ex}
 &\ProPhparttwo&\defi\\
{\bf h24}.&& \out{\ch}{(\sign{(\pharmID,\mpaIDph)}{\pharmSk}, \pharmID)}.\\
{\bf h25}.&&\readin{\ch}{\xmpaAuth}.\\
{\bf h26}.&& \tif \checkMpaAuth(\xmpaAuth,\pkmpaph)=\true \then\\
{\bf h27}.&&\out{\ch}{(\xprescProofph,\\
&&\hspace{8.5ex}\xvc_1,\xvc_2,\xvc_3,\xvc'_3,\xvc_4,\xc_5,\\
&&\hspace{8.5ex}\xReceptionAck)}
\end{array}
\end{math}
\end{specification}
\caption[The pharmacist process part \two: $\ProPhparttwo$.]
{The pharmacist process in the pharmacist-MPA sub-protocol $\ProPhparttwo$.}
\label{fig:proph2}
\end{figure}
\begin{figure}[!ht]
\begin{specification}
\begin{math}
\begin{array}{lrl}
\rule{0pt}{3.0ex}
&\ProMPApartone&\defi\\
{\bf m1}.&&\readin{\ch}{(\xpharmAuthmpa,\xpharmIDmpa)}.\\
{\bf m2}.&&\tlet \pkphmpa=\key(\xpharmIDmpa) \letin\\
{\bf m3}.&&\tif \checkPharmAuth(\xpharmAuthmpa,\pkphmpa)=\true \then\\
{\bf m4}.&&\tlet (=\xpharmIDmpa,=\mpaID)\\
&&\hfill=\getmsg(\xpharmAuthmpa,\pkphmpa) \letin\\
{\bf m5}.&&\out{\ch}{\sign{\mpaID}{\mpaSk}}.\\

{\bf m6}.&&\readin{\ch}{(\xprescProofmpa,\xvcmpa_1,\xvcmpa_2,\xvcmpa_3,\\
&&\hspace{7ex}\xvcmpa'_3,\xvcmpa_4,\xcmpa_5,\xReceptionAckmpa)}. \\
{\bf m7}.&&\tlet (\xdoctorCommitmpa,\xdoctorCredmpa)\\
&&\hfill =\getpubmsg(\xprescProofmpa) \letin\\
{\bf m8}.&&\tlet (\xprescTextmpa,\xprescIDmpa,=\xdoctorCommitmpa,\xpatientCommitmpa)\\
&&\hfill=\getmsg(\xprescProofmpa) \letin\\
{\bf m9}.&&\tif \prescProofVer(\xprescProofmpa,(\xdoctorCredmpa,\xprescTextmpa,\\
&&\hfill \xprescIDmpa,\xdoctorCommitmpa,\xpatientCommitmpa))=\true \then\\

{\bf m10}.&&\tlet (=\xpatientCredmpa,\xEncmpa_1)=\getpublic(\xvcmpa_1) \letin\\
{\bf m11}.&&\tif \CheckVEncHii(\xvcmpa_1,\\
&&\hfill(\xpatientCredmpa,\xEncmpa_1,\pkmpa))=\true \then\\
{\bf m12}.&&\tlet \xpatientHiimpa=\dec{\xEncmpa_1}{\mpaSk} \letin\\

{\bf m13}.&&\tlet (=\xprescProofmpa,\xEncmpa_2)\\
&&\hfill=\getpublic(\xvcmpa_2) \letin\\
{\bf m14}.&&\tif \CheckVEncDrnymMpa(\xvcmpa_2,\\
&&\multicolumn{1}{r}{(\xprescProofmpa,\xEncmpa_2,\pkmpa))=\true \then}\\
{\bf m15}.&&\tlet \xdoctorPseudompa=\dec{\xEncmpa_2}{\mpaSk} \letin\\

{\bf m16}.&&\tlet (=\xpatientCredmpa,\xEncmpa3)=\getpublic(\xvcmpa_3) \letin\\
{\bf m17}.&&\tif \CheckVEncPtnym(\xvcmpa_3,\\
&&\hfill(\xpatientCredmpa,\xEncmpa_3,\pksso))=\true \then\\

{\bf m19}.&&\tlet (=\xpatientCredmpa,\xEncmpa'_3)=\getpublic(\xvcmpa'_3) \letin\\
{\bf m20}.&&\tif \CheckVEncHii(\xvcmpa'_3,\\
&&\hfill(\xpatientCredmpa,\xEncmpa'_3,\pksso))=\true \then\\

{\bf m21}.&&\tlet (=\xpatientCredmpa,\xEncmpa_4)=\getpublic(\xvcmpa_4) \letin\\
{\bf m22}.&&\tif \CheckVEncPtnym(\xvcmpa_4,\\
&&\hfill(\xpatientCredmpa,\xEncmpa_4,\pkmpa))=\true \then\\
{\bf m23}.&&\tlet \xpatientPseudompa=\dec{\xEncmpa_4}{\mpaSk} \letin         \\

{\bf m24}.&&\tif \CheckReceptionAck(\xReceptionAckmpa,(\xpatientCredmpa,\\
&&\hspace{4ex}                 \xprescIDmpa,\xpharmIDmpa,\xvcmpa_1,\xvcmpa_2,\\
&&\hspace{4ex} \xvcmpa_3,\xvcmpa'_3,\xvcmpa_4,\xcmpa_5))=\true \then\ \ProMPAparttwo
\end{array}
\end{math}
\end{specification}
\caption[The MPA process part \one: $\ProMPApartone$.]
{The MPA process in the pharmacist-MPA sub-protocol $\ProMPApartone$.}
\label{fig:prompa1}
\end{figure}

As the pharmacist mostly forwards the information supplied by the
patient, this protocol greatly resembles the patient-pharmacist protocol
described above.
Each step is modelled in details as follows:
The pharmacist authenticates himself to his MPA by sending his identity and
the signed identities of the pharmacist and the MPA ({\bf h24}).
The MPA stores this authentication in $\xpharmAuthmpa$, and
stores the pharmacist's identity in $\xpharmIDmpa$ ({\bf m1}).
From the pharmacist's identity, the MPA obtains the pharmacist's public key ({\bf m2}).
Then the MPA
verifies the pharmacist's authentication against the pharmacist's public key ({\bf m3}).
If the verification succeeds, according to the corresponding rule in the equational theory,
and the MPA verifies that he is indeed the pharmacist's MPA ({\bf m4}),
the MPA then authenticates itself to the pharmacist by sending the
signature of his identity ({\bf m5}).
The pharmacist reads in the MPA's authentication in $\xmpaAuth$ ({\bf h25}), and verifies the authentication ({\bf h26}).
If the verification succeeds, the pharmacist
sends the following to the MPA:
prescription $\xprescProofph$, received receipt
$\xReceptionAck$, and verifiable encryptions $\xvc_1$,
$\xvc_2$, $\xvc_3$, $\xvc'_3$, $\xvc_4$, $\xc_5$ ({\bf h27}).
The MPA reads in the information ({\bf m6}) and verifies their correctness ({\bf m7-m24}).
If the verifications succeed,
the MPA decrypts the corresponding encryptions ($\xEncmpa_1$, $\xEncmpa_2$, and
$\xEncmpa_4$) embedded in $\xvcmpa_1, \xvcmpa_2, \xvcmpa_4$, and obtains the patient's HII ({\bf m12}), the doctor
pseudonym ({\bf m15}), the patient pseudonym ({\bf m23}).
Then the MPA continues the next sub-protocol behaving as in process $\ProMPAparttwo$.
The storing information to database by the MPA is beyond our concern.

\paragraph{Modelling the MPA-HII sub-protocol.}
This protocol covers the exchange of information between the
pharmacist's MPA, whose steps are labelled \textbf{m$i$} in
Figure~\ref{fig:prompa2} and the patient's HII, whose steps are
labelled \textbf{i$i$} in Figure~\ref{fig:prohii}.

\begin{figure}[!h]
\begin{specification}
\begin{math}
\begin{array}{lrl}
\rule{0pt}{3.0ex}
&\ProMPAparttwo&\defi\\
{\bf m25}.&&\out{\ch}{(\sign{\mpaID}{\mpaSk},\mpaID)}. \\

{\bf m26}.&&\readin{\ch}{\xhiiAuthmpa}.\\
{\bf m27}.&&\tlet \pkhiimpa=\key(\xpatientHiimpa) \letin\\
{\bf m28}.&&\tif \checkHiiAuth(\xhiiAuthmpa,\pkhiimpa)=\true \then\\
{\bf m29}.&&\tif \getsignedmessage(\xhiiAuthmpa,\pkhiimpa)=\xpatientHiimpa \then\\

{\bf m30}.&&\out{\ch}{(\xReceptionAckmpa,\dec{\xcmpa_5}{\mpaSk})}.\\
{\bf m31}.&&\readin{\ch}{\xinvoicempa}
\end{array}
\end{math}
\end{specification}
\caption[The MPA process part \two: $\ProMPAparttwo$.]
{The MPA process in the MPA-HII sub-protocol $\ProMPAparttwo$.}
\label{fig:prompa2}
\end{figure}
\begin{figure}[!h]
\begin{specification}
\begin{math}
\begin{array}{lrl}
\rule{0pt}{3.0ex}
&\ProHII&\defi\\
{\bf i1}.&&\readin{\ch}{(\xmpaAuthhii,\xmpaIDhii)}.\\
{\bf i2}.&&\tlet \pkmpahii=\key(\xmpaIDhii) \letin\\
{\bf i3}.&&\tif \checkMpaAuth(\xmpaAuthhii,\pkmpahii)=\true \then \\

{\bf i4}.&&\out{\ch}{\sign{\hiiID}{\hiiSk}}.\\

{\bf i5}.&&\readin{\ch}{(\xReceptionAckhii,\xvchii_5)}.\\
{\bf i6}.&&\tlet \xpatientCredhii=\getpubmsg(\xReceptionAckhii) \letin\\
{\bf i7}.&& \tlet(\xprescIDhii,\xpharmIDhii,\xvchii_1,\xvchii_2,\xvchii_3,\xvchii'_3,\\
&&\hfill\xvchii_4, \xchii_5)=\getmsg(\xReceptionAckhii) \letin\\
{\bf i8}.&&\tif \CheckReceptionAck(\xReceptionAckhii,(\xpatientCredhii,\\
&&\hspace{4ex}\xprescIDhii,\xpharmIDhii,\xvchii_1,\xvchii_2,\xvchii_3,\xvchii'_3,\\
&&\hspace{4ex}\xvchii_4,\xchii_5))  =\true \then\\

{\bf i9}.&&\tlet (=\xpatientCredhii,\xEnchii_5)=\getpublic(\xvchii_5) \letin\\
{\bf i10}.&&\tif \CheckVEncPtnym(\xvchii_5,(\xpatientCredhii,\xEnchii_5,\pkhii))=\true\
\mbox{then}\\
{\bf i11}.&&\tlet \xpatientPseudohii=\dec{\xEnchii_5}{\hiiSk} \letin\\

{\bf i12}.&&\out{\ch}{\invoice(\xprescIDhii)}
\end{array}
\end{math}
\end{specification}
\caption{The HII process $\ProHII$.}
\label{fig:prohii}
\end{figure}
The MPA sends his identity to the HII and authenticates to the HII using public
key authentication ({\bf m25}). The HII stores the MPA's identity in
$\xmpaIDhii$ and stores the authentication in $\xmpaAuthhii$ ({\bf i1}).
From the MPA's identity, the HII obtains the MPA's public key ({\bf i2}).
Then the HII verifies the MPA's authentication ({\bf i3}).
If the verification succeeds, the HII
authenticates to the MPA using public key authentication ({\bf i4}).
The MPA stores the authentication in $\xhiiAuthmpa$ ({\bf m26}). Then the
MPA obtains the HII's public key from the HII's identity ({\bf m27}) and
verifies the HII's authentication ({\bf m28}).
If the verification succeeds, and the MPA verifies that the authentication is
from the intended HII ({\bf m29}), the MPA
sends the receipt $\xprescProofmpa$ and the
patient pseudonym encrypted for the HII -- verifiable encryption
$\xvcmpa_5=\dec{\xcmpa_5}{\mpaSk}$ ({\bf m30}).
The HII receives the receipt as $\xReceptionAckhii$ and the encrypted patient
pseudonym for the HII as $\xvchii_5$ ({\bf i5}). The HII verifies the above
two pieces of information ({\bf i6-i10}). If the verifications succeed,
the HII decrypts the encryption $\xEnchii_5$ and obtains the patient's pseudonym ({\bf i11}).
Finally, the HII sends an invoice of the prescription identity to the
MPA ({\bf i12}). The MPA stores the invoice in $\xinvoicempa$ ({\bf m31}).
Afterwards, the HII pays the MPA and updates the patient account. As
before, handling payment and storing information are beyond the scope of
the DLV08 protocol and therefore, we do not model this stage.

\begin{figure}[!h]
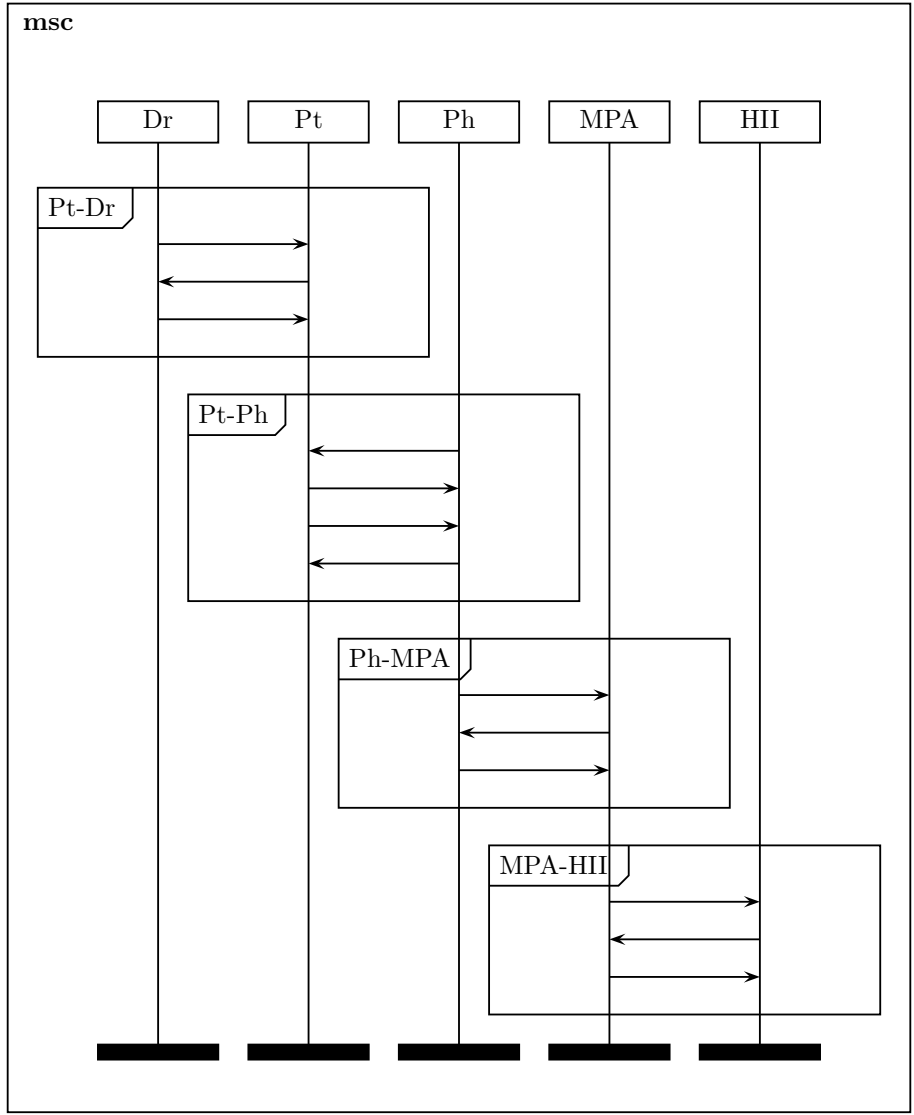

\begin{msc}{}
\setlength{\instdist}{2cm}
\declinst{Dr}{}{Dr}
\declinst{Pt}{}{Pt}
\declinst{Ph}{}{Ph}
\declinst{MPA}{}{MPA}
\declinst{HII}{}{HII}
\setlength{\inlineoverlap}{1.6cm}
\inlinestart{exp1}{Pt-Dr}{Dr}{Pt}
\nextlevel[1.5]
\mess{}{Dr}{Pt}
\nextlevel
\mess{}{Pt}{Dr}
\nextlevel
\mess{}{Dr}{Pt}
\nextlevel
\inlineend{exp1}
\nextlevel
\inlinestart{exp2}{Pt-Ph}{Pt}{Ph}
\nextlevel[1.5]
\mess{}{Ph}{Pt}
\nextlevel
\mess{}{Pt}{Ph}
\nextlevel
\mess{}{Pt}{Ph}
\nextlevel
\mess{}{Ph}{Pt}
\nextlevel
\inlineend{exp2}
\nextlevel
\inlinestart{exp3}{Ph-MPA}{Ph}{MPA}
\nextlevel[1.5]
\mess{}{Ph}{MPA}
\nextlevel
\mess{}{MPA}{Ph}
\nextlevel
\mess{}{Ph}{MPA}
\nextlevel
\inlineend{exp3}
\nextlevel
\inlinestart{exp4}{MPA-HII}{MPA}{HII}
\nextlevel[1.5]
\mess{}{MPA}{HII}
\nextlevel
\mess{}{HII}{MPA}
\nextlevel
\mess{}{MPA}{HII}
\nextlevel
\inlineend{exp4}
\end{msc}
\caption{Overview of DLV08 protocol.}
\label{msc:dlv}
\end{figure}

\paragraph{The full protocol.}
In summary, the DLV08 protocol is composed as shown in Figure~\ref{msc:dlv}.
The DLV08 protocol is modelled as the five roles $\Role_{\thedoctor}$,
$\Role_{\thepatient}$, $\Role_{\thepharm}$, $\Role_{\thempa}$, and
$\Role_{\thehii}$ running in parallel (Figure~\ref{fig:composedlv}).
\[
\begin{array}{rcl}
\dlvprotocol&\defi& \nu \vect{mc}. \init. (!\Role_{\thepatient}\mid!
\Role_{\thedoctor}\mid!\Role_{\thepharm}\mid!\Role_{\thempa}\mid!\Role_{\thehii})\\
\init&\defi&
    \tlet \pksso=\pk(\ssoSk) \letin\
         \out{\ch}{\pksso}
         \end{array}
\]
where $\nu \vect{mc}$ represents global secrets $\ssoSk$ and private channels
$\privchhiipt$, $\privchmpaph$, $\privchphpt$;
process $\init$ initialises the settings of the protocol -- publishing the
public key $\pksso$, so that the adversary knows it.
The roles $\Role_{\thedoctor}$, $\Role_{\thepatient}$,
$\Role_{\thepharm}$, $\Role_{\thempa}$ and $\Role_{\thehii}$ are obtained by
adding the settings of each role (see Section~\ref{ssec:setting}) to the
previously modelled corresponding process of the role as shown in
Figure~\ref{fig:composedr}, Figure~\ref{fig:composept}, Figure~\ref{fig:composeph},
Figure~\ref{fig:composempa} and Figure~\ref{fig:composehii}, respectivley.
\begin{figure}[!h]
\begin{specification}
\begin{math}
\begin{array}{rl}
\rule{0pt}{3.0ex}
\dlvprotocol\defi
&\left. \nu \ssoSk.
 \nu \privchhiipt.
 \nu \privchmpaph.
 \nu \privchphpt. \hspace{20ex} \right \} \nu\, \dataset\\

&\left. \hspace{-1ex}
\begin{array}{l}
\tlet \pksso=\pk(\ssoSk) \letin \\
\out{\ch}{\pksso}.
\end{array} \hspace{21ex} \right \} \init
\\
&(!(\Role_{\thedoctor})\mid
!(\Role_{\thepatient}) \mid
!(\Role_{\thepharm}) \mid
!(\Role_{\thempa}) \mid
!(\Role_{\thehii}))
\end{array}
\end{math}
\end{specification}
\caption{The process for the DLV08 protocol.}
\label{fig:composedlv}
\end{figure}

Each doctor has an identity $\doctorID$ ({\bf rd1}), a pseudonym $\doctorPseudo$ ({\bf rd2}) and behaves
like $\ProDr$ ({\bf rd3}) as shown in Figure~\ref{fig:composedr}.
The anonymous doctor credential is modelled by applying function $\drcred$ on $\doctorID$ and $\doctorPseudo$.
\begin{figure}[!h]
\begin{specification}
\begin{math}
\begin{array}{lrcl}
\rule{0pt}{3.0ex}
&\Role_{\thedoctor}&\defi& \\
{\bf rd1.} &&& \nu \doctorID.\\
{\bf rd2.} & &&\left.
    \nu \doctorPseudo.
  \hspace{42ex} \right \} \init_{\thedoctor}\\
{\bf rd3.} &   &&!(\ProDr)
\end{array}
\end{math}
\end{specification}
\caption{The process for role doctor $\Role_{\thedoctor}$.}
\label{fig:composedr}
\end{figure}

Each patient (as shown in Figure~\ref{fig:composept}) has an identity $\patientID$ ({\bf rt1}), a pseudonym $\patientPseudo$,
a social security status $\patientSSS$, a health expense account $\patientAcc$ ({\bf rt2}).
Unlike the identity and the pseudonym, which are attributes of a doctor, a
doctor's $\patientHii$ is an association relation, and thus is modelled by
reading in an HII identity to establish the relation ({\bf rt3}). In addition,
a patient communicates with a pharmacist in each session. Which pharmacist the
patient communicates with is decided by reading in a pharmacist's public key ({\bf rt4}).
From the pharmacist's public key, the patient can obtain the pharmacist's identity ({\bf rt5}).
Finally the patient behaves as $\ProPtpartone$ ({\bf rt6}).
\begin{figure}[!h]
\begin{specification}
\begin{math}
\begin{array}{lrcll}
\rule{0pt}{3.0ex}
&\Role_{\thepatient}&\defi&\\
{\bf rt1.} &&& \nu \patientID.\\
\multicolumn{4}{l}{ \left.
  \begin{array}{l}
\hspace{-1.5ex}{\bf rt2.} \hspace{13ex} \nu \patientPseudo.
  \nu \patientSSS.
  \nu \patientAcc.  \\
\hspace{-1.5ex}{\bf rt3.} \hspace{13ex}\readin{\privchhiipt}{\patientHii}.
  \tlet \pkhiipt=\key(\patientHii) \letin
  \end{array}
 \hspace{22ex} \right \} \init_{\thepatient}}\\
\multicolumn{4}{l}{\left.
 \begin{array}{l}
\hspace{-1.5ex}{\bf rt4.} \hspace{13ex}!(\readin{\privchphpt}{\xpkph}.\\
\hspace{-1.5ex}{\bf rt5.} \hspace{15ex}\tlet \pkphpt=\xpkph \letin\
     \tlet \pharmID=\host(\xpkph) \letin\\
\hspace{-1.5ex}{\bf rt6.} \hspace{15ex}\ProPtpartone)
\end{array}
\right \} !(\ProPt)}
\end{array}
\end{math}
\end{specification}
\caption{The process for role patient $\Role_{\thepatient}$.}
\label{fig:composept}
\end{figure}

Each pharmacist has a secret key $\pharmSk$ ({\bf rh1}), a public key
$\pkph$ ({\bf rh2}) and an identity $\pharmID$ ({\bf rh3}) as shown in
Figure~\ref{fig:composeph}. The public key of a pharmacist is published over
channel $\ch$, so that the adversary knows it ({\bf rh4}). In addition, the public key
is sent to the patients via private channel $\privchphpt$ ({\bf rh4}), so that the patients
can choose one to communicate with.
In each session, the pharmacist communicates with an MPA. Which MPA the
pharmacist communicates with is decided by reading in a public key of MPA ({\bf rh5}).
From the public key, the pharmacist can obtain the identity of the MPA ({\bf rh6}).
Finally, the pharmacist behave as $\ProPhpartone$.
\begin{figure}[!ht]
\begin{specification}
\begin{math}
\begin{array}{lrclll}
\rule{0pt}{3.0ex}
&\hspace{5.5ex}\Role_{\thepharm}&\defi&\\
\multicolumn{4}{l}{
\left.
\hspace{-1.5ex}\begin{array}{lrcl}
{\bf rh1.}\hspace{7ex}&  &&\nu \pharmSk.\\
{\bf rh2.}&&&\tlet \pkph=\pk(\pharmSk) \letin\\
{\bf rh3.}&&&\tlet \pharmID=\host(\pkph) \letin\\
{\bf rh4.}&&&\out{\ch}{\pkph}.\out{\privchphpt}{\pkph}.
\end{array}\hspace{16ex}\right\}\init_{\thepharm}}\\
\multicolumn{4}{l}{
\left.
\begin{array}{l}
\hspace{-1.20ex}{\bf rh5.} \hspace{14ex}!(\readin{\privchmpaph}{\pkmpaph}.\\
\hspace{-1.2ex}{\bf rh6.}\hspace{15.5ex} \tlet \mpaIDph=\host(\pkmpaph) \letin\ \ProPhpartone)
\end{array}\right\}!(\ProPh)}
\end{array}
\end{math}
\end{specification}
\caption{The process for role pharmacist $\Role_{\thepharm}$.}
\label{fig:composeph}
\end{figure}

Each MPA has a secret key $\mpaSk$ ({\bf rm1}), a public key $\pkmpa$ ({\bf rm2})
and an identity $\mpaID$ ({\bf rm3}). The MPA publishes his public key as well
as sends his public key to pharmacists ({\bf rm4}), and behaves as $\ProMPApartone$ ({\bf rm5})
as shown in Figure~\ref{fig:composempa}.
\begin{figure}[!ht]
\begin{specification}
\begin{math}
\begin{array}{lrclll}
\rule{0pt}{3.0ex}
&\Role_{\thempa}&\defi&\\
\multicolumn{4}{l}{
\left.\hspace{-1.5ex}
\begin{array}{lrcl}
{\bf rm1.} \hspace{8ex}&&& \nu \mpaSk.\\
{\bf rm2.} &&&  \tlet \pkmpa=\pk(\mpaSk) \letin\\
{\bf rm3.} &&&\tlet \mpaID=\host(\pkmpa) \letin\\
{\bf rm4.} &&&
\out{\ch}{\pkmpa}.\out{\privchmpaph}{\pkmpa}.
\end{array} \hspace{12ex}\right\}\init_{\thempa}}\\
{\bf rm5.}&&&!(\ProMPApartone)
\end{array}
\end{math}
\end{specification}
\caption{The process for role MPA $\Role_{\thempa}$.}
\label{fig:composempa}
\end{figure}

Similar to MPA, each HII (Figure~\ref{fig:composehii}) has a secret key $\hiiSk$
({\bf ri1}), a public key $\pkhii$ ({\bf ri2}) and an identity $\hiiID$ ({\bf ri3}).
The public key is revealed to the adversary via channel $\ch$ and sent to the
patients via channel $\privchhiipt$ ({\bf ri4}). Then the HII behaves as $\ProHII$ ({\bf ri5}).
\begin{figure}[!ht]
\begin{specification}
\begin{math}
\begin{array}{lrclll}
\rule{0pt}{3.0ex}
&\Role_{\thehii}&\defi& \\
\multicolumn{4}{l}{
\left.\hspace{-1.5ex}
\begin{array}{lrcl}
{\bf ri1.} \hspace{7ex}&&& \nu \hiiSk.\\
{\bf ri2.} &&&  \tlet \pkhii=\pk(\hiiSk) \letin\\
{\bf ri3.} &&&\tlet \hiiID=\host(\pkhii) \letin\\
{\bf ri4.} &&&\out{\ch}{\pkhii}.\out{\privchhiipt}{\hiiID}.
\end{array} \hspace{18ex}\right\}\init_{\thehii}}\\
{\bf ri5.} &&&!(\ProHII)
\end{array}
\end{math}
\end{specification}
\caption{The process for role HII $\Role_{\thehii}$.}
\label{fig:composehii}
\end{figure}

\section{Analysis of DLV08}
\label{sec:analyse}
In this section, we analyse whether DLV08 satisfies the following properties:
\begin{itemize}
\item \emph{secrecy of patient and doctor information},
\item \emph{authentication},
\item \emph{(strong) patient and doctor anonymity},
\item \emph{(strong) patient and doctor untraceability},
\item \emph{(enforced) \docunlink}, and
\item \emph{independence of (enforced) \docunlink}.
\end{itemize}
The properties \emph{doctor anonymity} and \emph{untraceability} are not
required by the protocol but are still interesting to analyse. The
verification is supported by the automatic verification tool
\emph{ProVerif}~\cite{Blanchet01,Blanchet02,Blanchet04}. The tool has been
used to verify many secrecy, authentication and privacy properties,
e.g., see~\cite{AB05,ABF07,LCPD07,BC08,DJP11}.
The verification results for secrecy are summarised in
Table~\ref{tab:secrecy}, and those for authentication in
Table~\ref{tab:auth}. As we are foremost interested in privacy
properties, the verification results for privacy properties, and
suggestions for improvements are discussed in
Section~\ref{sec:dlv08fixes}. Table~\ref{tab:result} summarises those
results, causes of privacy weaknesses, suggested improvements, and the
effect of the improvements. In this section, we show the verification results
of properties from basic to more complicated. A flaw which fails a basic property is likely to
fail a more complicated property as well. Thus we first show flaws of basic properties and how to
fix them, then we show new flaws of complicated properties based on the fixed model.

\subsection{ProVerif}
ProVerif takes a protocol and a property modelled in the applied pi
calculus as input (the input language (untyped version) differs slightly
from applied pi, see~\cite{proverif12}), and returns either a proof of
correctness or potential attacks. A protocol modelled in the
applied pi calculus is translated to Horn clauses~\cite{Horn51}. The
adversary's capabilities are added as Horn clauses as well. Using
these clauses, verification of secrecy and authentication is equivalent
to determining whether a certain clause is derivable from the set of
initial clauses.

Secrecy of a term is defined as the adversary cannot obtain the term by
communicating with the protocol and/or applying cryptography on the
output of the protocol~\cite{AB05}. The secrecy property is modelled as
a predicate in ProVerif: the query of secrecy of term $M$ is
$``attacker: M"$~\cite{Blanchet01}. ProVerif determines whether the term
$M$ can be inferred from the Horn clauses representing the adversary
knowledge.

Authentication is captured by correspondence properties of events in
processes: if one event happens the other event must have happened
before~\cite{ABF07,Blanchet09}. Events are tags which mark important
stages reached by the protocol. Events have arguments, which allow us to
express relationships between the arguments of events. A correspondence
property is a formula of the form: $\mi{ev}: \eventf{M}
\correspondenceto \mi{ev}: \eventg{N}$. That is, in any process, if
event $\eventf{M}$ has been executed, then the event $\eventg{N}$ must
have been previously executed, and any relationship between $M$ and $N$
must be satisfied. To capture stronger authentication, where an
injective relationship between executions of participants is required,
an injective correspondence property $\mi{evinj}: \eventf{M}
\correspondenceto \mi{evinj}: \eventg{N}$ is defined: in any process,
for each execution of event $\eventf{M}$, there is a distinct earlier
execution of the event $\eventg{N}$, and the relationship between $M$
and $N$ is satisfied.


In addition, ProVerif provides automatic verification of labelled
bisimilarity of two processes which differ only in the choice of some
terms~\cite{BAF08}. An operation $``choice[a, b]"$ is introduced to
model the different choices of a term in the two processes. Using this
operation, the two processes can be written as one process -- a
\emph{bi-process}. 
\begin{example}
To verify the equivalence
\[ \nu \name{a}. \nu \name{b}. \out{\pubchannel}{\name{a}}. \out{\pubchannel}{\name{e}}\symbol{symbol:pubchannel}
\eq
 \nu \name{a}. \nu \name{b}. \out{\pubchannel}{\name{b}}. \out{\pubchannel}{\name{d}}
\]
where $\pubchannel$ is a public channel, $\name{e}$ and $\name{d}$ are two free names,
we can query the following bi-process in ProVerif:
\[
\prs\defi \nu \name{a}. \nu \name{b}. \out{\pubchannel}{\choice{\name{a}}{\name{b}}}. \out{\pubchannel}{\choice{\name{e}}{\name{d}}}.
\]
\vspace{-4ex}
\label{ex:biprocess}
\end{example}
Using the first parameter of all $``choice"$ operations in a 
bi-process $\prs$, we obtain one side of the equivalence (denoted as 
$\fst{\prs}$); using the second parameters, we obtain the other side 
(denoted as $\snd{\prs}$).
\begin{example}
For the bi-process in Example~\ref{ex:biprocess}, using the first parameter to 
replace each $``choice"$ operation, we obtain 
$$\nu \name{a}. \nu \name{b}. \out{\pubchannel}{\name{a}}. \out{\pubchannel}{\name{e}},$$ 
which is the left-hand side of the equivalence in Example~\ref{ex:biprocess}; 
using the second parameter to replace each $``choice"$ operation, we obtain 
$$\nu \name{a}. \nu \name{b}. \out{\pubchannel}{\name{b}}. \out{\pubchannel}{\name{d}},$$ 
which is the right-hand side of the equivalence.
\end{example}

Given a bi-process $\prs$, ProVerif tries to prove that
$\fst{\prs}$ is labelled bisimilar to $\snd{\prs}$. The fundamental idea
is that ProVerif reasons on traces of the bi-process $\prs$: the
bi-process $\prs$ reduces when $\fst{\prs}$ and $\snd{\prs}$ reduce in
the same way; when $\fst{\prs}$ and $\snd{\prs}$ do something that may
differentiate them, the bi-process is stuck. Formally, ProVerif shows
that the bi-process $\prs$ is uniform, that is, if $\fst{\prs}$ can do a
reduction to some $\Prs_1$, then the bi-process can do a reduction to
some bi-process $\Prs$, such that $\fst{\Prs}\steq \Prs_1$ and
symmetrically for $\snd{\prs}$ taking a reduction to $\Prs_2$. When the
bi-process $\prs$ always remains uniform after reduction and addition of
an adversary, $\fst{\prs}$ is labelled bisimilar to $\snd{\prs}$.

\label{ssec:proverif}

\subsection{Secrecy of patient and doctor information}\label{sec:sec}

The DLV08 protocol claims to satisfy the following requirement: any
party involved in the prescription processing workflow should not know
the information of a patient and a doctor unless the information is
intended to be revealed in the protocol. In~\cite{DLVV08}, this
requirement is considered as an access control requirement. We argue
that ensuring the requirement with access control is not sufficient when
considering a communication network. A dishonest party could potentially
act as an attacker from the network (observing the network and
manipulating the protocol) and obtain information which he should not
access. It is not clearly stated which (if any) of the involved parties
are honest. We find that in such a way, some patient and doctor
information may be revealed to parties who should not know the
information.

We formalise the requirement as standard secrecy of patient and doctor
information with respect to the Dolev-Yao adversary. Standard secrecy of
a term captures the idea that the adversary cannot access to that term
(see Section~\ref{ssec:proverif}). If a piece of information is known to
the adversary, a dishonest party acting like the adversary can access to
the information. We do not consider strong secrecy, as it is unclear
whether the information is guessable. Recall that standard secrecy of a
term $M$ is formally defined as a predicate $``attacker: M"$ (see
Section~\ref{ssec:proverif}). By replacing $M$ with the listed private
information, we obtain the formal definition of the secrecy of patient
and doctor information. The list of private information of patients and
doctors that
needs to be protected is:
patient identity ($\patientID$),
doctor identity ($\doctorID$),
patient pseudonym ($\patientPseudo$),
doctor pseudonym ($\doctorPseudo$),
a patient's social security status ($\patientSSS$), and
a patient's health insurance institute ($\patientHii$).
Although DLV08 does not explicitly require it, we additionally analyse
secrecy of the health expense account $\patientAcc$ of a patient.
\[
\begin{array}{lll}
query\ attacker: \patientID&
query\ attacker: \doctorID&
query\ attacker: \patientPseudo\\
query\ attacker: \doctorPseudo&
query\ attacker: \patientSSS&
query\ attacker: \patientHii\\
query\ attacker: \patientAcc
\end{array}
\]

\paragraph{Verification result.}
We query the standard secrecy of the set of private information using ProVerif~\cite{Blanchet01}.
The verification results (see Table~\ref{tab:secrecy}) show that
a patient's identity, pseudonym, health expense
account, health insurance institute and identity of a doctor
($\patientID$, $\patientPseudo$, $\patientHii$
$\patientAcc$, $\doctorID$) satisfy standard secrecy;
a patient's social security status $\patientSSS$ and a doctor's pseudonym
$\doctorPseudo$ do not satisfy standard secrecy.
The $\patientSSS$ is revealed by the proof of social security status from the
patient to the pharmacist. The $\doctorPseudo$ is revealed by the revealing of
both the commitment of the patient's pseudonym and the open key to the
commitment during the communication between the patient and
the doctor.

Fixing secrecy of a patient's social security status requires
that the proof of social security status only reveals the status to the
pharmacist. Since how a social security status is represented and what the
pharmacist needs to verify are not clear,
we cannot give explicit suggestions. However, if the social security status
is a number, and the pharmacist
only needs to verify that the number is higher than a certain threshold, the
patient can prove it using zero-knowledge proof without revealing the number;
if the pharmacist needs to verify the exact value of the status, one way to fix
its secrecy is that the pharmacist and the patient agree on a session key and the
status is encrypted using the key.
Similarly, a way to fix the secrecy of $\doctorPseudo$
is to encrypt the opening information using the agreed session key.

\begin{table}[!t]
\centering
\begin{tabular}{| l | c | c | c |}
\hline
{\bf checked Security property} & {\bf initial model}  & {\bf cause(s)} &
{\bf improvement} \\
\hline
Secrecy of $\patientID$ & $\ok$ & & \\
Secrecy of $\patientPseudo$ & $\ok$  && \\
Secrecy of $\patientSSS$ & $\fail$ &revealed& session key \\
Secrecy of $\patientHii$ & $\ok$&&  \\
Secrecy of $\patientAcc$& $\ok$ && \\
Secrecy of $\doctorID$& $\ok$ && \\
Secrecy of $\doctorPseudo$& $\fail$ & revealed& session key   \\
\hline
\end{tabular}
\caption{Verification results of secrecy for patients and doctors.}
\label{tab:secrecy}
\end{table}

\subsection{Patient and doctor authentication}\label{sec:auth}
The protocol claims that all parties should be able to properly authenticate each other.
Compared to authentications between public entities, pharmacists, MPA and HII,
we focus on authentications between patients and doctors, as patients and
doctors use anonymous authentication.
Authentications between patients and pharmacists are sketched as well.

The DLV08 claims that no party should be able to succeed in claiming a false
identity, or false information about his identity. That is the adversary cannot
pretend to be a patient or a doctor.
\paragraph{Authentication from a patient to a doctor.}
The authentication from a patient to a doctor is defined as when the doctor finishes his process
and believes that
he prescribed medicine for a patient, then the patient did ask the doctor
for prescription.
To verify the authentication of a patient,
we add an event \(\EndDr(\xpatientCred,\xpatientCommit)\) at
the end of the doctor process (after line {\bf d10}), meaning the doctor believes that he
prescribed medicine for a patient who has a credential $\xpatientCred$ and committed
$\xpatientCommit$;
and add an event
\(\StartPt(\ptcred(\patientID,\patientPseudo,\patientHii,\patientSSS, \patientAcc),
\commit(\patientID, \patientOpenInfo))\) in the patient process (between line {\bf t4} and line {\bf t5}),
meaning that the patient did ask for a
prescription. The definition is captured by the following correspondence property:
 \(\mi{ev(inj)}: \EndDr(x,y) ==> \mi{ev(inj)}: \StartPt(x,y),\) meaning
that when the event $\EndDr$ is executed, there is
a (unique) event $\StartPt$ has been executed before.

\paragraph{Authentication from a doctor to a patient.}
Similarly, the authentication from a doctor to a patient is defined as when the patient believes that he
visited a doctor, the doctor did prescribe medicine for the patient.
To authenticate a doctor, we add to the patient process an event
\(\EndPt(\xdoctorCred,\xdoctorCommit,\xprescText,\xprescID)\) (after line {\bf t9}), and add an event
\(\StartDr(\drcred(\doctorPseudo,\doctorID),
\commit(\doctorPseudo,\doctorOpenInfo),\prescText,\prescID)\) in the
doctor process (between line {\bf d9} and line {\bf d10}), then
query \(\mi{ev(inj)}: \EndPt(x,y,z,t) ==> \mi{ev(inj)}: \StartDr(x,y,z,t).\)

\paragraph{Authentication from a patient to a pharmacist.}
The authentication from a patient to a doctor is defined as when the pharmacist
finishes a session and believes that he communicates with a patient, who is identified with the
credential $\xpatientCredph$, then the patient with the credential did communicate
with the pharmacist. to verify the authentication of a patient, we add to the
pharmacist process the event $\EndPh(\xpatientCredph)$ (after line {\bf h23}),
add to the patient process $\StartPtph(\ptcred(\patientID,\patientPseudo,
\patientHii,\patientSSS, \patientAcc))$ (between line {\bf t13} and line {\bf t14}),
and query $\mi{ev(inj)}: \EndPh(x)==>\mi{ev(inj)}:\StartPtph(x)$.

\paragraph{Authentication from a pharmacist to a patient.}
The authentication from a pharmacist to a patient is defined as when the patient
finishes a session and believes that he communicates with a pharmacist with the
identity $\xpharmID$, then the pharmacist is indeed the one who communicated
with the patient. To verify this authentication, we add the event
$\EndPtph(\xpharmID)$ into the patient process (after line {\bf t26}), add
the event $\StartPh(\pharmID)$ into the pharmacist process (between line {\bf h1}
and line {\bf h2}), and query $\mi{ev(inj)}:\EndPtph(x) ==> \mi{ev(inj)}:\StartPh(x)$.
In addition, we add the conditional evaluation $\tif \xinvoice=\inv{\xprescID} \then$
before the end $\EndPtph(\xpharmID)$ in the patient process
to capture that the patient checks the correctness of the invoice.

\paragraph{Verification results.}
The queries are verified using ProVerif.
The verification results show that
doctor authentication, both injective and non-injective, succeed;
non-injective patient authentication succeeds and
injective patient authentication fails.
The failure is caused by a replay attack from the adversary. That is,
the adversary can impersonate a patient by replaying old messages from the
patient. This authentication flaw leads to termination of the successive
procedure, the patient-pharmacist sub-process.
We verified authentication between patients and pharmacists as well.
Non-injective patient authentication succeeds, whereas injective patient authentication fails.
This means that the messages received by a pharmacist are from the correct patient, but
not necessarily from this communication session.
Neither non-injective nor injective pharmacist authentication succeeds:
the adversary can record and replay the first message which is sent from a pharmacist to a patient,
and pass the authentication by pretending to be that pharmacist. In addition,
the adversary can prepare the second message sending from a pharmacist to a patient, and thus does not need to replay the second message.
Since the adversary alters messages, non-injective pharmacist authentication fails. The verification results are summarised in Table~\ref{tab:auth}.

The reason that injective patient authentication fails for both doctors
and pharmacists is that they suffer from replay attacks. One possible
solution approach is to add a challenge sent from the doctor
(respectively, the pharmacist) to the patient. Then, when the patient
authenticates to the doctor or pharmacist, the patient includes this challenge
in the proofs. This approach assures that the proof is freshly generated.
Therefore, this prevents the adversary replaying old messages.

The reason that (injective and non-injective) authentication from a
pharmacist to a patient fails is that
the adversary can generate an invoice to replace the one from the real
pharmacist. One solution is for
the pharmacist to sign the invoice.
\begin{table}
\centering
\begin{tabular}{| l | c | c | c |}
\hline
{\bf checked Auth} & {\bf initial model}  & {\bf cause(s)} &
{\bf improvement} \\
\hline
dr to pt (inject)& $\ok$ &  &  \\
dr to pt (non-inject)& $\ok$ && \\
pt to dr (inject)& $\fail$ & replay attack& add challenge  \\
pt to dr (non-inject)& $\ok$ &  & \\
ph to pt (inject)& $\fail$ & adv. can replay 1st message,&  sign the invoice \\
 &&compute 2nd message & \\
ph to pt (non-inject)& $\fail$ &adv. can replay 1st message,  & sign the invoice  \\
&&compute 2nd message&\\
pt to ph (inject)& $\fail$ & replay attack& add challenge  \\
pt to ph (non-inject)& $\ok$ &  & \\
\hline
\end{tabular}
\caption{Verification results of authentication of patients and doctors.}
\label{tab:auth}
\end{table}

\paragraph{Authentications between public entities.}
The public entities -- pharmacists, MPAs, HIIs, authenticate each other using
public key authentication. The authentication is often used to agree on a way
for the later communication. Since it is not mentioned in the original protocol
that a key or a communication channel is established during authentication, we
assume that the later message exchanges are over public channels, to model
the worst case. In this model, the authentications between public entities are
obviously flawed, since the adversary can reuse messages from other sessions.
The flaws are confirmed by the verification results using ProVerif.

\subsection{(Strong) patient and doctor anonymity}

The DLV08 protocol claims that no party should be able to determine the identity
of a patient. We define (strong) patient anonymity to capture the requirement. Note
that in the original paper of the DLV08 protocol, the terminology of the
privacy property for capturing this requirement is patient untraceability.
Our definition of untraceability (Definition~\ref{def:untra}) has different
meaning from theirs (for details, see Section~\ref{sec:untrace}).
Also note that the satisfaction of standard secrecy of patient identity does
not fully capture this requirement, as
the adversary can still guess about it.

\paragraph{Patient and doctor anonymity.}
Doctor anonymity is defined as in Definition~\ref{def:ano}. Patient
anonymity can be defined in a similar way by replacing the role of
doctor with the role of patient.
\[
\begin{array}{rl}
\contexthealth{
\init_{\thepatient}\substitution{\ptA}{\patientIDv}.
!\ProPt\substitution{\ptA}{\patientIDv}
}
\eq
\contexthealth{
\init_{\thepatient}\substitution{\ptB}{\patientIDv}.
!\ProPt\substitution{\ptB}{\patientIDv}
}.
\end{array}
\]
To verify doctor/patient anonymity, is to check the satisfiability of
the corresponding equivalence between processes in the definition. This is done
by modelling the two processes on two sides of the equivalence as a
bi-process, and verify the bi-process using ProVerif.
Recall that a bi-process models two processes sharing the same structure and
differing only in terms or destructors.
The two processes are written as one process with choice-constructors which
tells ProVerif the spots where the two processes differ.
The bi-process for verifying doctor anonymity is
\[
\nu \vect{mc}. \init.
(!\Role_{\thepatient}\mid
!\Role_{\thedoctor}\mid
!\Role_{\thepharm}\mid
!\Role_{\thempa}\mid
!\Role_{\thehii}\mid
(\nu \doctorPseudo.
\tlet \doctorID=\choice{\drA}{\drB} \letin\ !\ProDr),
\]
and the bi-process for verifying patient anonymity is
\[
\begin{array}{l}
\nu \vect{mc}. \init.
(!\Role_{\thepatient}\mid
!\Role_{\thedoctor}\mid
!\Role_{\thepharm}\mid
!\Role_{\thempa}\mid
!\Role_{\thehii}\mid
(\tlet \patientID=\choice{\ptA}{\ptB} \letin\\
 \hspace{9ex}\nu \patientPseudo.
  \nu \patientSSS.
  \nu \patientAcc.
  \readin{\privchhiipt}{\patientHii}.
  \tlet \pkhiipt=\key(\patientHii) \letin\
!\ProPt).
\end{array}
\]
Since the doctor identity is a secret information, we define $\drA$ and $\drB$ as private
names $\private\ \free\ \drA. $\linebreak$\private\ \free\ \drB$. In addition, we
consider a stronger version, in which the adversary knows the two doctor
identities a priori, i.e., we verify whether the adversary can distinguish two known doctors
as well. This is modelled by defining the two doctor identities as free names,
$\free\ \drA. \free\ \drB$. Similarly, we verified two versions of patient anonymity
- in one version, the adversary does not know the two patient identities, and
in the other version, the adversary initially knows the two patient identities.
\paragraph{Strong patient and doctor anonymity.}
Strong doctor anonymity
is defined as in Definition~\ref{def:sano}. By replacing the role of doctor
with the role of patient, we obtain the definition of strong patient anonymity.
The bi-process for verifying strong doctor anonymity is
\[
\begin{array}{l}
\free\ \drA;\\
\nu \vect{mc}. \init. (!\Role_{\thepatient}\mid!\Role_{\thedoctor}\mid!
\Role_{\thepharm}\mid!\Role_{\thempa}\mid!\Role_{\thehii}\mid\\
 \hspace{9ex}(\nu \drB.
\nu \ndoctorPseudo.
\tlet \doctorPseudo=\ndoctorPseudo \letin\ !(\tlet \doctorID=\choice{\drB}{\drA} \letin\
 \ProDr))),
\end{array}
\]
and the bi-process for verifying strong patient anonymity is
\[
\begin{array}{l}
\free\ \ptA;\\
\nu \vect{mc}. \init. (!\Role_{\thepatient}\mid!\Role_{\thedoctor}\mid!
\Role_{\thepharm}\mid!\Role_{\thempa}\mid!\Role_{\thehii}\mid
 (
\nu \ptB.
\nu \patientPseudo.
  \nu \patientSSS.
  \nu \patientAcc.\\
\hspace{9ex}\readin{\privchhiipt}{\patientHii}.
  \tlet \pkhiipt=\key(\patientHii) \letin\
  !(\tlet \patientID=\choice{\ptB}{\ptA} \letin\
     \ProPt))).
\end{array}
\]
Note that by definition, the identities $\drA$ and $\ptB$ is known by the adversary.

In the first bi-process, by choosing $\drB$, we obtain
\[
\begin{array}{l}
\free\ \drA;\\
\nu \vect{mc}. \init. (!\Role_{\thepatient}\mid!\Role_{\thedoctor}\mid!
\Role_{\thepharm}\mid!\Role_{\thempa}\mid!\Role_{\thehii}\mid\\
 \hspace{9ex}(\nu \drB.
\nu \ndoctorPseudo.
\tlet \doctorPseudo=\ndoctorPseudo \letin\ !(\tlet \doctorID=\drB \letin\
 \ProDr))).
\end{array}
\]
Since $\drA$ never appears in the remaining process, removing the declaration ``$\free\ \drA;$" does not affect the process.
Since process ``$\nu \drB.\nu \ndoctorPseudo.
\tlet \doctorPseudo=\ndoctorPseudo \letin\ !(\tlet \doctorID=\drB \letin\
 \ProDr)$" essentially renames the doctor role process ``$\Role_{\thedoctor}$" (Figure~\ref{fig:composedr}) - renaming $\doctorID$ as $\drB$ and renaming $\doctorPseudo$ as $\ndoctorPseudo$, we have that the above
 process is structurally equivalent to (using rule $\REPL$)
\[
\begin{array}{l}
\nu \vect{mc}. \init. (!\Role_{\thepatient}\mid!\Role_{\thedoctor}\mid!
\Role_{\thepharm}\mid!\Role_{\thempa}\mid!\Role_{\thehii}),
\end{array}
\]
which is the left-hand side of Definition~\ref{def:sano} - the $\dlvprotocol$ in the case study.

On the other hand, by choosing $\drA$, we obtain process
\[
\begin{array}{l}
\free\ \drA;\\
\nu \vect{mc}. \init. (!\Role_{\thepatient}\mid!\Role_{\thedoctor}\mid!
\Role_{\thepharm}\mid!\Role_{\thempa}\mid!\Role_{\thehii}\mid\\
 \hspace{9ex}(\nu \drB.
\nu \ndoctorPseudo.
\tlet \doctorPseudo=\ndoctorPseudo \letin\ !(\tlet \doctorID=\drA \letin\
 \ProDr))).
\end{array}
\]
Since $\drB$ only appears in the sub-process ``$\nu \drB.$" which generates $\drB$ and never appears in the remaingin process, the process is structurally equivalent to (applying rule $\NEWPar$)
\[
\begin{array}{l}
\free\ \drA;\\
{\bf \nu \drB.}
\nu \vect{mc}. \init. (!\Role_{\thepatient}\mid!\Role_{\thedoctor}\mid!
\Role_{\thepharm}\mid!\Role_{\thempa}\mid!\Role_{\thehii}\mid\\
 \hspace{9ex}(
\nu \ndoctorPseudo.
\tlet \doctorPseudo=\ndoctorPseudo \letin\ !(\tlet \doctorID=\drA \letin\
 \ProDr))).
\end{array}
\]
The above process is structurally equivalent to (proved later)
\[
\begin{array}{l}
\free\ \drA;\\
\nu \vect{mc}. \init. (!\Role_{\thepatient}\mid!\Role_{\thedoctor}\mid!
\Role_{\thepharm}\mid!\Role_{\thempa}\mid!\Role_{\thehii}\mid\\
 \hspace{9ex}(
\nu \ndoctorPseudo.
\tlet \doctorPseudo=\ndoctorPseudo \letin\ !(\tlet \doctorID=\drA \letin\
 \ProDr))),
\end{array}
\]
which is the right-hand side of Definition~\ref{def:sano}, where $\drA$ is a free name.
This structural equivalent relation is proved as follows.
\proof Assuming the above process is $P$ (i.e., $P=\nu \vect{mc}. \init. (!\Role_{\thepatient}\mid!\Role_{\thedoctor}\mid!
\Role_{\thepharm}\mid!\Role_{\thempa}\mid!\Role_{\thehii}\mid(
\nu \ndoctorPseudo.
\tlet \doctorPseudo=\ndoctorPseudo \letin\ !(\tlet \doctorID=\drA \letin\
 \ProDr)))$), by applying rule $\PARnull$, we have
$P\steq P\mid 0$. By rule $\NEWnull$, $\nu \drB.0 \steq 0$. Thus, $P\steq P\mid \nu \drB.0$.
Since $\drB$ never appears in the process $P$, i.e., $\drB \not\in \fn{P} \cup \fv{P}$, by
applying rule $\NEWPar$, we have $P\mid \nu \drB.0 \steq \nu \drB. (P\mid 0) \steq \nu \drB. P$. Therefore, $P\steq \nu \drB. P$.
\paragraph{Verification result.}
The bi-processes are verified using ProVerif.
The verification results show that patient anonymity (with and without revealed patient
identities a priori) and strong patient anonymity are satisfied; doctor anonymity
is satisfied; neither doctor anonymity with revealed doctor identities nor strong doctor
anonymity is satisfied.

For strong doctor anonymity, the adversary can distinguish a
process initiated by an unknown doctor and a known doctor.
Given a doctor process, where the doctor has identity $\drA$,
pseudonym $\doctorPseudo$, and credential $\drcred(\doctorPseudo,\drA)$, the terms
$\doctorPseudo$ and
\(
\drcred(\doctorPseudo,\drA)
\) are revealed. We assume that the adversary knows another
doctor identity $\drB$. The adversary
can fake an anonymous authentication
by faking the zero-knowledge proof as
\(
\zk((\doctorPseudo,\drB),\drcred(\doctorPseudo,\drA))
\).
If the zero-knowledge proof passes the corresponding verification $\doctorAuthVer$ by the patient,
then the adversary knows that the doctor process is executed by
the doctor $\drB$. Otherwise, not.

For the same reason, doctor anonymity fails the verification.
Both flaws can be fixed by requiring a doctor to generate a new credential in each session ({\bf s4'}).

\subsection{(Strong) patient and doctor untraceability}
\label{sec:untrace}
Even if a user's identity is not revealed, the adversary may be able to trace
a user by telling whether two executions are done by the same user.
The DLV08 protocol claims that prescriptions issued to the same patient should not be linkable
to each other. In other words, the situation in which a patient executes the protocol twice should be
indistinguishable from the situation in which two different patients execute the protocol individually.
To satisfy this requirement, patient untraceability is required.
(Remark that the original DLV08 paper calls this untraceability ``patient
unlinkability''.)

\paragraph{Patient and doctor untraceability.}
Doctor untraceability has been defined in Definition~\ref{def:untra},
and patient untraceability can be defined in a similar style.
The bi-process for verifying doctor untraceability is
\[
\begin{array}{l}
\nu \vect{mc}. \init.
(!\Role_{\thepatient}\mid
!\Role_{\thedoctor}\mid
!\Role_{\thepharm}\mid
!\Role_{\thempa}\mid
!\Role_{\thehii}\mid
(\nu \ndoctorPseudo. \nu \wdoctorPseudo.\\
 ((\tlet \doctorID=\drA \letin\
  \tlet \doctorPseudo=\ndoctorPseudo \letin\
   \ProDr )
\mid \\
\hspace{1ex}(\tlet \doctorID=\choice{\drA}{\drB} \letin\
 \tlet \doctorPseudo=\choice{\ndoctorPseudo}{\wdoctorPseudo} \letin\
\ProDr)))),
\end{array}
\]
and the bi-process for verifying patient untraceability is
\[
\begin{array}{l}
\nu \vect{mc}. \init.
(!\Role_{\thepatient}\mid
!\Role_{\thedoctor}\mid
!\Role_{\thepharm}\mid
!\Role_{\thempa}\mid
!\Role_{\thehii}\mid \\
(\nu \npatientPseudo.
  \nu \npatientSSS.
  \nu \npatientAcc.
 \nu \wpatientPseudo.
  \nu \wpatientSSS.
  \nu \wpatientAcc.\\
\hspace{1ex}\readin{\privchhiipt}{\npatientHii}.
\readin{\privchhiipt}{\wpatientHii}.\\
\hspace{1ex}  \tlet \npkhiipt=\key(\npatientHii) \letin\
  \tlet \wpkhiipt=\key(\wpatientHii) \letin\\
(\tlet \patientHii=\npatientHii \letin\
 \tlet \pkhiipt=\npkhiipt \letin\
 \tlet \patientID=\ptA \letin\\
\ \tlet \patientPseudo=\npatientPseudo \letin\
 \tlet \patientSSS=\npatientSSS \letin\
 \tlet \patientAcc=\npatientAcc \letin\
\ProPt) \mid\\

(
\tlet \patientHii=\choice{\npatientHii}{\wpatientHii} \letin\
 \tlet \pkhiipt=\choice{\npkhiipt}{\wpkhiipt} \letin\\
\ \tlet \patientID=\choice{\ptA}{\ptB} \letin\
\tlet \patientPseudo=\choice{\npatientPseudo}{\wpatientPseudo} \letin\\
\ \tlet \patientSSS=\choice{\npatientSSS}{\wpatientSSS} \letin\
 \tlet \patientAcc=\choice{\npatientAcc}{\wpatientAcc} \letin\
\ProPt)
)).
\end{array}
\]
We verified two versions of doctor and patient untraceability,
- in one version, the adversary does not know the two doctor/patient identities, and
in the other version, the adversary initially knows the two doctor/patient identities.
\paragraph{Strong patient and doctor untraceability.}
Strong untraceability is modelled as a patient
executing the protocol repeatedly is indistinguishable from different patients
executing the protocol each once. Strong doctor untraceability
is defined as in Definition~\ref{def:suntra} and strong patient untraceability
can be defined in the same manner.
The bi-process for verifying strong doctor untraceability is
\[
\begin{array}{l}
\nu \vect{mc}. \init.
(!\Role_{\thepatient}\mid
!\Role_{\thepharm}\mid
!\Role_{\thempa}\mid
!\Role_{\thehii}\mid!(\nu \ndoctorID. \nu \ndoctorPseudo.
 !(\nu \wdoctorID. \nu \wdoctorPseudo. \\
\ \tlet \doctorID=\choice{\ndoctorID}{\wdoctorID} \letin\
\tlet \doctorPseudo=\choice{\ndoctorPseudo}{\wdoctorPseudo} \letin\ \ProDr))),
\end{array}
\]
and the bi-process for verifying strong patient untraceability is
\[
\begin{array}{l}
\nu \vect{mc}. \init.
(!\Role_{\thedoctor}\mid
!\Role_{\thepharm}\mid
!\Role_{\thempa}\mid
!\Role_{\thehii}\mid
!(\nu \npatientID. \nu \npatientPseudo. \nu \npatientSSS. \nu \npatientAcc.
 \readin{\privchhiipt}{\npatientHii}.\\
\ !(\nu \wpatientID. \nu \wpatientPseudo. \nu \wpatientSSS. \nu \wpatientAcc.\\
\ \ \ \tlet \patientID=\choice{\npatientID}{\wpatientID} \letin\
 \tlet \patientPseudo=\choice{\npatientPseudo}{\wpatientPseudo} \letin\\
\ \ \ \tlet \patientSSS=\choice{\npatientSSS}{\wpatientSSS} \letin\
\tlet \patientAcc=\choice{\npatientAcc}{\wpatientSSS} \letin\\
\ \ \ \readin{\privchhiipt}{\wpatientHii}.
\tlet \patientHii=\choice{\npatientHii}{\wpatientHii} \letin\
\tlet \pkhiipt=\key(\patientHii) \letin\
 \ProPt))).
\end{array}
\]
This definition does not involve a specific doctor/patient, and thus needs not to
distinguish whether the adversary knows the identities a priori.
\paragraph{Verification result.}
The bi-processes are verified using ProVerif.
The verification results show that the DLV08 protocol does not satisfy
patient/doctor untraceability (with/without revealed identities), nor strong untraceability.

The strong doctor untraceability fail, because the adversary can distinguish sessions
initiated by one doctor and by different doctors. The doctor's pseudonym is
revealed and a doctor uses the same pseudonym in all sessions.
Sessions with the same doctor pseudonyms are initiated by the same doctor.
For the same reasons, doctor untraceability without revealing doctor identities also fails.
Both of them can be fixed by requiring the representation of a
doctor's pseudonym ($\patientSSS$) differ in each session ({\bf s3'}).

However, assuming {\bf s3'} (doctor pseudonym is fresh in every sessions)
is not sufficient for satisfying doctor anonymity with
doctor identities revealed. The adversary can still distinguish two sessions
initiated by one doctor or by two different doctors, by comparing the
anonymous authentications of the two sessions. From the communication in the two
sessions, the
adversary is able to learn two doctor pseudonyms $\doctorPseudo'$ and $\doctorPseudo''$,
two doctor credentials $\doctorCred'$ and $\doctorCred''$ and two anonymous authentications
$\doctorAuth'$ and $\doctorAuth''$.
Since the adversary knows $\drA$ and $\drB$ in advance, he could construct
the eight anonymous authentications by applying the zero-knowledge proof function,
i.e., $\zk((\vdoctorPseudo,\doctorIDv),\doctorCred)$,
where $\vdoctorPseudo=\doctorPseudo'$ or $\vdoctorPseudo=\doctorPseudo''$,
$\doctorIDv=\drA$ or $\doctorIDv=\drB$, $\doctorCred=\doctorCred'$ or
$\doctorCred=\doctorCred'$.
By comparing the constructed anonymous authentications with the observed ones,
the adversary is able to tell who generated which anonymous authentication, and
thus is able to tell whether the two sessions are initiated by the same doctor
or different doctors.
This can be fixed by additionally requiring that the doctor anonymous
authentication differs in every session ({\bf s4'}).

For strong patient untraceability, the adversary can distinguish sessions initiated by one patient (with
identical social security statuses) and initiated by different patients
(with different social security statuses).
Second, the adversary can distinguish sessions initiated by
one patient (with identical cipher texts $\enc{\patientPseudo}{\pksso}$ and identical
cipher texts $\enc{\patientHii}{\pksso}$) and initiated by different patients (with different
cipher texts $\enc{\patientPseudo}{\pksso}$ and different cipher texts $\enc{\patientHii}{\pksso}$).
Third, since the patient credential is the same in all sessions and is revealed,
the adversary can also trace a patient by the patient's credential.
Fourth, the adversary can distinguish sessions using the same HII and sessions using different HIIs.
For the same reasons, patient untraceability fails.
Both flaws can be fixed by requiring that the representation of a patient's social security
status to be different in each session ({\bf s5'}), the encryptions are probabilistic ({\bf s2'}),
a patient freshly generates a credential in each session ({\bf s4''}), and
patients who shall not be distinguishable share the same HII ({\bf s6'}).

\subsection{\Docunlink}

\Docunlink\ has been defined in Definition~\ref{def:drpriv}.
To verify the \docunlink\ is to check the satisfaction of the equivalence in the
definition.
The bi-process for verifying the equivalence is
\[
\begin{array}{ll}
\multicolumn{2}{l}{(\private) \free\ \drA.
(\private) \free\ \drB.
\free\ \pa.
\free\ \pb.}\\
\nu \vect{mc}. \init. (!\Role_{\thepatient}\mid!\Role_{\thedoctor}\mid!
\Role_{\thepharm}\mid
& (\nu \ndoctorPseudo. \nu \wdoctorPseudo.\\
&\hspace{1ex}\tlet \doctorID=\choice{\drA}{\drB} \letin\\
&\hspace{1ex}\tlet \doctorPseudo=\choice{\ndoctorPseudo}{\wdoctorPseudo} \letin\\
&\hspace{1ex}\tlet \prescText=\pa \letin\ \vProDr)\mid\\
& (\nu \ndoctorPseudo. \nu \wdoctorPseudo.\\
&\hspace{1ex}\tlet \doctorID=\choice{\drB}{\drA} \letin\\
&\hspace{1ex}\tlet \doctorPseudo=\choice{\ndoctorPseudo}{\wdoctorPseudo} \letin\\
&\hspace{1ex}\tlet \prescText=\pb \letin\ \vProDr)).
\end{array}
\]
Similarly, we verified two versions - in one version, the adversary does not know
$\drA$ and $\drB$, and in the other, the adversary knows $\drA$ and $\drB$.
\paragraph{Verification result.}
The verification, using ProVerif, shows that the DLV08 protocol satisfies \docunlink\
when the adversary does not know the doctor identities a priori, and does not satisfy \docunlink\
when the adversary knows the doctor identities a priori, i.e.,
the adversary can distinguish whether a prescription
is prescribed by doctor $\drA$ or doctor $\drB$, given the adversary knows $\drA$ and $\drB$.
In the prescription proof, a prescription is linked to a doctor credential.
And a doctor credential is linked to a doctor identity. Thus, the adversary can link
a doctor to his prescription. To break the link, one way is to make sure that the
adversary cannot link a doctor credential to a doctor identity. This can be
achieved by adding randomness to the credential ({\bf s4'}).

\subsection{\Docrf}\label{ssec:docrf}
The definition of \docrf\ is
modelled as the existence of a process $\ProDr'$,
such that the two equivalences in Definition~\ref{def:drf} are satisfied.
Due to the existential quantification, we cannot verify the property directly using ProVerif.

Examining the DLV08 protocol, we find an attack on \docrf, even with
assumption {\bf s4'} (after fixing \docunlink\ with doctor ID revealed).
A bribed doctor is able
to prove to the adversary of his prescription as follows:
\begin{enumerate}
\item A doctor communicates with the adversary to agree on a bit-commitment
	that he will use, which links the doctor to the commitment.
\item The doctor uses the agreed bit-commitment in the communication
	with his patient. This links the bit-commitment to a prescription.
\item Later, when the patient uses this prescription to get medicine from a
	pharmacist, the adversary can observe the prescription being
	used. This proves that the doctor has really prescribed the medicine.
\end{enumerate}
We formally confirm the attack using ProVerif, i.e., we show that in the protocol
model, if a doctor reveals all his
information to the adversary,
the doctor's \docunlink\ is broken. The same attack exists for \docrfm\ as well --
a bribed doctor is able to prove his prescriptions by agreeing with the adversary
on the bit-commitments in each session.

\begin{theorem}[\docrf]\label{theorem:docrf}
The DLV08 protocol fails to satisfy \docrf\ under both the standard
assumption \textbf{s4} (a doctor has the same credential in every session),
and also under assumption \textbf{s4'} (a doctor generates a new credential
for each session).
\end{theorem}
Formal proof of the theorem can be found in Appendix~\ref{sec:proof1}.

\subsection{Independency of (enforced) \docunlink}

To determine whether the doctor's \docunlink\ is independent of the
pharmacist, we replace regular pharmacist role $\Role_{i}$ with
collaborating role $\Role_{\thepharm}$ in Definition~\ref{def:dpi}. The
bi-process for verifying the property is:
\[
\begin{array}{ll}
\multicolumn{2}{l}{(\private) \free\ \drA.
(\private) \free\ \drB.
\free\ \pa.
\free\ \pb.}\\
\nu \vect{mc}. \init.
(!\Role_{\thepatient}\mid
!\Role_{\thedoctor}\mid
!(\Role_{\thepharm})^{\chc}\mid
&(\nu \ndoctorPseudo. \nu \wdoctorPseudo.\\
&\hspace{1ex}\tlet \doctorID=\choice{\drA}{\drB} \letin\\
&\hspace{1ex}\tlet \doctorPseudo=\choice{\ndoctorPseudo}{\wdoctorPseudo} \letin\\
&\hspace{1ex}\tlet \prescText=\pa \letin\ \vProDr)\mid\\
&(\nu \ndoctorPseudo. \nu \wdoctorPseudo.\\
&\hspace{1ex}\tlet \doctorID=\choice{\drB}{\drA} \letin\\
&\hspace{1ex}\tlet \doctorPseudo=\choice{\ndoctorPseudo}{\wdoctorPseudo} \letin\\
&\hspace{1ex}\tlet \prescText=\pb \letin\ \vProDr)).
\end{array}
\]
Verification using ProVerif shows that the protocol (the original version where
the adversary does not know $\drA$ and $\drB$, and the version after fixing the
flaw on \docunlink\ with assumption {\bf s4'} where the adversary knows $\drA$
and $\drB$) satisfies this property.

The case of pharmacist-independent \docrf\ is treated analogously.
We replace regular pharmacist $\Role_i$ with $\Role_{\thepharm}$ in
Definition~\ref{def:drfi}.
The flaw described in Section~\ref{ssec:docrf} also surfaces here.
This was expected: when a doctor can prove his prescription
without the pharmacist sharing information with the adversary,
the doctor can also prove this when the pharmacist
genuinely cooperates with the adversary.

\subsection{Dishonest users}\label{ssec:dishonest}

So far, we have considered security and privacy with respect to a
Dolev-Yao style adversary (see Section~\ref{ssub:adversary}).
The initial knowledge of the adversary was modelled such, that the
adversary could not take an active part in the execution of the protocol.
This constitutes the basic DY adversary, as shown in Table~\ref{tab:adversary}.
In more detail, for secrecy of private doctor and patient information
(see Table~\ref{tab:secrecy}),
\cite{DLVV08} claims that no third party (including the basic DY
adversary) shall be able to know a patient's or
doctor's private information (refer to the beginning of
Section~\ref{sec:dlv08}). Similarly, the verification of authentication
properties in Table~\ref{tab:auth} is also with respect to the basic DY
adversary. This captures that no third party that does not participate
in the execution shall be able to impersonate any party (involved in the
execution). The same basic DY adversary model is used to verify anonymity,
untraceability and \docunlink. The exceptions are (1) for verifying
\docrf\ and independency of enforced \docunlink, the basic DY adversary
is extended with information from the targeted doctor; (2) for
verifying independency of \docunlink\ and independency of enforced
\docunlink, the basic DY adversary is extended with information
from pharmacists.

\begin{table}[!h]
\begin{tabular}{|l|l|}
\hline
Properties & Adversary\\
\hline
Secrecy, Authentication, & basic DY\\
Anonymity, Untraceability, \Docunlink & \\
\hline
\Docrf & basic DY + info. from doctor\\
\hline
Independency of \docunlink & basic DY + info. from pharmacy \\
\hline
Independency of enforced \docunlink & basic DY + info. from doctor and pharmacy\\
\hline
\end{tabular}
\caption{Summary of the respected adversary}\label{tab:adversary}
\end{table}

In this section, we consider dishonest users, that is, malicious users
that collaborate with the adversary and are
part of the execution, into consideration.
For each property previously verified, we analyse the
result once again with respect to each dishonest role.

Dishonest users are
modelled by providing the adversary certain initial knowledge such that the
adversary can take part in the protocol.
To execute the protocol as a doctor, i.e., to instantiate the doctor process
$\proc_{\thedoctor}$, the adversary only needs to have an identity and a pseudonym.
Since the adversary is able to generate data, the adversary can create his own
identity $\doctorIDatt$ and pseudonym $\doctorIDatt$. However, this is not
sufficient, because an legitimate doctor has a credential issued by authorities.
The credential is captured by the private function $\drcred$. The adversary
cannot obtain this credential, since he cannot apply the function $\drcred$.
When the function $\drcred$ is modelled as public, the adversary is able
to obtain his credential $\drcred(\doctorIDatt,\doctorIDatt)$, and thus
has the ability to behave like a doctor. Hence, by modelling the
function $\drcred$ as public, we allow the adversary to have the ability
of dishonest doctors.
Note that an honest doctor's identity is secret (see
Table~\ref{tab:secrecy}). The attacker thus
cannot forge credentials of honest doctors, as the doctor's identity
must be known for this. Thus, making the function $\drcred$
public does not bestow extra power on the attacker.

Similarly, by allowing the adversary to have patient credentials, we strengthen
the adversary with the ability to control dishonest patients. This is modelled by
changing the private functions $\ptcred$ to be public.

Each public entity (pharmacist, MPA or HII) has a secret key as distinct identifier, i.e., its public key
and identity can be derived from the secret key. The adversary can create such
a secret key by himself. However, only the legitimate entities can participate
in the protocol. This is modelled using private channels -- only the honest
entities are allowed to publish their information to the channels, and participants
only read in entities, which they are going to communicate with, from the private channels.
By changing the private channels to be public, the adversary is able to behave as dishonest public
entities. Note that when considering the adversary only controlling dishonest
pharmacists among the public entities, for the sake of simplicity of modelling,
a dishonest pharmacist is modelled as $\Role_{\thepharm}^{\chc}$ (same as in \docindep)
in which the pharmacist shares all his information with the adversary.

The above modelling of dishonest users captures the following scenarios.
For verifying secrecy of doctor/paitent information,
\begin{itemize}
\item
the dishonest doctor/patient models other doctors/patients that may break
secrecy of the target doctor/patient;
\item
the dishonest patient/doctor models the patients/doctors that may communicate with the target doctor/patient;
\item
the dishonest pharmacist, MPA and HII may participate in the same execution as the target doctor/patient.
\end{itemize}
Since secrecy is defined as no other party
(including dishonest users) should be able to know a doctor's/patient's information, unless the
information is intended to be revealed, we would not consider it as an attack if the dishonest user intends to
receive the private information, for details, see Section~\ref{sec:dishonest_secrecy}.

For verifying authentication properties, for example, a doctor authenticates a patient, a dishonest doctor is not
the doctor directly communicating with the patient, and a dishonest patient is not the one who directly communicates with the doctor, because it does not make sense to analyse a dishonest user authenticates or authenticates to another user. Instead, the dishonest doctors and patients are observers who may participate in other execution sessions.
\begin{itemize}
\item
In general, if user $A$ authenticates to user $B$, the dishonest users taking the same role of $A$ or $B$ are observers participating in different sessions, i.e., cannot be $A$ or $B$.
\item
For the dishonest pharmacists, MPAs and HIIs, since they are not the authentication parties, they can be users participating in the same session.
\end{itemize}

For verifying \docunlink\ and \docrf,
\begin{itemize}
\item
the dishonest doctors are other doctors that aim to break the target doctor's privacy;
\item
the dishonest patients can be patients communicating with the target doctor;
\item
dishonest pharmacists, MPAs and HIIs can be users participating in the same session.
\end{itemize}
Note that the dishonest doctor differs from the bribed doctor, as the bribed doctor tries to break his own privacy, while dishonest doctor tries to break others' privacy.

For verifying independency of \docunlink\ and independency of enforced \docunlink,
\begin{itemize}
\item
the dishonest doctors (not the target doctor) try to break the target doctor's privacy;
\item
the dishonest patients may directly communicate with the target doctor;
\item
the dishonest pharmacists are the same as the bribed pharmacists, since 1) the bribed pharmacists genuinely forward information to the adversary and 2) all the actions that a dishonest pharmacists can do can be simulated by the basic DY with the received information from the bribed pharmacists, i.e., there is no private functions or private channels that the dishonest pharmacists can use but the adversary with bribed pharmacists information cannot;
\item
the dishonest MPAs and HIIs can participate in the same session as the target doctor.
\end{itemize}

The verification with dishonest users shows similar results as the verification without dishonest users. The reason is that if there is an attack with respect to the basic DY attacker when verifying a property, then the property is also broken when additionally considering dishonest users. The exceptions (i.e., the additional identified attacks) are shown in Table~\ref{tab:add_attacks}, and the details of the additional attacks are shown in the remaining part of this section.
\begin{table}[!h]
\begin{tabular}{|l|l|}
\hline
Additional attacks & Dishonest users\\
\hline
secrecy of patient $\ptB$'s pseudonym & basic DY + dishonest patient $\ptA$\\
\hline
secrecy of a patient's pseudonym & basic DY + dishonest pharmacists\\
\hline
(strong) patient anonymity when using different HIIs & basic DY + dishonest HIIs\\
\hline
(strong) patient anonymity when using different HIIs & basic DY + dishonest MPAs\\
\hline
\end{tabular}
\caption{Additional attacks when considering dishonest users}\label{tab:add_attacks}
\end{table}

\subsubsection{Secrecy}\label{sec:dishonest_secrecy}
When considering dishonest doctors and pharmacists, secrecy results in Table~\ref{tab:secrecy} do not change,
since doctors do not receive any information that the adversary does not know (see Figure~\ref{msc:dp}).

When considering dishonest patients, an additional attack is found.
When a patient $\ptA$ is dishonest, he can obtain another patient $\ptB$'s
pseudonym by doing the following:
\begin{enumerate}
\item $\ptA$ observes $\ptB$'s communication and reads in $\vc_4$ (the verifiable
encryption which encrypts $\ptB$'s pseudonym with the public key of an MPA).
\item Hence, from $\vc_4$, $\ptA$ can obtain the cipher-text $\enc{\patientPseudoB}{\pkmpapt}$, where $\patientPseudoB$ is $\ptB$'s pseudonym and $\pkmpapt$ is the public key of the MPA.
\item $\ptA$ initiates the protocol with his own data.
\item In the communication with a pharmacist, $\ptA$ replaces his $\C_5$ (which should be
a verifiable encryption, containing a cipher-text from $\ptA$ encrypted with the
public key of the MPA) with $\vc_4$.
\item On receiving $\vc_4$ (the fake $\C_5$), the pharmacist sends it to the MPA, and the MPA
decrypts the cipher-text $\enc{\patientPseudoB}{\pkmpapt}$, embedded in $\vc_4$ and sends the decryption result to HII.
\item $\ptA$ observes the communication between the MPA and HII, and reads the
decrypted text of the fake $\C_5$ (i.e., $\patientPseudoB$), which is $\ptB$'s pseudonym.
\end{enumerate}
This attack does not exist when the attacker cannot participant as a patient.
Because the dishonest patient has to replace $\C_5$ in his own communication.
If an hones patient's $\C_5$ is replaced by the adversary, the pharmacist would detect
it by verifying the receipt $\ReceptionAck$, which should contain the correct $\C_5$.
This attack can be addressed by explicitly ask the MPA to verify the decrypted
message of $\C_5$ to be a verifiable encryption before sending it out. Alternatively,
if the communication between MPA and the HII is secured after authentication,
the adversary would not be able to observe $\ptB$'s pseudonym, and thus the attack
would not happen.

When considering dishonest pharmacists, an additional attack may exist on a patient's
pseudonym. The dishonest pharmacist has/creates a secret key $y$ and obtains its
corresponding public key $\pk(y)$. The pharmacist creates a fake MPA identity
using the secret key and public key, i.e., $\host(\pk(y))$.
The dishonest pharmacist provides the patient a fake MPA identity, from
which the patient obtains the MPA public key $\pk(y)$. Later, the patient encrypts his
pseudonym using the fake MPA's public key $\enc{\patientPseudo}{\pk(y)}$, and provides a verifiable encryption,
$\vc_4$ in particular. The verifiable encryption is sent to the pharmacist, from
which he pharmacist can read the cipher-text $\enc{\patientPseudo}{\pk(y)}$.
Since the pharmacist knows the secret key $y$, he can decrypt the cipher-text
and obtains the patient's pseudonym $\patientPseudo$. Note that we model that the
patient obtains the MPA of the pharmacist and the MPA's public key from the pharmacist, thus the attack may
not happen if the patient initially knows the MPA of any pharmacist, or the patient
has the ability to immediately check whether the MPA provided by a pharmacist
is indeed an legitimate MPA.

When considering dishonest MPAs, the adversary additionally knows a patient's pseudonym.
However, this can hardly be an attack, as the patient pseudonym is intended to be known
by the MPA. Similarly, the HII is the intended receiver of a patient's pseudonym.
Other than a patient's pseudonym, the dishonest MPA and HII do not know any information
that the adversary does not know
without controlling dishonest agents. Note that in reality, the MPA and HII may
know more sensitive information, for example, from the pseudonym, the HII is able
to obtain the patient's identity, and a dishonest MPA can claim that a prescription
has medical issues and obtains the doctor identity in a procedure, which is
beyond the scope of this protocol.

\subsubsection{Authentication}

When considering dishonest users, the verification results of the
authentication remain the same, except the authentication from the dishonest
user to other parties. For example, when doctors are dishonest, we do not
need to consider the authentication from doctors to a patient, since the dishonest
users are part of the adversary. Similarly, when a
patient is dishonest, the authentication from a patient to a doctor or a pharmacist
is obviously unsatisfied, other authentication verification results remain the same.
When pharmacists, MPAs or HIIs are dishonest, the verification results remain
unchanged.

\subsubsection{Privacy properties}
For those privacy properties which are not satisfied with respect to the adversary
controlling no dishonest agents, the properties are not satisfied when considering
the adversary who controls dishonest users. Thus, we only need to analyse the property that
are satisfied with respect to the adversary controlling no dishonest agents,
i.e. (Strong) patient anonymity. Obviously, two patients
can be distinguished by the adversary who controls dishonest HIIs, when the two patients
use different HIIs, because the patients
use different HII public keys to encrypt his pseudonym. When the two patients use
different HIIs, and the HIIs are honest, the adversary, who controls dishonest
MPAs, can still distinguish them, because a patient's HII is intended to be known
by the MPA. Finally, (strong) patient anonymity is satisfied with respect to the adversary
controlling dishonest doctors and dishonest pharmacists.

\section{Addressing the flaws of the DLV08 protocol}
\label{sec:dlv08fixes}
To summarise, we present updates to the assumptions
of Section~\ref{sec:ambiguities} to fix the flaws found in our analysis of the privacy properties.
\begin{itemize}
\item {\bf s2'}
The encryptions are probabilistic.
\item {\bf s3'}
The value of $\doctorPseudo$ is freshly generated in every session.
\item {\bf s4'}
A doctor freshly generates an unpredictable credential in each session.
We model this with another parameter (a random number) of the credential.
Following this, anonymous authentication
using these credentials proves
knowledge of the used randomness.
\item {\bf s4''}
A patient freshly generates an unpredictable credential in each session. Similar
to {\bf s4'}, this can be achieved by add randomness in the credential. The
anonymous authentication using the credentials proves
the knowledge of the used randomness.
\item {\bf s5'}
The values of $\patientSSS$ differ in sessions.
\item {\bf s6'}
The value of $\patientHii$ shall be the same for all patients.
\end{itemize}
The proposed assumptions are provided on the model level. Due to the ambiguities in the
original protocol (e.g., it is not clear how a social security status is
represented), it is difficult to propose detailed solutions.
To implement a proposed assumption, one only needs to capture its properties.

To capture {\bf s2'}, the encryption scheme can be ElGamal cryptosystem, or RSA
cryptosystem with encryption padding, which are probabilistic. In some systems,
deterministic encryption, e.g., RSA without encryption padding, may be more useful
than probabilistic encryption, for example for database searching of encrypted data.
In such systems, designers need to carefully distinguish which encryption
scheme is used in which part of the system.

{\bf s3'} can be achieved by directly requiring a doctor's pseudonym to be fresh in every session,
for example, a doctor generates different pseudonyms in sessions and keeps the authorities, who
maintain the relation between the doctor identity and pseudonyms, updated in a secure way; or
before every session the doctor requests a pseudonym from the authorities.
Alternatively, it can be achieved by changing the value of $\doctorPseudo$.
Assuming the authorities share a key with each doctor; instead of
directly using pseudonym in a session, the doctor encrypts his pseudonym with the key using
probabilistic encryption. That is, the value of $\doctorPseudo$ is a cipher-text which differs in
sessions. When an MPA wants to find out the doctor of a prescription, he
can contact the authorities to decrypt the $\doctorPseudo$ and finds out the
pseudonym of the doctor or the identity of the doctor directly.

{\bf s4'} and {\bf s4''} together form the updates to {\bf s4}. We separate the
update to the doctor credential in {\bf s4'} and the update to the patient credential in {\bf s4''}
for the convenience of referring to them individually in other places.

Similar to {\bf s3'}, {\bf s5'} can be achieved by directly requiring that
a patient's social security status is different in each session, e.g., by
embedding a timestamp in the status. Alternatively,
the value of $\patientSSS$ can be a cipher-text which is a probabilistic encryption of a patient's social
security status with the pharmacist's public key, since the social security status
is used for the pharmacist to check the status of the patient.

{\bf s6'} can be achieved by directly requiring that all patients share
the same HII. In the case of multiple HIIs, different HIIs should not be
distinguishable, for example, HIIs may cooperate together and provide a uniformed
reference (name and key). In fact, if patients are satisfied with untraceability within a group
of a certain amount of patients, patient untraceability can be satisfied as long as
each HII has more patients than the expected size of the group. If only untraceability is
required (instead of strong untraceability), the use of a group key of all HIIs
is sufficient. The common key among HIIs can be established
by using asymmetric group key agreement. In this way, the HIIs cannot be
distinguished by their keys. In addition, the identities of HIIs are not revealed,
and thus cannot be used to distinguish HIIS. Hence, the common key ensures
that two patients executing the protocol once and one patient executing the protocol twice
cannot be distinguished by their HIIs.

The modified protocol was verified again using ProVerif. The verification results show that
the protocol with revised assumptions satisfies doctor anonymity, strong doctor
anonymity, and \docunlink, as well as untraceability and strong untraceability
for both patient and doctor.

However, the modified protocol model does not satisfy \docrf, to make the protocol satisfy \docrf,
we apply the following assumption on communication channels.
\begin{itemize}
\item {\bf s8'}
Communication channels are untappable (i.e., the adversary does not observe anything from the channel), except those used for
authentication, which remain public.
\end{itemize}

\begin{table}
\centering
\begin{tabular}{| l | c | c | c |  c |}
\hline
{\bf checked privacy property} & {\bf initial} & {\bf cause(s)} &
	{\bf improvement} & {\bf revised}\\
 & {\bf model} &&& {\bf model}\\
\hline
\docunlink\ & $\ok$ & &  & $\ok$\\
\docunlink\ w. ID reveal & $\fail$ &{\bf s4} & {\bf s4'} & $\ok$\\
\docrfsh & $\fail$ (with {\bf s4'} ) && {\bf s8'} & $\ok$\\
\docindepshshort & $\ok$ (with {\bf s4'})&& & $\ok$\\
\docrfindepshshort & $\fail$(with {\bf s4'}) && {\bf s8'} & $\fail$\\
pt. anonym. (w. ID reveal) & $\ok$ && & $\ok$\\
strong pt. anonym.& $\ok$ && & $\ok$\\
dr. anonym. & $\ok$&&&$\ok$\\
dr. anonym. w. ID reveal& $\fail$ & {\bf s4} & {\bf s4'} & $\ok$\\
strong dr. anonymity& $\fail$ & {\bf s4} & {\bf s4'} & $\ok$\\
pt. untrace. (w. ID reveal)& $\fail$ &\rule{.25em}{0pt}{\bf s2}, {\bf s4}, {\bf s5}, {\bf s6}\rule{.25em}{0pt} &\rule{.25em}{0pt} {\bf s2'}, {\bf s4''}, {\bf s5'}, {\bf s6'}\rule{.25em}{0pt} & $\ok$\\
strong pt. untrace.& $\fail$ & {\bf s2}, {\bf s4}, {\bf s5}, {\bf s6}\ &\ {\bf s2'}, {\bf s4''}, {\bf s5'}, {\bf s6'} & $\ok$\\
dr. untrace. & $\fail$ & {\bf s3} & {\bf s3'} & $\ok$\\
dr. untrace. w. ID reveal& $\fail$ & {\bf s3},{\bf s4} & {\bf s3'},{\bf s4'} & $\ok$\\
strong dr. untrace.& $\fail$ &{\bf s3} & {\bf s3'} & $\ok$\\
\hline
\end{tabular}
\caption{Verification results of privacy properties and revised assumptions.}
\label{tab:result}
\end{table}

Our model of the protocol is accordingly modified as follows: replacing
channel $\ch$ in lines {\bf d10}, {\bf t6} with an untappable channel
$\privchdrpt$, replacing channel $\ch$ in lines {\bf t23}, {\bf t26},
{\bf h5}, {\bf h22} with an untappable channel $\privchptph$, and
replacing channel $\ch$ in lines {\bf t24}, {\bf h21} with an untappable
channel $\privchphpt$. The untappable channels are modelled as global
private channels. We prove that the protocol (with {\bf s4'} and {\bf
s8'}) satisfies \docrf\ by showing the existence of a process $\ProDr'$
(as shown in Figure~\ref{fig:prodradd}) such that the equivalences in
Definition~\ref{def:drf} are satisfied. This was verified using ProVerif (for verification code, see~\cite{DJPa12}).
\begin{figure}[!ht]
\begin{specification}
\begin{math}
\begin{array}{ll}
\rule{0pt}{3.0ex}
\init_{\thedoctor}\substitution{\drA}{\doctorIDv}.
(!\ProDr\substitution{\drA}{\doctorIDv}\mid
  \ProDr'\substitution{\drA}{\doctorIDv}) \defi \\

\tlet \doctorID=\drA \letin\
\nu \doctorPseudo.\\
\begin{array}{rl}
(!\ProDr &(*\text{the $\ProDr$ has assumptions {\bf s4'} and {\bf s8'}}*)\\
\mid& (\out{\chc}{\doctorID}. \\
&\hspace{1ex} \out{\chc}{\doctorPseudo}.\\
&\hspace{1ex} \nu \noncedr.
             \out{\chc}{\noncedr}. \\
&\hspace{1ex} (*\text{{\bf s4'}: creating a nonce and adding it in \zk\ and \spk}*)\\
&\hspace{1ex}\out{\ch}{\zk((\doctorPseudo, \doctorID, \noncedr),
                \drcred(\doctorPseudo, \doctorID, \noncedr))}.\\

&\hspace{1ex}\readin{\ch}{(\xpatientAuth,\xpatientProof)}.   \\
&\hspace{1ex}\out{\chc}{(\xpatientAuth,\xpatientProof)}.  \\
&\hspace{1ex}\tlet \xpatientCred=\getpublic(\xpatientAuth) \letin \\
&\hspace{1ex}\tlet (\xpatientCommit,=\xpatientCred)=\getpublic(\xpatientProof) \letin   \\
&\hspace{1ex}\tif \patientAuthVer(\xpatientAuth,\xpatientCred)=\true \then \\
&\hspace{1ex}\tif \patientProofVer(\xpatientProof,(\xpatientCommit,\xpatientCred))=\true \then \\

&\hspace{1ex}\out{\chc}{\pa}.\\
&\hspace{1ex}\nu \doctorOpenInfo.\\
&\hspace{1ex}\out{\chc}{\doctorOpenInfo}.\\
&\hspace{1ex}\tlet \prescID=\hash(\pb,\xpatientCommit,
                       \comt(\doctorPseudo,\doctorOpenInfo)) \letin \\

&\hspace{1ex}\out{\privchdrpt}{(\spk((\doctorPseudo, \doctorOpenInfo, \doctorID, \noncedr),\\
&\hspace{15.5ex}  (\comt(\doctorPseudo,\doctorOpenInfo),
                  \drcred(\doctorPseudo, \doctorID, \noncedr)),\\
&\hspace{15.5ex}  (\pb,\prescID,
                   \comt(\doctorPseudo,\doctorOpenInfo),\xpatientCommit)),\\
&\hspace{12.5ex}            \doctorOpenInfo)}.\\
&\hspace{1ex}\out{\chc}{(\spk((\doctorPseudo, \doctorOpenInfo, \doctorID, \noncedr),\\
&\hspace{15ex}  (\comt(\doctorPseudo,\doctorOpenInfo),
                  \drcred(\doctorPseudo, \doctorID, \noncedr)),\\
&\hspace{15ex}  (\pa,\hash(\pa,\xpatientCommit,
                       \comt(\doctorPseudo,\doctorOpenInfo)),\\
&\hspace{16ex}         \comt(\doctorPseudo,\doctorOpenInfo),\xpatientCommit)),\\
&\hspace{11.5ex}            \doctorOpenInfo)}))
\end{array}
\end{array}
\end{math}
\end{specification}
\caption{The doctor process $\ProDr'$ (using untappable channels).}
\label{fig:prodradd}
\end{figure}

Messages over untappable channels are assumed to
be perfectly secret to the adversary (for example, the channels assumed in~\cite{DKR09,HNB11}).
Thus, the security and classical privacy properties, which are satisfied
in the model with public channels only, are also satisfied when
replacing some public channels with untappable channels.
Similar to other proposed assumptions, the assumption of untappable
channels is at the model level. This is a strong assumption, as the
implementation of an untappable channel is difficult~\cite{HNB11}.
However, as this assumption is often used in literature to achieve
privacy in the face of bribery and coercion
(e.g.~\cite{Okamoto96,AS02,HNB11}), we feel that its use here is
justifiable.

However, even with the above assumptions the DLV08 protocol does not
satisfy \docrfindep.
The proof first shows that $\ProDr'$ is not sufficient for proving this
with ProVerif. Then we prove (analogous to the proof in
Section~\ref{ssec:docrf}) that there is no alternative process $\ProDr'$
which satisfies Definition~\ref{def:drfi}. Intuitively, all information
sent over untappable channels is received by pharmacists and can be
genuinely revealed to the adversary (no lying
assumption). Hence, the links between a doctor, his
nonces, his commitment, his credential and his prescription can still be
revealed when the
doctor is bribed/coerced to reveal those nonces.

\begin{theorem}[\docrfindep]\label{theorem:docrfindep}
The DLV08 protocol fails to satisfy \docrfindep.
\end{theorem}
Formal proof the theorem can be found in Appendix~\ref{sec:proof2}.

Intuitively, a bribed doctor is linked to the nonces he sent to the
adversary. The nonces are linked to the doctor's prescription in a
prescription proof. A doctor's prescription proof is sent over
untappable channels first to a patient and later from the patient to a
pharmacist. Malicious pharmacists reveal the prescription proof to the
adversary. If a bribed doctor lied about his prescription, the
adversary can detect it by checking the doctor's corresponding
prescription proof revealed by the pharmacist. The untappable channel
assumption enables the protocol to satisfy \docrf\ but not \docrfindep\
because untappable channels enable a bribed doctor to hide his
prescription proof and thus allow the doctor to lie about his
prescription, however the pharmacist gives the prescription proof away,
from which the adversary
can detect whether the doctor lied about the prescription.

\section{Conclusions}
In this paper, we have studied security and privacy properties, particular enforced privacy, in the e-health domain.
We identified the requirement that doctor privacy should be enforced to
prevent doctor bribery by, for example, the pharmaceutical industry. To capture
this requirement, we first formalised the classical privacy property, i.e.,
\docunlink, and its enforced privacy counterpart, i.e., \docrf. The
cooperation between the bribed doctor and the adversary is formalised in
the same way as in receipt-freeness in e-voting. However, the
formalisation of \docrf\ differs from receipt-freeness, due
to the domain requirement that only part of the doctor's process needs
to share information with the adversary.

Next, we noted that e-health systems involve not necessarily trusted
third parties, such as pharmacists. Such parties should not be able
to assist an adversary in breaking doctor privacy. To capture this
requirement, we formally defined \docindep. Moreover, this new requirement
must hold, even if the doctor is forced to help the adversary. To
capture that, we formally defined \docrfindep.

These formalisations were
validated in a case study of the DLV08 protocol. The protocol was
modelled in the applied pi calculus and verified with the help of the
ProVerif tool. In addition to the (enforced) doctor privacy properties,
we also analysed secrecy, authentication, anonymity and untraceability for
both patients and doctors. Ambiguities in the original description of the protocol
which may lead to flaws were found and addressed. 

We notice that the property \docrfindep\ is not satisfied in the 
case study protocol, and we were not able to propose a reasonable fix for it. 
Thus, it is interesting for us to design a new protocol
to satisfy such strong property in the future. Furthermore, when considering dishonest 
users, we did not consider one dishonest user taking multiple roles. Thus, it would be
interesting to analyse the security and privacy properties with respect to 
dishonest users taking various combination of roles.

\bibliographystyle{alpha}
\bibliography{ehealth}
\appendix
\appendixpage
\addappheadtotoc

\section{Proof of Theorem~\ref{theorem:docrf}}\label{sec:proof1}

{\bf Theorem 1} (\docrf).
{\it The DLV08 protocol fails to satisfy \docrf\ under both the standard
assumption \textbf{s4} (a doctor has the same credential in every session),
and also under assumption \textbf{s4'} (a doctor generates a new credential
for each session).}

It is obvious that the DLV08 protocol fails to satisfy \docrf\ under assumption 
\textbf{s4} (a doctor has the same credential in every session), 
since DLV08 does not even satisfy \docunlink\ with assumption \textbf{s4}. The reasoning is as follows: since 
the adversary can link a prescription to a doctor without additional 
information from the bribed doctor, he can also link a prescription to a doctor
when he has additional information from the bribed doctor.
Therefore, the adversary can always tell whether a bribed doctor lied. 

Next we prove that the DLV08 protocol fails to satisfy \docrf\ under 
assumption \textbf{s4'} (a doctor generates a new credential
for each session).
That is to prove that there exists no indistinguishable process in which the doctor
lies to the adversary. To do so, we assume that there exists such a process
$\ProDr'$ which satisfies the definition of \docrf, and then derive some
contradiction. In generic terms, the proof runs as follows: a bribed
doctor reveals the nonces used in the commitment and the credential to
the adversary. This allows the adversary to link a bribed doctor to his
commitment and credential. In the prescription proof, a prescription is
linked to a doctor's commitment and credential. Suppose there exists a
process $\ProDr'$ in which the doctor lies to the adversary that he
prescribed $\pa$, while the adversary observes that the commitment or
the credential is linked to $\pb$. The adversary can detect that the
doctor has lied.

\proof 
Assume there exist process $\init_{\thedoctor}'$ and $P_{\thedoctor}'$, so that 
the two equivalences in the definition of \docrf\ are satisfied, i.e., $\exists$ $\init_{\thedoctor}'$ and $P_{\thedoctor}'$ satisfying
\[
\begin{array}{lrl}
1.\ &&\contexthealth{
\big(\init_{\thedoctor}'.
 (!\ProDr\substitution{\drA}{\doctorIDv}
\mid 
\ProDr')\big)
\mid \\
&&\hspace{4ex} 
\big(\init_{\thedoctor}\substitution{\drB}{\doctorIDv}.
 (! \ProDr\substitution{\drB}{\doctorIDv}
\mid 
  \vProDr\dsub{\drB}{\doctorIDv}{\pa}{\prescTextv})\big)}\\
&\eq
&\contexthealth{
\big((\init_{\thedoctor}\substitution{\drA}{\doctorIDv})^{\chc}.
(!\ProDr\substitution{\drA}{\doctorIDv}
\mid 
(\vProDr\dsub{\drA}{\doctorIDv}{\pa}{\prescTextv})^{\chc})\big)
\mid\\
&&\hspace{4ex} 
\big(\init_{\thedoctor}\substitution{\drB}{\doctorIDv}.
(!\ProDr\substitution{\drB}{\doctorIDv}
\mid
 \vProDr\dsub{\drB}{\doctorIDv}{\pb}{\prescTextv})\big)};  \vspace{2mm} \text{and}\\ 
2.\ &&
\contexthealth{\big(
(\init_{\thedoctor}'.(!\ProDr\substitution{\drA}{\doctorIDv}\mid \ProDr'))^{\backslash \out{\chc}{\cdot}}\big)
\mid \\
&&\hspace{3.9ex} 
\big(\init_{\thedoctor}\substitution{\drB}{\doctorIDv}.
 (! \ProDr\substitution{\drB}{\doctorIDv}
\mid 
  \vProDr\dsub{\drB}{\doctorIDv}{\pa}{\prescTextv})\big)}\\
&\eq 
&
\contexthealth{\big(
\init_{\thedoctor}\substitution{\drA}{\doctorIDv}.
(!\ProDr\substitution{\drA}{\doctorIDv}\mid\vProDr\dsub{\drA}{\doctorIDv}{\pb}{\prescTextv})\big)
\mid \\
&&\hspace{3.9ex} 
\big(\init_{\thedoctor}\substitution{\drB}{\doctorIDv}.
 (! \ProDr\substitution{\drB}{\doctorIDv}
\mid 
  \vProDr\dsub{\drB}{\doctorIDv}{\pa}{\prescTextv})\big)}.
\end{array}
\]

According to the definition of labelled bisimilarity (Definition~\ref{def:labelled_bisimilarity}),
if process $A$ can reach $A'$ ($A \lduc{\alpha*} A'$) and $A\eq B$, then 
$B$ can reach $B'$ ($B \lduc{\alpha*} B'$) and $A'\eq B'$. Vice versa.
Note that we use $\lduc{\alpha*}$ to denote one or more internal and/or labelled 
reductions.

According to Definition~\ref{def:labelled_bisimilarity}, if $A'\eq B'$ then $A' \seq B'$.

According to the definition of static equivalence (Definition~\ref{def:static_equivalence}), 
if two processes are static equivalent $A' \seq B'$, then $\frm{A'}\seq \frm{B'}$. Thus we have that
$\forall M, N$, $(M =_E N)\frm{A'}$ iff $(M =_E N)\frm{B'}$. 

Let $A$ be the right-hand side of the first equivalence and $B$ be the left-hand side, i.e.,
\[
\begin{array}{rcl} 
A &=& \contexthealth{
\big((\init_{\thedoctor}\substitution{\drA}{\doctorIDv})^{\chc}.
(!\ProDr\substitution{\drA}{\doctorIDv} \mid 
(\vProDr\dsub{\drA}{\doctorIDv}{\pa}{\prescTextv})^{\chc})\big)\mid\\
&&\hspace{4ex}  \big(\init_{\thedoctor}\substitution{\drB}{\doctorIDv}.
(!\ProDr\substitution{\drB}{\doctorIDv} \mid
 \vProDr\dsub{\drB}{\doctorIDv}{\pb}{\prescTextv})\big)}\\
 
 B &=& \contexthealth{\big(\init_{\thedoctor}'. 
(!\ProDr\substitution{\drA}{\doctorIDv} \mid \ProDr')\big) \mid \\ 
&&\hspace{4ex} \big(\init_{\thedoctor}\substitution{\drB}{\doctorIDv}.
 (! \ProDr\substitution{\drB}{\doctorIDv} \mid 
  \vProDr\dsub{\drB}{\doctorIDv}{\pa}{\prescTextv})\big)}.
 
\end{array}
\]
 
On the right-hand side of the first equivalence (process $A$), there exists an output 
of a prescription proof $\prescProofr$ (together with the open information of the doctor commitment $\doctorOpenInfor$), over public channels, 
from the process $(\vProDr\dsub{\drA}{\doctorIDv}{\pa}{\prescTextv})^{\chc}$ initiated by doctor $\drA$.
Formally,
$A \lduc{\alpha*} A_i = \contexti{\out{ch}{(\prescProofr,\doctorOpenInfor)}} \structeq 
\nu x.(\contexti{\out{ch}{x}} \mid \substitution{(\prescProofr,\doctorOpenInfor)}{x})\\
\lduc{\nu x. \out{ch}{x}} \contexti{0}\mid \substitution{(\prescProofr,\doctorOpenInfor)}{x}$. Let $A''=\contexti{0}$, we have  
$A\lduc{\alpha*} \lduc{\nu x. \out{ch}{x}} A'' \mid \substitution{(\prescProofr,\doctorOpenInfor)}{x}$.
Let $A'=A''\mid \substitution{(\prescProofr,\doctorOpenInfor)}{x}$, we have $A\lduc{\alpha*} \lduc{\nu x. \out{ch}{x}} A'$.

Since $A\eq B$, we have that $B\lduc{\alpha*} \lduc{\nu x. \out{ch}{x}} B'$ and $A'\eq B'$.
Hence, $A'\seq B'$ and thus $\frm{A'}\seq \frm{B'}$. 
Since $\frm{A'}=\frm{A''}\mid\substitution{(\prescProofr,\doctorOpenInfor)}{x}$,
the adversary can obtain the prescription $\pa$: $\pa=\frt{\getmsg(\frt{x})}$, where function $\frtname$ returns the first 
element of a tuple or a pair. Since $\frm{A'}\seq \frm{B'}$, we should have the same relation $\pa=\frt{\getmsg(\frt{x})}$ in 
$\frm{B'}$.

Intuitively, since on the right-hand side, the adversary can obtain the prescription $\pa$ from $\prescProofr$, due to
\(
(\pa, \prescID^{r}, \doctorCommitr, \xpatientCommitr)=\getmsg(\prescProofr),
\)
on the left-hand side of the first equivalence, there should also 
exist an output of a prescription proof $\prescProofl$ over public channels,
from which the adversary can obtain a prescription $\pa$, following the same relation:
\[
(\pa, \prescID^{l}, \doctorCommitl, \xpatientCommitl)=\getmsg(\prescProofl).
\]

Next, we prove that the corresponding prescription proof $\prescProofl$ is indeed the prescription proof in the doctor sub-process
$\init_{\thedoctor}'$ or
$\ProDr'$ in process $B$, rather than other sub-processes. Formally, the action $\lduc{\nu x. \out{ch}{x}}$ in process $B$ happens in sub-process $\init_{\thedoctor}'$ or $\ProDr'$.

On the right-hand side (process $A$),
the doctor pseudonym $\doctorPseudor$ and the nonce for doctor commitment
$\doctorOpenInfor$ and the nonce for doctor credential
$\noncedrr$ (used for assumption {\bf s4'}) 
are revealed to the adversary on $\chc$ channel.
Formally, 
\[
\begin{array}{rll}
A&\lduc{\alpha*}\lduc{\nu x_1. \out{chc}{x_1}}&A_1\mid\substitution{\doctorPseudor}{x_1}\\
&\lduc{\alpha*}\lduc{\nu x_2. \out{chc}{x_2}}&A_2\mid\substitution{\doctorPseudor}{x_1}\mid\substitution{\doctorOpenInfor}{x_2}\\
&\lduc{\alpha*}\lduc{\nu x_3. \out{chc}{x_3}}&A_3\mid\substitution{\doctorPseudor}{x_1}\mid\substitution{\doctorOpenInfor}{x_2}\mid\substitution{\noncedrr}{x_3}\\
&\lduc{\alpha*}\lduc{\nu x. \out{ch}{x}} & A_4\mid\substitution{\doctorPseudor}{x_1}\mid\substitution{\doctorOpenInfor}{x_2}\mid\substitution{\noncedrr}{x_3}\mid \substitution{(\prescProofr,\doctorOpenInfor)}{x}\structeq A'.
\end{array}
\]
Since $A\eq B$, we have that 
\[
\begin{array}{rll}
B&\lduc{\alpha*}\lduc{\nu x_1. \out{chc}{x_1}}&B_1\mid\substitution{\doctorPseudol}{x_1}\\
&\lduc{\alpha*}\lduc{\nu x_2. \out{chc}{x_2}}&B_2\mid\substitution{\doctorPseudol}{x_1}\mid\substitution{\doctorOpenInfol}{x_2}\\
&\lduc{\alpha*}\lduc{\nu x_3. \out{chc}{x_3}}&B_3\mid\substitution{\doctorPseudol}{x_1}\mid\substitution{\doctorOpenInfol}{x_2}\mid\substitution{\noncedrl}{x_3}\\
&\lduc{\alpha*}\lduc{\nu x. \out{ch}{x}} &B_4\mid\substitution{\doctorPseudol}{x_1}\mid\substitution{\doctorOpenInfol}{x_2}\mid\substitution{\noncedrl}{x_3}\mid \substitution{(\prescProofl,\doctorOpenInfol)}{x} \structeq B'.
\end{array}
\]
That is, on the left-hand side of the first equivalence, to be equivalent to 
the right-hand side, there also exist sub-processes which output messages on 
$\chc$ channel. Such sub-processes can only be 
$\init_{\thedoctor}'$ and 
$\ProDr'$, because there is no output on $\chc$ 
in other sub-processes in the left-hand side process (process $B$).

In $\frm{A'}=\frm{A_4}\mid\substitution{\doctorPseudor}{x_1}\mid\substitution{\doctorOpenInfor}{x_2}\mid\substitution{\noncedrr}{x_3}\mid \substitution{(\prescProofr,\doctorOpenInfor)}{x}$, we have the following relation between two terms,
\[
\begin{array}{rcl}
\frt{x}&=&\spk((x_1,x_2,\drA,x_3),\\
&&\hspace{4ex}(\comt(x_1,x_2),\drcred(x_1,\drA,x_3)),\\
&&\hspace{4ex}(\pa,\prescID^{r}, \comt(x_1,x_2),\xpatientCommitr)).
 \end{array}
\]
Since $A'\seq B'$, we should have the same relation in $\frm{B'}$.
\[
\begin{array}{rcl}
\frt{x}&=&\spk((x_1,x_2,\drA,x_3),\\
&&\hspace{4ex}(\comt(x_1,x_2),\drcred(x_1,\drA,x_3)),\\
&&\hspace{4ex}(\pa,\prescID^{l}, \comt(x_1,x_2),\xpatientCommitl)).
 \end{array}
\]

On the left-hand side (process $B$), 
the terms sent by process $\init_{\thedoctor}'$ and/or $\ProDr'$ over $\chc$ -- the doctor pseudonym and the nonces $\doctorPseudol$, $\doctorOpenInfol$ and $\noncedrl$
(corresponding to $\doctorPseudor$, $\doctorOpenInfor$ and $\noncedrr$ on the right-hand side), are essential to compute $\prescProofl$. Thus, 
process $\init_{\thedoctor}'$ and/or $\ProDr'$ is able to compute and thus output the prescription proof $\prescProofl$, given that the coerced doctor $\drA$ has the knowledge of $\pa$, by applying the following function:
\[
\begin{array}{rcl}
\prescProofl&=&\spk((\doctorPseudol, \doctorOpenInfol, \drA, \noncedrl),\\
&&\hspace{4ex}      (\comt(\doctorPseudol,\doctorOpenInfol),
                    \drcred(\doctorPseudol, \drA, \noncedrl)),\\
&&\hspace{4ex}      (\pa,\prescID^{l}, 
                   \comt(\doctorPseudol,\doctorOpenInfol),\xpatientCommitl).
                   \end{array}
\] 

Now we have proved that the action of revealing $(\prescProofl, \doctorOpenInfol)$ ($\lduc{\nu x. \out{ch}{x}}$) can be taken in process $\init_{\thedoctor}'$ and/or $\ProDr'$ in process $B$.
Next we show that sub-processes except $\init_{\thedoctor}'$ and 
$\ProDr'$ in $B$, cannot take the action of revealing $(\prescProofl, \doctorOpenInfol)$, given the process $\init_{\thedoctor}'$ and 
$\ProDr'$ does not replay the message $(\prescProofl, \doctorOpenInfol)$ in an honest doctor process.

By examining process $B$, the sub-processes which send out a pair, the first element of which is a signed proof of knowledge, can only be doctor processes and the MPA processes, i.e., sub-processes that may send out a message $x$ potentially satisfying $\pa=\frt{\getmsg(\frt{x})}$ can only be doctor processes (at {\bf d10}) or MPA processes (at {\bf m30}). 

\begin{itemize}
\item Case 1:
considering that message $x$ should also satisfy $\open(\third{\getmsg(\frt{x})},\snd{x})=\doctorPseudol$, the processes revealing $x$ can only be 
doctor processes, because the second element in the message sent out at line {\bf m30} of an MPA process is a zero-knowledge proof, and thus cannot be used as a nonce to open a commitment \third{\getmsg(\frt{x})}.
\item Case 2:
considering that the message $x$ should satisfy $\frt{x}=\spk((x_1,x_2,\drA,x_3),\linebreak(\comt(x_1,x_2),$$\drcred(x_1,\drA,x_3)),$$(\pa,\prescID^{l}, $$\comt(x_1,x_2),\xpatientCommitl))$, where $x_1$ is the doctor pseudonym, $x_2$ and $x_3$ are nonces, and the adversary receive $x_1$, $x_2$, $x_3$ from $\chc$ channel, doctor sub-processes (except $\init_{\thedoctor}'$ and $\ProDr'$) cannot reveal the message $x$. 
Because these doctor sub-processes model honest doctor sessions, and thus use their own generated nonces to compute the signed proofs of knowledge (at line {\bf t23}). Such nonces are not sent to the adversary over $\chc$ channel, since these doctor processes are not bribed or coerced. Thus, the signed proofs of knowledge generated by these honest doctor prepossess cannot be the first element of the message $x$, unless the process $\init_{\thedoctor}'$ and/or $\ProDr'$ reuses one of the signed proofs of knowledge. 
\end{itemize}

In the case that the process $\init_{\thedoctor}'$ and/or $\ProDr'$ replay the message $(\prescProofl, \doctorOpenInfol)$ of an honest doctor process, $x_1$ needs to be the corresponding doctor pseudonym, and $x_2$ and $x_3$ need to be the corresponding nonces for the reused signed proof of knowledge the message. Otherwise, $(\prescProofl, \doctorOpenInfol)$ will be detected as a fake message. Thus, the message indeed represents the actual prescription in process $\init_{\thedoctor}'$ and/or $\ProDr'$. Although the action of revealing message $(\prescProofl, \doctorOpenInfol)$ may be taken in an honest doctor process, the same action will be eventually taken in process $\init_{\thedoctor}'$ or $\ProDr'$. 

Therefore, the process $\init_{\thedoctor}'$ and/or
$\ProDr'$ indeed outputs the prescription proof
$\prescProofl$ on the left-hand side of the first equivalence, i.e., $\init_{\thedoctor}'$/$\ProDr'\lduc{\alpha*}\lduc{\nu x. \out{ch}{x}} P\mid\substitution{(\prescProofl,\doctorOpenInfol)}{x}$. 

Let $C$ be the left-hand side of the second equivalence, and $D$ be the right-hand side.
\[
\begin{array}{lrl}

C&=&\contexthealth{\big(
(\init_{\thedoctor}'.(!\ProDr\substitution{\drA}{\doctorIDv}\mid \ProDr'))^{\backslash \out{\chc}{\cdot}}\big)
\mid \\
&&\hspace{3.9ex} 
\big(\init_{\thedoctor}\substitution{\drB}{\doctorIDv}.
 (! \ProDr\substitution{\drB}{\doctorIDv}
\mid 
  \vProDr\dsub{\drB}{\doctorIDv}{\pa}{\prescTextv})\big)}\\
D&=&
\contexthealth{\big(
\init_{\thedoctor}\substitution{\drA}{\doctorIDv}.
(!\ProDr\substitution{\drA}{\doctorIDv}\mid\vProDr\dsub{\drA}{\doctorIDv}{\pb}{\prescTextv})\big)
\mid \\
&&\hspace{3.9ex} 
\big(\init_{\thedoctor}\substitution{\drB}{\doctorIDv}.
 (! \ProDr\substitution{\drB}{\doctorIDv}
\mid 
  \vProDr\dsub{\drB}{\doctorIDv}{\pa}{\prescTextv})\big)}
  \end{array}
  \]
In process $C$, the sub-process $\init_{\thedoctor}.
(!\ProDr\substitution{\drA}{\doctorIDv}\mid\ProDr')^{\backslash \out{\chc}{\cdot}}\defi \nu \chc. (\init_{\thedoctor}'.(!\ProDr\substitution{\drA}{\doctorIDv}\mid\ProDr')\mid!\readin{\chc}{y})$, according to the definition of $\prs^{\backslash \out{\chc}{\cdot}}$.
Since $\init_{\thedoctor}'/\ProDr'$ may take the action $\lduc{\nu x. \out{ch}{x}}$ where $\substitution{(\prescProofl,\doctorOpenInfol)}{x}$, we have $\init_{\thedoctor}'.
(!\ProDr\substitution{\drA}{\doctorIDv}\mid\ProDr') \lduc{\alpha*}\lduc{\nu x. \out{ch}{x}} P'\mid\substitution{(\prescProofl,\doctorOpenInfol)}{x}$. By filling it in the context $\nu \chc.(\hole\mid!\readin{\chc}{y})$, we have $\nu \chc. ((\init_{\thedoctor}'.\linebreak
(!\ProDr\substitution{\drA}{\doctorIDv}\mid\ProDr')\mid!\readin{\chc}{y})\lduc{\alpha*} \lduc{\nu x. \out{ch}{x}}\nu \chc. ((P'\mid\substitution{(\prescProofl,\doctorOpenInfol)}{x})\mid!\readin{\chc}{y})$,
and thus $(\init_{\thedoctor}'.
(!\ProDr\substitution{\drA}{\doctorIDv}\mid\ProDr')^{\backslash \out{\chc}{\cdot}}\lduc{\alpha*}\lduc{\nu x. \out{ch}{x}}\nu \chc. (P'\mid\substitution{(\prescProofl,\doctorOpenInfol)}{x}\mid!\readin{\chc}{y})$. 
The sub-process $\vProDr\dsub{\drB}{\doctorIDv}{\pa}{\prescTextv})$ also outputs a signed proof of knowledge from which the adversary obtains $\pa$, i.e., $\vProDr\dsub{\drB}{\doctorIDv}{\pa}{\prescTextv})\lduc{\alpha*} \lduc{\nu x. \out{ch}{x}} C_1\mid\substitution{(\prescProof,\doctorOpenInfo)}{x_1}$ and $\pa=\frt{\getmsg(\frt{x_1})}$.
Thus, in process $C$, there are two outputs of a signed proof of knowledge, from which the adversary obtains $\pa$. Other signed proofs of knowledge will not lead to $\pa$ or $\pb$, as the prescription in process $! \ProDr\substitution{\drA}{\doctorIDv}$ and $! \ProDr\substitution{\drB}{\doctorIDv}$ are freshly generated.

In process $D$, the sub-process $\vProDr\dsub{\drA}{\doctorIDv}{\pb}{\prescTextv}$ outputs a prescription proof from which the adversary knows $\pb$, sub-process $\vProDr\dsub{\drB}{\doctorIDv}{\pa}{\prescTextv}$ outputs a prescription proof from which the adversary knows $\pa$, the prescriptions from the sub-process $!\ProDr\substitution{\drA}{\doctorIDv}$ and $!\ProDr\substitution{\drA}{\doctorIDv}$ are fresh names and thus cannot be $\pa$ or $\pb$.   
The adversary can detect that the process $C$ and $D$ are not equivalent: in process $C$, the adversary obtains two $\pa$, and in process $D$, the adversary obtains one $\pa$ and one $\pb$. This contradicts the assumption that $C\eq D$.
\qed

\section{Proof of Theorem~\ref{theorem:docrfindep}}\label{sec:proof2}

{\bf Theorem 2} (\docrfindep).
{\it The DLV08 protocol fails to satisfy \docrfindep.}

\proof 
Assume the DLV08 protocol satisfies \docrfindep. That is, $\exists$ $\init_{\thedoctor}'$ and $P_{\thedoctor}'$ satisfying
the following two equivalences in the definition of \docrfindep\ (Definition~\ref{def:drfi}).
\[
\begin{array}{lrl}
1.\ &&\contexthealth{!\Role_{\thepharm}^{\chc}\mid
\big(\init_{\thedoctor}'.
(!\ProDr\substitution{\drA}{\doctorIDv}
\mid 
\ProDr')\big)\mid\\
&&\hspace{11ex} \big(\init_{\thedoctor}\substitution{\drB}{\doctorIDv}.
(!\ProDr\substitution{\drB}{\doctorIDv}
\mid
 \vProDr\dsub{\drB}{\doctorIDv}{\pa}{\prescTextv})\big)}\\
&\eq
&\contexthealth{!\Role_{\thepharm}^{\chc}\mid
\big((\init_{\thedoctor}\substitution{\drA}{\doctorIDv})^{\chc}.\\
&&\hspace{12.5ex}(!\ProDr\substitution{\drA}{\doctorIDv}\mid
(\vProDr\dsub{\drA}{\doctorIDv}{\pa}{\prescTextv})^{\chc})\big)\mid\\
&&\hspace{11ex}  \big(\init_{\thedoctor}\substitution{\drB}{\doctorIDv}.
(!\ProDr\substitution{\drB}{\doctorIDv}
\mid
\vProDr\dsub{\drB}{\doctorIDv}{\pb}{\prescTextv})\big)}; \vspace{2mm} \text{and}\\
2.\ &&
\contexthealth{!\Role_{i}^{\chc}\mid
\big((\init_{\thedoctor}'.
(!\ProDr\substitution{\drA}{\doctorIDv}\mid\ProDr'))^{\backslash \out{\chc}{\cdot}}\big)\mid\\
&&\hspace{11ex} \big(\init_{\thedoctor}\substitution{\drB}{\doctorIDv}.
(!\ProDr\substitution{\drB}{\doctorIDv}
\mid
 \vProDr\dsub{\drB}{\doctorIDv}{\pa}{\prescTextv})\big)}\\
& \eq &
\contexthealth{!\Role_{i}^{\chc}\mid
\big(\init_{\thedoctor}\substitution{\drA}{\doctorIDv}.
(!\ProDr\substitution{\drA}{\doctorIDv}\mid\vProDr\dsub{\drA}{\doctorIDv}{\pb}{\prescTextv})\big)\mid\\
&&\hspace{11ex}  \big(\init_{\thedoctor}\substitution{\drB}{\doctorIDv}.
(!\ProDr\substitution{\drB}{\doctorIDv}
\mid
\vProDr\dsub{\drB}{\doctorIDv}{\pa}{\prescTextv})\big)}.
\end{array}
\]
We prove that this assumption leads to contradictions.

Similar to the proof of Theorem~\ref{theorem:docrf}, according to the definition of labelled bisimilarity (Definition~\ref{def:labelled_bisimilarity}) and static equivalence (Definition~\ref{def:static_equivalence}), 
Given $A\eq B$, if $A\lduc{\alpha*}A'$ and $M =_E N\frm{A'}$, then $B\lduc{\alpha*}B'$ and $M =_E N\frm{B'}$. Vice versa. 

Let $A$ be the right-hand side of the first equivalence, $B$ be the left-hand side.
\[
\begin{array}{lrl}
A&=
&\contexthealth{!\Role_{\thepharm}^{\chc}\mid
\big((\init_{\thedoctor}\substitution{\drA}{\doctorIDv})^{\chc}.\\
&&\hspace{12.5ex}(!\ProDr\substitution{\drA}{\doctorIDv}\mid
(\vProDr\dsub{\drA}{\doctorIDv}{\pa}{\prescTextv})^{\chc})\big)\mid\\
&&\hspace{11ex}  \big(\init_{\thedoctor}\substitution{\drB}{\doctorIDv}.
(!\ProDr\substitution{\drB}{\doctorIDv}
\mid
\vProDr\dsub{\drB}{\doctorIDv}{\pb}{\prescTextv})\big)}\\
B &=&\contexthealth{!\Role_{\thepharm}^{\chc}\mid
\big(\init_{\thedoctor}'.
(!\ProDr\substitution{\drA}{\doctorIDv}
\mid 
\ProDr')\big)\mid\\
&&\hspace{11ex} \big(\init_{\thedoctor}\substitution{\drB}{\doctorIDv}.
(!\ProDr\substitution{\drB}{\doctorIDv}
\mid
 \vProDr\dsub{\drB}{\doctorIDv}{\pa}{\prescTextv})\big)}
\end{array}
\]

In process $A$, the doctor $\drA$ 
computed a signed proof of knowledge $\prescProofr$
in the sub-process $(\vProDr\dsub{\drA}{\doctorIDv}{\pa}{\prescTextv})^{\chc}$.
The signed proof of knowledge is sent to a patient over private channel. In addition, the signed proof of knowledge is also sent to the adversary over $\chc$ together with a nonce (the message sent over $\chc$ is $(\prescProofr,\doctorOpenInfor)$). On receiving the signed proof of knowledge, the patient sends it together with other information to a pharmacist over a private channel. On receiving the message from the patient over private channel, the pharmacist forwards the message to the adversary over $\chc$ (the message sent over $\chc$ is $(\prescProofr,\patientspkr,\vc_1^r,\vc_2^r,\vc_3^r,\vc^{r'}_3,\vc_4^r,\C_5^r)$). 
Another sub-process $\vProDr\dsub{\drB}{\doctorIDv}{\pb}{\prescTextv}$ also generates a signed proof of knowledge $\prescProofzr$. This signed proof of knowledge 
is sent to a patient in a message over private channel (but it is not sent to the adversary over $\chc$, as this sub-process is not bribed or coerced), and then sent to a pharmacist in another message via private channel. Finally, the pharmacist, who receives the message containing $\prescProofzr$, sends the message to the adversary over channel $\chc$ (the message sent to $\chc$ is $(\prescProofzr,\patientspkzr,\vc_1^{zr},\linebreak\vc_2^{zr},\vc_3^{zr},\vc^{zr'}_3,\vc_4^{zr},\C_5^{zr})$).
Formally, there is a trace in process $A$ as follows.
\[
\begin{array}{rll}
A 
&\lduc{\alpha*}\lduc{\nu x. \out{\chc}{x}} &A_1\mid \substitution{(\prescProofr,\doctorOpenInfor)}{x}\\
&\lduc{\alpha*}\lduc{\nu y. \out{\chc}{y}}& A_2\mid \substitution{(\prescProofr,\doctorOpenInfor)}{x}\mid\\
&&\hspace{20px}\substitution{(\prescProofr,\patientspkr,\vc_1^r,\vc_2^r,\vc_3^r,\vc^{r'}_3,\vc_4^r,\C_5^r)}{y}\\
&\lduc{\alpha*}\lduc{\nu z. \out{\chc}{z}}& A_3\mid \substitution{(\prescProofr,\doctorOpenInfor)}{x}\mid\\
&&\hspace{20px}\substitution{(\prescProofr,\patientspkr,\vc_1^r,\vc_2^r,\vc_3^r,\vc^{r'}_3,\vc_4^r,\C_5^r)}{y}\mid\\
&&\hspace{20px}\substitution{(\prescProofzr,\patientspkzr,\vc_1^{zr},\vc_2^{zr},\vc_3^{zr},\vc^{zr'}_3,\vc_4^{zr},\C_5^{zr})}{z}
\end{array}
\]
Let $A'=A_3\mid\substitution{(\prescProofr,\doctorOpenInfor)}{x}\mid\substitution{(\prescProofr,\patientspkr,\vc_1^r,\vc_2^r,\vc_3^r,\vc^{r'}_3,\vc_4^r,\C_5^r)}{y}\mid\linebreak \substitution{(\prescProofzr,\patientspkzr,\vc_1^{zr},\vc_2^{zr},\vc_3^{zr},\vc^{zr'}_3,\vc_4^{zr},\C_5^{zr})}{z}$. We have $\pa=\frt{\getmsg(\frt{x})}=\frt{\getmsg(\frt{y})}$ and $\pb=\frt{\getmsg(\frt{z})}$ at frame $\frm{A'}$.

Since $A\eq B$, we should have that 
\[
\begin{array}{rll}
B
&\lduc{\alpha*}\lduc{\nu x. \out{\chc}{x}} &B_1\mid \substitution{(\prescProofl,\doctorOpenInfol)}{x}\\
&\lduc{\alpha*}\lduc{\nu y. \out{\chc}{y}}& B_2\mid \substitution{(\prescProofl,\doctorOpenInfol)}{x}\mid\\
&&\hspace{20px} \substitution{(\prescProofl,\patientspkl,\vc_1^l,\vc_2^l,\vc_3^l,\vc^{l'}_3,\vc_4^l,\C_5^l)}{y}\mid\\
&\lduc{\alpha*}\lduc{\nu z. \out{\chc}{z}}& B_3\mid \substitution{(\prescProofl,\doctorOpenInfol)}{x}\mid\\
&&\hspace{20px}\substitution{(\prescProofl,\patientspkl,\vc_1^l,\vc_2^l,\vc_3^l,\vc^{l'}_3,\vc_4^l,\C_5^l)}{y}\mid\\
&&\hspace{20px}\substitution{(\prescProofzl,\patientspkzl,\vc_1^{zl},\vc_2^{zl},\vc_3^{zl},\vc^{zl'}_3,\vc_4^{zl},\C_5^{zl})}{z}.
\end{array}
\]
Let $B'=B_3\mid \substitution{(\prescProofl,\doctorOpenInfol)}{x}\mid\substitution{(\prescProofl,\patientspkl,\vc_1^l,\vc_2^l,\vc_3^l,\vc^{l'}_3,\vc_4^l,\C_5^l)}{y}\mid \substitution{(\prescProofzl,\patientspkzl,\vc_1^{zl},\vc_2^{zl},\vc_3^{zl},\vc^{zl'}_3,\vc_4^{zl},\C_5^{zl})}{z}$. We should have $A'\eq B'$, and thus, $\pa=\frt{\getmsg(\frt{x})}=\frt{\getmsg(\frt{y})}$ and $\pb=\frt{\getmsg(\frt{z})}$ at frame $\frm{B'}$.

In process $B$, the sub-process $\vProDr\dsub{\drB}{\doctorIDv}{\pa}{\prescTextv}$ generates a signed proof of knowledge $f$ and $\pa=\frt{\getmsg(f)}$. This signed proof of knowledge will be eventually revealed by a pharmacist, $B\lduc{\alpha*}\lduc{\nu h. \out{\chc}{h}}B_4\mid\substitution{(f,\patientspkf,\vc_1^{f},\vc_2^{f},\vc_3^{f},\vc^{f'}_3,\vc_4^{f},\C_5^{f})}{h}$ and $\pa=\frt{\getmsg(\frt{h})}=\frt{\getmsg(f)}$.

By examining process $B$, we observe that $y=h$. The reason is as follows: since $\pa=\frt{\getmsg(\frt{y})}$, sub-process $!\ProDr\substitution{\drB}{\doctorIDv}$ generate fresh prescriptions and thus cannot be $\pa$, therefore, $!\ProDr\substitution{\drB}{\doctorIDv}$ does not generate a prescription which eventually leads to the action of sending $y$. Thus the possible sub-process which generates the prescription $\pa$ and potentially leads to sending $y$ can only be $\init_{\thedoctor}'$, $\ProDr'$ or $\vProDr\dsub{\drB}{\doctorIDv}{\pa}{\prescTextv}$.
Assume $\init_{\thedoctor}'$ or $\ProDr'$ generates prescription $\pa$ which leads to the action of sending $y$ and $y\neq h$, then, the adversary obtains three $\pa$ in process $B$: one from $\pa=\frt{\getmsg(\frt{h})}$, one from $\pa=\frt{\getmsg(\frt{y})}$, and one from $\pa=\frt{\getmsg(\frt{x})}$. However, in process $A$, the adversary can only observe two $\pa$: one from $\pa=\frt{\getmsg(\frt{x})}$ and one from $\pa=\frt{\getmsg(\frt{y})}$. This contradicts the assumption that $A\eq B$. 
Therefore, the prescription $\pa$ which leads to the action of revealing $y$ is generated in sub-process $\vProDr\dsub{\drB}{\doctorIDv}{\pa}{\prescTextv}$, and thus $y=h$.

In addition, in process $B$, we observe that the prescription $\pb$ is generated in process $\init_{\thedoctor}'$ or $\ProDr'$. As sub-process $!\ProDr\substitution{\drB}{\doctorIDv}$ generates fresh prescriptions and thus cannot be $\pb$, and sub-process $\vProDr\dsub{\drB}{\doctorIDv}{\pa}{\prescTextv}$ generates $\pa$ ($\pa\neq\pb$), the only sub-process can generate $\pb$ is $\init_{\thedoctor}'$ or $\ProDr'$. The second equivalence also confirms this observation. 
Let $C$ be the left-hand side of the second equivalence, and $D$ be the right-hand side.
\[
\begin{array}{lrl}
C &=&
\contexthealth{!\Role_{i}^{\chc}\mid
\big((\init_{\thedoctor}'.
(!\ProDr\substitution{\drA}{\doctorIDv}\mid\ProDr'))^{\backslash \out{\chc}{\cdot}}\big)\mid\\
&&\hspace{11ex} \big(\init_{\thedoctor}\substitution{\drB}{\doctorIDv}.
(!\ProDr\substitution{\drB}{\doctorIDv}
\mid
 \vProDr\dsub{\drB}{\doctorIDv}{\pa}{\prescTextv})\big)}\\
D& = &
\contexthealth{!\Role_{i}^{\chc}\mid
\big(\init_{\thedoctor}\substitution{\drA}{\doctorIDv}.
(!\ProDr\substitution{\drA}{\doctorIDv}\mid\vProDr\dsub{\drA}{\doctorIDv}{\pb}{\prescTextv})\big)\mid\\
&&\hspace{11ex}  \big(\init_{\thedoctor}\substitution{\drB}{\doctorIDv}.
(!\ProDr\substitution{\drB}{\doctorIDv}
\mid
\vProDr\dsub{\drB}{\doctorIDv}{\pa}{\prescTextv})\big)}.
\end{array}
\]
In process $D$, the adversary can obtain one $\pa$ and one $\pb$. Since $C\eq D$, in process $D$, the adversary should also obtain one $\pa$ and one $\pb$. Since sub-process $\vProDr\dsub{\drB}{\doctorIDv}{\pa}{\prescTextv}$ generates $\pa$ and sub-process $!\ProDr\substitution{\drA}{\doctorIDv}$ and $!\ProDr\substitution{\drB}{\doctorIDv}$ cannot generates $\pb$, it must be process $\init_{\thedoctor}'$ or $\ProDr'$ who generates $\pb$. 

As the generated $\pb$ in process $\init_{\thedoctor}'$ or $\ProDr'$ is first sent to patient, then sent to a pharmacist and thus leads to a message sending over $\chc$. The message revealed by the pharmacist is $z$, because $\pb=\frt{\getmsg(\frt{z})}$, and on other process can generate $\pb$ in process $B$.

By examining process $B$, the only sub-process which can take the action of sending $x$ is process $\init_{\thedoctor}'$ or $\ProDr'$, as process $!\Role_{\thepharm}^{\chc}$ does not send a message $x$ which is a pair and thus satisfies $x=\pair(\frt{x}, \snd{x})$, and other processes does not involving using channel $\chc$. 

Intuitively, in process $B$, sub-process $\init_{\thedoctor}'$ and/or $\ProDr'$ sends $\pb$ to the patient which leads to the action of sending $z$; meanwhile, the sub-process $\init_{\thedoctor}'$ and/or $\ProDr'$ lies to the adversary that the singed proof of knowledge for prescription is $\prescProofl$ by sending $x$.

In process $A$, in addition to $\prescProofr$, process $(\vProDr\dsub{\drA}{\doctorIDv}{\pa}{\prescTextv})^{\chc}$ also sends other information over channel $\chc$. Formally,
\[
\begin{array}{rll}
A&\lduc{\alpha*}\lduc{\nu x_1. \out{chc}{x_1}}&A^1\mid\substitution{\doctorPseudor}{x_1}\\
&\lduc{\alpha*}\lduc{\nu x_2. \out{chc}{x_2}}&A^2\mid\substitution{\doctorPseudor}{x_1}\mid\substitution{\doctorOpenInfor}{x_2}\\
&\lduc{\alpha*}\lduc{\nu x_3. \out{chc}{x_3}}&A^3\mid\substitution{\doctorPseudor}{x_1}\mid\substitution{\doctorOpenInfor}{x_2}\mid\substitution{\noncedrr}{x_3}\\
&\lduc{\alpha*}\lduc{\nu x. \out{ch}{x}} & A_1\mid\substitution{\doctorPseudor}{x_1}\mid\substitution{\doctorOpenInfor}{x_2}\mid\substitution{\noncedrr}{x_3}\mid \substitution{(\prescProofr,\doctorOpenInfor)}{x},
\end{array}
\]
and $x_1$, $x_2$, $x_3$ and $x$ satisfy
\[
\begin{array}{rcl}
\frt{x}&=&\spk((x_1,x_2,\drA,x_3),\\
&&\hspace{4ex}(\comt(x_1,x_2),\drcred(x_1,\drA,x_3)),\\
&&\hspace{4ex}(\pa,\snd{\getmsg(\frt{x})}, \comt(x_1,x_2),\fourth(\getmsg(\frt{x})))).
 \end{array}
\]
Since $A\eq B$, we should have that 
\[
\begin{array}{rll}
B&\lduc{\alpha*}\lduc{\nu x_1. \out{chc}{x_1}}&B^1\mid\substitution{\doctorPseudol}{x_1}\\
&\lduc{\alpha*}\lduc{\nu x_2. \out{chc}{x_2}}&B^2\mid\substitution{\doctorPseudol}{x_1}\mid\substitution{\doctorOpenInfol}{x_2}\\
&\lduc{\alpha*}\lduc{\nu x_3. \out{chc}{x_3}}&B^3\mid\substitution{\doctorPseudol}{x_1}\mid\substitution{\doctorOpenInfol}{x_2}\mid\substitution{\noncedrl}{x_3}\\
&\lduc{\alpha*}\lduc{\nu x. \out{ch}{x}} &B_1\mid\substitution{\doctorPseudol}{x_1}\mid\substitution{\doctorOpenInfol}{x_2}\mid\substitution{\noncedrl}{x_3}\mid \substitution{(\prescProofl,\doctorOpenInfol)}{x},
\end{array}
\]
and the same relation holds between $x_1$, $x_2$, $x_3$ and $x$,
\[
\begin{array}{rcl}
\frt{x}&=&\spk((x_1,x_2,\drA,x_3),\\
&&\hspace{4ex}(\comt(x_1,x_2),\drcred(x_1,\drA,x_3)),\\
&&\hspace{4ex}(\pa,\snd{\getmsg(\frt{x})}, \comt(x_1,x_2),\fourth(\getmsg(\frt{x})))). \tag{eq1}
 \end{array}
\]

The sub-process $\init_{\thedoctor}'$ and $\ProDr'$ do not know $\doctorPseudol, \noncedrl$ since the two information satisfies secrecy. Thus sub-process $\init_{\thedoctor}'$ and/or $\ProDr'$ cannot send $x_1$ and $x_3$ over $\chc$ channel. Even assume the process 
$\init_{\thedoctor}'$ and $\ProDr'$ know the private information $\doctorPseudol, \doctorOpenInfol, \noncedrl$ for constructing $\prescProofl$,
since $\prescProofl$ is actually generated by $\vProDr\dsub{\drB}{\doctorIDv}{\pa}{\prescTextv}$, we have the following relation:
\[
\begin{array}{rcl}
f=\prescProofl&=&\spk((\doctorPseudol, \doctorOpenInfol, \drB, \noncedrl),\\
&&\hspace{4ex}      (\comt(\doctorPseudol,\doctorOpenInfol),
                    \drcred(\doctorPseudol, \drB, \noncedrl)),\\
&&\hspace{4ex}      (\pa,\prescID^{l}, 
                   \comt(\doctorPseudol,\doctorOpenInfol),\xpatientCommitl).
                   \end{array}
\] 
that is, 
\[
\begin{array}{rcl}
\frt{x}&=&\spk((x_1,x_2,\drB,x_3),\\
&&\hspace{4ex}(\comt(x_1,x_2),\drcred(x_1,\drB,x_3)),\\
&&\hspace{4ex}(\pa,\snd{\getmsg(\frt{x})}, \comt(x_1,x_2),\fourth(\getmsg(\frt{x})))).\tag{eq2}
 \end{array}
\]
and thus, the adversary can detect that $\prescProofl$ is generated by $\drB$ by telling the difference between (eq1) and (eq2).
This contradicts the assumption that $A\eq B$.
\qed

\end{document}